\documentclass[twoside]{article}
\usepackage{graphicx,amssymb,amsfonts,latexsym,amsmath,amsthm,times}
\usepackage[section]{placeins}
\usepackage{subfigure}
\usepackage{fancyhdr}
\usepackage{epsfig}
\usepackage{MathShorthand}
\usepackage[english]{babel}
\usepackage{setspace}
\usepackage{mathrsfs}
\usepackage{graphicx}
\usepackage{epstopdf}
\usepackage{titlesec}
\usepackage[font=small,labelfont=bf]{caption}
\usepackage[margin=1in]{geometry}

\titleformat{\section}{\large\bfseries}{\thesection}{1em}{}
\titleformat{\subsection}{\normalsize\bfseries}{\thesubsection}{1em}{}

\DeclareMathAlphabet{\mathpzc}{OT1}{pzc}{m}{it}

\renewcommand\thesection{\arabic{section}}
\renewcommand\thesubsection{\thesection.\arabic{subsection}}

\renewcommand{\thefootnote}{\fnsymbol{footnote}}
\input{tcilatex}
\renewcommand{\thefootnote}{\alph{footnote})}
\begin{document}

\title{Mean first passage time for a small rotating trap inside a \\ reflective
disk}
\date{\today }
\author{ \renewcommand{\thefootnote}{\alph{footnote}} J.~C.~Tzou\qquad
T.~Kolokolnikov \\
\\
{\small \emph{Department of Mathematics and Statistics, Dalhousie
University, Halifax, Nova Scotia, B3H 3J5 Canada}}\\
[3ex]}
\maketitle

\begin{abstract}
We compute the mean first passage time (MFPT) for a Brownian particle inside
a two-dimensional disk with reflective boundaries and a small interior trap
that is rotating at a constant angular velocity. The inherent symmetry of
the problem allows for a detailed analytic study of the situation. For a
given angular velocity, we determine the optimal radius of rotation that minimizes the
average MFPT over the disk. Several distinct regimes are observed, depending on the ratio
between the angular velocity $\omega$ and the trap size $\varepsilon$, and
several intricate transitions are analyzed using the tools of asymptotic
analysis and Fourier series. For $\omega \sim \mO(1)$, we compute a
critical value $\omega_c>0$ such that the optimal trap location is at the
origin whenever $\omega<\omega_c$, and is off the origin for $\omega>\omega_c
$. In the regime $1 \ll \omega \ll \mO(\varepsilon^{-1})$ the optimal trap path
approaches the boundary of the disk. However as $\omega$ is further
increased to $\mathcal{O}(\varepsilon^{-1})$, the optimal trap path ``jumps'' closer to
the origin. Finally for $\omega \gg \mO(\varepsilon^{-1})$ the optimal trap path
subdivides the disk into two regions of equal area. This simple geometry
provides a good test case for future studies of MFPT with more complex trap
motion.
\end{abstract}

\thispagestyle{plain}


\baselineskip=12pt

\vspace{16pt}


\baselineskip=16pt

\setcounter{equation}{0}

\singlespacing

\section{Introduction}

Numerous problems in nature can be formulated in terms of mean escape time
of Brownian particles in the presence of small traps. This is often referred
to as the mean first passage time (MFPT) or  the narrow escape problem, and
there is a large and growing literature on the subject; see for example
reviews \cite{redner2001guide, chevalier2011first, choufirst,
bray2013persistence, drewitz2012survival, schuss2012narrow, holcman2014time}
and references therein. Examples where first-passage problems arise
include:\ oxygen transport in muscle tissue \cite{titcombe2000asymptotic}, 
cold atoms in optical traps \cite{barkai2013transport}, molecular
self-assembly \cite{yvinec2012first}, the protein target site location in
DNAs \cite{mirny2009protein, benichou2009searching}, signal transduction and immune cell activation \cite{coombs2009diffusion}, search and rescue \cite%
{moreau2003pascal, oshanin2002trapping, drewitz2012survival} and
predator-prey interactions \cite{komarov2013capturing, oshanin2009survival,
oshanin2002trapping, gabel2012can, kehagias2013cops}. See a recent review of the narrow escape problem (\cite{holcman2014narrow}) and references therein for more applications and associated methods.

Generally speaking, MFPT problems fall into two classes:\ either the trap is
stationary or it is moving. In the case of a stationary traps, very precise
information can be obtained, in particular when the traps have small area 
\cite{holcman2014time, schuss2007narrow, singer2006narrow,
chevalier2011first, pillay2010asymptotic, cheviakov2010asymptotic,
kolokolnikov2005optimizing, coombs2009diffusion}. A scenario involving moving traps was introduced in \cite{toussaint1983particle} in the context of an annihilation process $A + B \to 0$. While originally motivated by the annihilation of monopole-antimonopole pairs in the early universe, the annihilation reaction may also serve as a model in chemical kinetics and collision-induced quenching of excited-state particles \cite{szabo1988diffusion,rice1985diffusion}. Subsequent studies \cite{bramson1988asymptotic, bray2002exact} have addressed the asymptotics of the long time survival probability of a particle diffusing in a continuum distribution of traps.  There is also an extensive literature on searching strategies, where a moving trap represents a
searcher (e.g. police)\ and Brownian particles are sought (e.g. drunken
robbers). See for example \cite{komarov2013capturing, benichou2005optimal,
oshanin2009survival, gabel2012can, chung2011search, kehagias2013cops}. In
some of this literature, the seeker is assumed to follow some kind of random
strategy. For example, in \cite{benichou2005optimal} it was shown that an
intermittent searching strategy consisting of large jumps and random walks
works best under many circumstances where the seeker does not know anything
about the target. Other works study pursuit problems where either the seeker
or the target have some (or full) information about the other party, and can
adjust their strategy accordingly.

In a recent review article of first passage problems in finite domains \cite{benichou2014first}, it is remarked that although problems involving stationary traps have been well-studied, the case of mobile traps still remains largely unexplored. In particular, the only studies to have considered mobile traps in confined geometries have done so in one dimension \cite{giuggioli2013encounter, holcman2009probability, tejedor2011residual, tzou2014drunken}. Mobile traps are not only more realistic in many applications, but can significantly increase or decrease MFPT depending on parameters of their motion. The goal of this paper is to illustrate these effects in a confined two-dimensional domain.

Let us first briefly review the derivation of the continuum equations for the MFPT
as outlined in \cite{redner2001guide}, page 31. We first consider the
simplified situation of a particle undergoing a discrete random walk moving in
one dimension with a stationary trap located at $x = x_0$. Assume that within $\Delta t$ time, the particle jumps a
distance $\Delta x$ with equal probability to the left and to the right, and
let $v(x)$ denote the mean first passage time of a particle initially located at $x$. Then the MFPT at location $x$ may be expressed in terms of the MFPT of its two neighboring locations as

\begin{equation}
v(x)=\frac{1}{2}\left\{v(x+\Delta x)+ v(x-\Delta x) \right\} + \Delta t \,; \qquad v=0 \enspace \mbox{at} \enspace x = x_{0}\,,  \label{constitutive}
\end{equation}

\noI where the condition $v(x_0) = 0$ indicates that a particle whose starting location coincides with the trap location is expected to survive for precisely zero units of time. Taking the limit $\Delta t,\Delta x\rightarrow 0$ and expanding (\ref%
{constitutive})\ in Taylor series, we obtain the continuum equation

\BE \label{meanfield}
 Dv_{xx}+1 = 0 \,, \quad v(x_0) = 0 \,; \qquad D \equiv \frac{(\Delta x)^2}{2\Delta t} \,,
\EE

\noI subject to appropriate boundary conditions. Here, $D$ is the diffusion rate, which
can be non-dimensionalized to 1. In Figure \ref{simulreflect}, we illustrate a scenario in which a trap is located at $x = 1/2$ on a domain with reflecting boundaries at $x = 0$ and $x = 1$. The solid curve denotes the MFPT as obtained from a Monte Carlo simulation of 5000 individual agents undergoing an unbiased random walk starting from location $x_i \in (0,1)$. At each interval of time $\Delta t$, each agent takes one step of size $\Delta x$ to the left or right with equal probability. The quantities are such that $(\Delta x)^2/(2\Delta t) = 1$. An agent that steps outside the domain is reflected back into the domain. The time for each agent to hit the trap is recorded, then averaged over all agents. Repeating the procedure for a set of points on the interval $(0,1)$, we obtain an approximation for the MFPT as a function of starting location $x_i$. The dashed curve represents the true MPFT obtained by solving \eqref{meanfield} with $D = 1$, trap location $x_0 = 1/2$, and pure Neumann boundary conditions $v_x(0) = v_x(1) = 0$. Excellent agreement is observed between the Monte Carlo simulation and the exact solution.

\begin{empty}\begin{figure}[htbp]
  \begin{center}
    \mbox{
    \subfigure[reflecting boundaries] 
        {\label{simulreflect}
        \includegraphics[width=.4\textwidth]{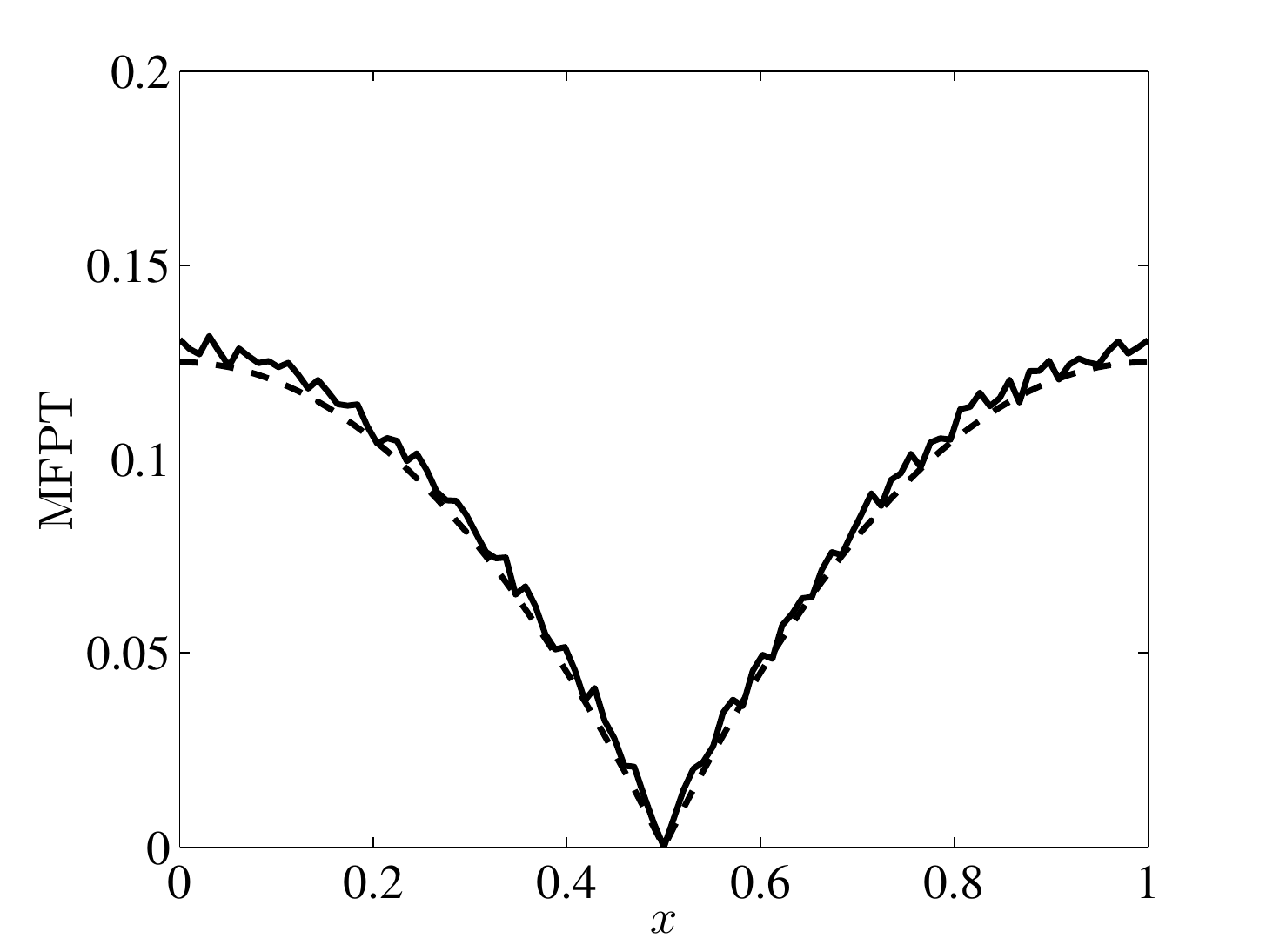}
        }   \hspace{1.5cm}
    \subfigure[periodic boundaries] 
        {\label{simulperiod}
        \includegraphics[width=.4\textwidth]{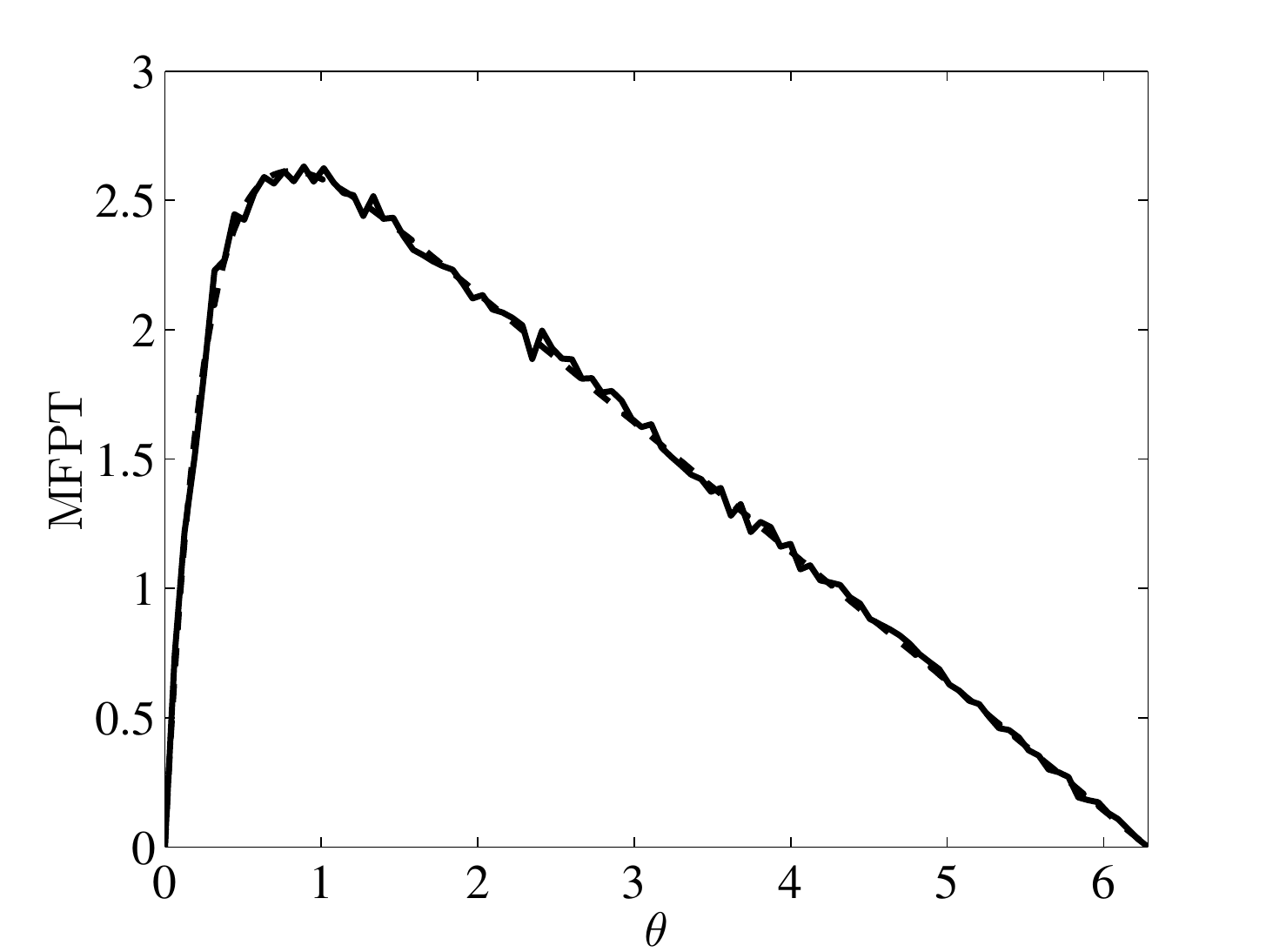}
        }}
    \caption{(a) MFPT on a domain of unit length with reflecting boundaries. The dashed curve is an approximation of the MFPT obtained from a Monte Carlo simulation with $\Delta x = \sqrt{2}/100$ and $\Delta t = 1\times 10^{-4}$. The diffusion coefficient $D$ defined in \eqref{meanfield} is then $D = 1$. For each grid point in $x$, an average of capture times of $5000$ agents was used to generate the MFPT. The dashed curve represents the true MPFT obtained by solving \eqref{meanfield} with $D = 1$ and pure Neumann boundary conditions. (b) MFPT on a one-dimensional circle with trap rotating clockwise at constant angular velocity $\omega > 0$. The plot represents the MFPT for all locations on the circle at the instant when the trap is located at $\theta = 0$. For the Monte Carlo simulation (solid line), $\Delta \theta = 0.01$, $\Delta t = 1\times 10^{-4}$, and $\omega = 2$. The results are an average over $500$ agents at each grid point. The dashed line is the solution of \eqref{meanfielddrift} with $D = 0.5$ and $\omega = 2$.}
     \label{fig:simul}
  \end{center}
\end{figure}\end{empty}

A similar derivation may be used to obtain an ODE describing MFPT on a one-dimensional circle $\theta \in \lbrack 0, 2\pi)$ with a \textit{moving} trap traveling with constant velocity. Consider a trap rotating clockwise on the circle at constant angular velocity $\omega > 0$. At the instant when the trap is located at $\theta = \theta_0$, the MFPT for a particle with initial location $\theta$ can be expressed in terms of the MFPT of neighbors of the site equidistant from the trap at the previous time step. Since the trap is displaced by an angle of $-\omega \Delta t$ each time step, we have

\BE \label{constitutivedrift}
	v(\theta) = \frac{1}{2} \left\{v(\theta + \omega \Delta t - \Delta \theta )+ v(\theta + \omega \Delta t + \Delta \theta )\right\} + \Delta t ; \qquad v=0 \enspace \mbox{at} \enspace \theta=\theta_0, 
\EE
 
\noI where we have assumed a jump of $\Delta \theta$ per time step with equal probability in each direction. Expanding \eqref{constitutivedrift} to leading order, noting that $\mO(\Delta \theta) \sim \mO(\sqrt{\Delta t})$, we obtain the ODE for MFPT

\BE \label{meanfielddrift}
D v_{\theta\theta} + \omega v_\theta + 1 = 0 \quad v(\theta_0) = 0 \,; \qquad D = \frac{(\Delta \theta)^2}{2\Delta t} \,,
\EE

\noI with periodic boundary conditions. The diffusion coefficient in \eqref{meanfielddrift} may be scaled to unity, leaving a non-dimensional angular velocity in front of the advection term. The Monte Carlo simulation may be performed in the same way as in the case of the stationary trap. On a periodic domain of length $2\pi$, we initialize $500$ agents at location $\theta_i \in [0, 2\pi)$ with the trap located at $\theta_0 = 0$. For each time step $\Delta t$, we allow each agent to move clockwise or counterclockwise with equal probability, while also advancing the location of the trap by $-\omega \Delta t$, where $\omega$ is the speed of the trap. The time required for each agent to be captured is recorded, then averaged over all agents. Repeating the procedure for a discrete set of points on the interval $[0, 2\pi)$, we obtain Figure \ref{simulperiod}. The MFPT in Figure \ref{simulperiod} therefore represents the MFPT for a random walker starting at location $\theta$ at the instant in time when the trap is located at $\theta_0 = 0$. In Figure \ref{simulperiod}, we observe excellent agreement between the solution of \eqref{meanfielddrift} and a Monte Carlo simulation. The trap is located at $\theta_0 = 0$, and is moving to the left, re-entering at $\theta = 2\pi$ by periodicity. Note that, as expected, the MFPT in front of the rotating trap is lower than that behind the trap.

The mean-field equation \eqref{meanfielddrift} bears close relation to the parabolic PDE

\BE \label{meanfieldpde}
 u_t = Du_{\phi\phi} + 1 \,, \qquad u(\bmod(\omega t, 2\pi), t) = 0 \,,
\EE

\noI with periodic boundary boundary conditions and appropriate initial conditions. Applying the transformation $\theta = \phi - \omega t$ and $u(\phi, t) = v(\theta)$ to \eqref{meanfieldpde}, one recovers \eqref{meanfielddrift} with $\theta_0 = 0$. Note, however, that with $\omega > 0$ in both \eqref{meanfielddrift} and \eqref{meanfieldpde}, the trap in \eqref{meanfielddrift} rotates clockwise while it rotates counterclockwise in \eqref{meanfieldpde}. The quantity $u$ in \eqref{meanfieldpde} is thus different from the MFPT interpretation of $v$ in \eqref{meanfielddrift}. We interpret $u$ as the rescaled continuum limit of a quantity that satisfies the discrete equation

\BE \label{constitutivepde}
	u(\phi, t+\Delta t) = \frac{1}{2}\left\{u(\phi+\Delta \phi, t) + u(\phi - \Delta \phi, t) \right\} + r \Delta t \,; \qquad u=0 \enspace \mbox{at} \enspace \phi =\bmod(\omega t, 2\pi) \,.
\EE

\noI A simple interpretation for $u$ in \eqref{constitutivepde} is that of a concentration of particles that undergo an unbiased random walk with a constant external feed rate $r$, which can be normalized to unity. The rotating Dirichlet trap acts to remove particles from the domain. It may also be interpreted as a temperature, with the Dirichlet trap acting to cool a domain subject to uniform external heat influx. For the same set of parameters, the solution for $u$ at a specific instant when the trap is located at $\phi = 2\pi$ is given by Figure \ref{simulperiod}. The trap, however, is to be assumed to be traveling to the right, re-entering at $\phi = 0$. As expected, the concentration or temperature behind the trap is lower than that in front.

In this paper, we examine the MFPT\ for a moving circular trap of small radius $%
\varepsilon $ inside a unit disk. The trap is assumed to rotate clockwise at a
constant rate $\omega $ along a circle of radius $r_{0}<1$ concentric with
the unit disk. That is, the location of the center of the trap is given by

\BE \label{trapdyn}
(x_0, y_0) = (r_0\cos\omega t, -r_0\sin\omega t) \,.
\EE

\noI The derivation for the elliptic PDE describing the MFPT with this geometry follows closely to that leading to \eqref{constitutivedrift}. That is, for a trap of radius $\varepsilon$ centered at $(r_0, \theta_0)$ in polar coordinates, the MFPT for a particle initially located at $(r,\theta)$ may be expressed in terms of the MFPT of the neighbors of the point $(r, \theta + \omega t)$ at the previous time step. As in the case of the rotating trap on a one-dimensional circle, this equivalence may be attributed the dependence of MFPT on only the relative starting location of the particle with respect to the trap. In Cartesian coordinates, this may be expressed as

\begin{equation} \label{constitutivedisk}
\begin{gathered}
 v(x,y) = \frac{1}{4} \left\{v(x_p + \Delta x, y_p) +  v(x_p - \Delta x, y_p) + v(x_p, y_p + \Delta y) + v(x_p, y_p - \Delta y) \right\} + \Delta t \,, \\ v = 0 \enspace \mbox{when} \enspace |(x,y) - (x_0, y_0)| \leq \varepsilon \,,
 \end{gathered}
\end{equation}

\noI where we have assumed that, at each time step, the particle may move one step on a square lattice with equal probability in all four directions. The condition $v = 0$ when $|(x,y) - (x_0, y_0)| \leq \varepsilon$ states that the MFPT of a particle starting inside or on the trap centered at $(x,y) = (x_0,y_0)$ is exactly zero. Since the angular coordinate of the trap location decreases by $\omega \Delta t$ each time step, the location $(x_p, y_p)$ is

\BE \label{xp}
	(x_p, y_p) = (r\cos(\theta + \omega \Delta t), r\sin(\theta + \omega \Delta t)) \,.
\EE

\noI Expanding \eqref{xp} for small $\Delta t$ and using $(x,y) = (r\cos\theta, r\sin\theta)$, we calculate

\BE \label{xptaylor}
	(x_p, y_p) = (x - \omega y\Delta t, y + \omega x\Delta t) \,.
\EE

\noI Substituting \eqref{xptaylor} into \eqref{constitutivedisk} and expanding to leading order, noting that $\mO(\Delta x) \sim \mO(\Delta y) \sim \mO(\sqrt{\Delta t})$, we obtain

\BE \label{continuumcart}
\begin{gathered}
	\frac{(\Delta x)^2}{4\Delta t}v_{xx} + \frac{(\Delta y)^2}{4\Delta t}v_{yy} + \omega( x v_y - yv_x)  + 1 = 0 \,, \\
	v = 0 \enspace \mbox{when} \enspace |(x,y) - (x_0, y_0)| \leq \varepsilon \,.
	\end{gathered}
\EE

\noI Letting $\Delta x = \Delta y = \Delta \ell$ and $D \equiv (\Delta \ell)^2/(4\Delta t)$ in \eqref{continuumcart} we obtain in polar coordinates

\BE \label{continuumpol}
	D\Delta v + \omega v_\theta + 1 = 0 \,; \qquad v_r(1) = 0 \,, \quad v = 0 \enspace \mbox{when} \enspace |(x,y) - (x_0, y_0)| \leq \varepsilon \,,
\EE

\noI where we have used in \eqref{continuumpol} that $xv_y - yv_x = v_\theta$. Both the radius of the disk and the diffusion coefficient $D$ may be scaled to unity without loss of generality. The pure Neumann boundary condition indicates a disk with a reflecting wall. To illustrate the theory, we compare the MFPT obtained from a Monte Carlo simulation (Figure \ref{simul2D}) to that of a numerical solution of \eqref{continuumpol} with $D = 1$, $r_0 = 0.6$, $\omega = 200$, and $\varepsilon = 0.1$ (Figure \ref{pdesolvssimul}). The simulations were performed in the same way as that on the one-dimensional circle, with $\Delta \ell$ and $\Delta t$ set such that the diffusion coefficient was unity. That is, with a trap of radius $\varepsilon$ centered at $(x,y) = (r_0, 0)$, we initialize $1000$ agents at a location $(x_i, y_i)$ in the unit disk. We then evolve each agent according to a nearest neighbor random walk, as well as the trap according to \eqref{trapdyn}. For each agent, we record the time elapsed before it comes within $\varepsilon$ distance of the trap center. The MFPT at point $(x_i, y_i)$ is then approximated by the average capture time of the $1000$ particles. Repeating over a grid of points inside the unit disk, we generate Figure \ref{simul2D}. We observe excellent qualitative agreement between the simulation result and PDE solution. In both figures, the regions with the darkest shade of red have a value of approximately $0.13$, indicating also quantitative agreement. Observe that, similar to the case of a rotating trap on a one-dimensional circle, the MFPT is lower in front of the clockwise-rotating trap than it is behind it.

\begin{empty}\begin{figure}[htbp]
  \begin{center}
    \mbox{
    \subfigure[MFPT from Monte Carlo simulation] 
        {\label{simul2D}
        \includegraphics[width=.4\textwidth]{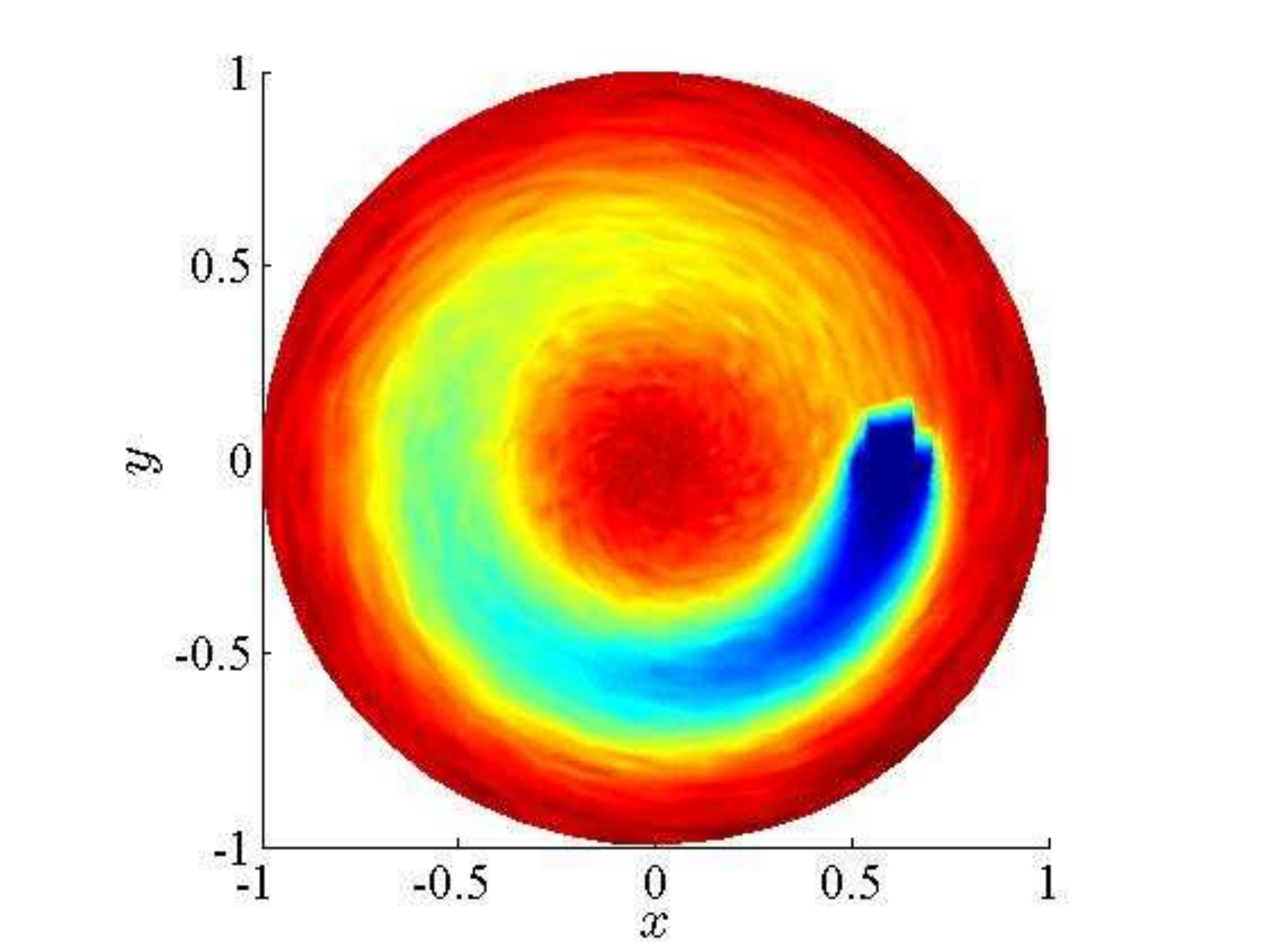}
        }   \hspace{1.5cm}
    \subfigure[MFPT from solution of PDE] 
        {\label{pdesolvssimul}
        \includegraphics[width=.4\textwidth]{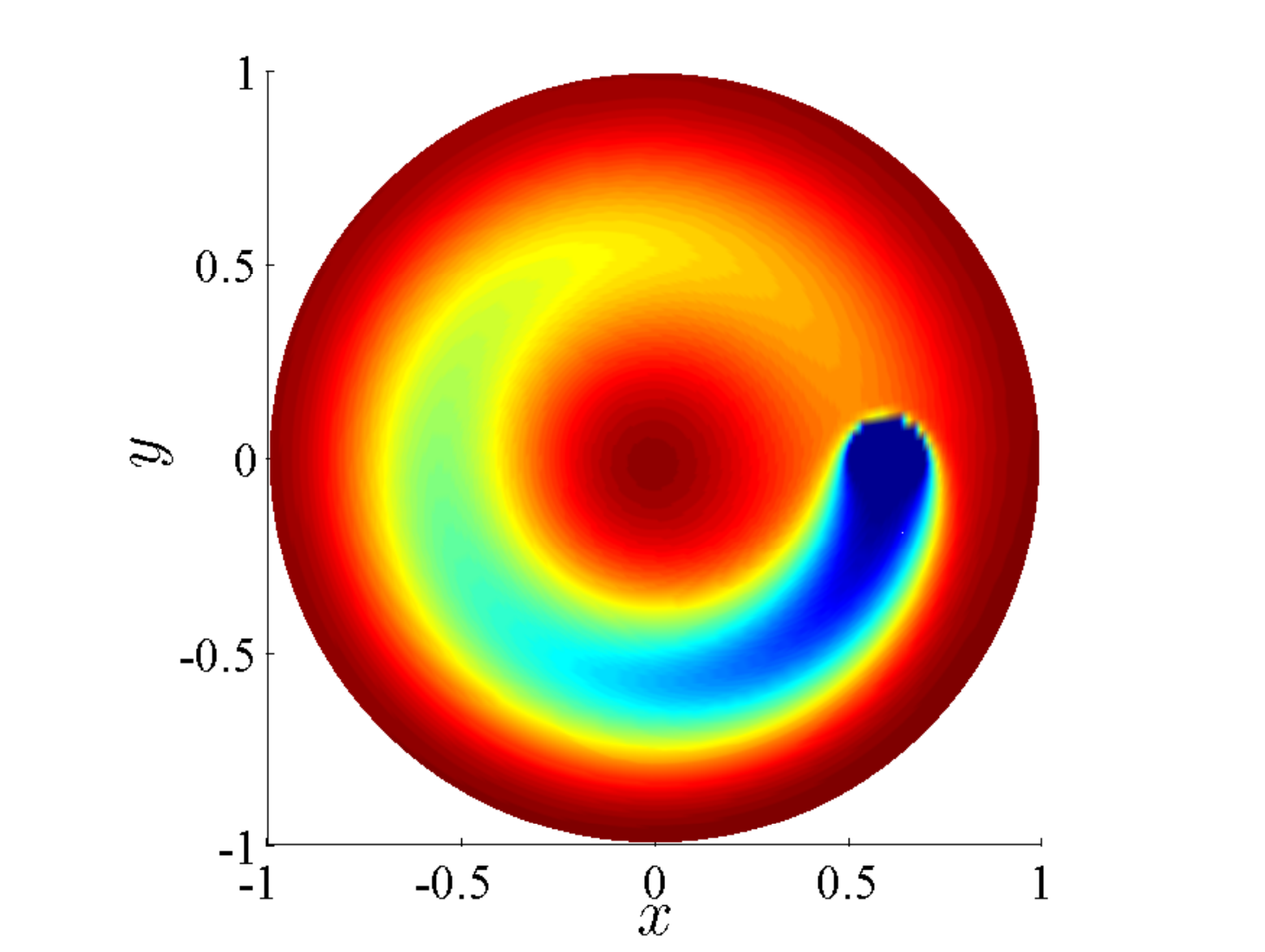}
        }}
    \caption{(a) Monte Carlo approximation of MFPT on a unit disk with trap located at $(x,y) = (0.6, 0)$ rotating clockwise with angular velocity of $\omega = 200$. Red (blue) regions indicate large (small) values of MFPT. The parameters of the random walk are such that $D = 1$. $1000$ trials per grid point were used to obtain an average approximation. (b) Numerical solution of \eqref{continuumpol} with $D = 1$ and $\omega = 200$. In both figures, the regions shaded in dark red have a value of approximately $0.13$. Observe that the MFPT is lower in front of the trap than it is behind it.}
     \label{fig:simul2D}
  \end{center}
\end{figure}\end{empty}

In the same way that the time-independent problem \eqref{meanfielddrift} may be interpreted as a transformation of the time-dependent problem \eqref{meanfieldpde} into a rotating frame, the time-dependent analog of \eqref{continuumpol} may be formulated as

\begin{empty}\begin{equation}\label{timedep}
\begin{gathered}
	v_t = \Delta v + 1 \,, \quad \mathbf{x} \in \Omega \setminus \Omega_\varepsilon(\mathbf{x}_0(t)) \,; \\
	 \partial_n v = 0 \,, \quad \mathbf{x} \in \partial \Omega \,;  \qquad v = 0\,, \quad \mathbf{x} \in \partial \Omega_\varepsilon(\mathbf{x}_0(t)) \,; \qquad \mathbf{x}_0(t) = \left(r_0 \cos \omega t, r_0 \sin \omega t\right) \,,
\end{gathered}
\end{equation}
\end{empty}

\noI supplemented by appropriate initial conditions. Here, $\mathbf{x}=(x,y)$
is a two-dimensional vector in Cartesian coordinates, $v=v(x,y,t)$, $\Omega $
denotes the unperturbed unit disk, $\Omega _{\varepsilon }(\mathbf{x}_{0})$
the circular trap of radius $\varepsilon $ centered at $\mathbf{x}=\mathbf{x}%
_{0}$, and $\partial _{n}v$ the normal derivative of $v$ on $\partial \Omega 
$. With $\omega > 0$, the trap rotates counterclockwise. As in \eqref{constitutivepde}, $v$ may be interpreted as a concentration of particles or a temperature, with the trap acting to remove the quantity from the domain subject to a constant uniform influx. In Figures \ref{simul2D} and \ref{pdesolvssimul}, we thus observe the counterclockwise-rotating hole leaving a region of low particle concentration or temperature in its trail.

Our goal is to describe the \textquotedblleft optimal\textquotedblright\
radius $r_{0}=r_{0}^{opt}$ as a function of both $\omega $ and $\varepsilon$. For such a rotating trap, we define the optimal radius as the one that minimizes the MFPT \emph{averaged} over all points in the domain. MFPT optimization problems with absorbing boundaries (stationary traps) were considered for particles in one dimension under the presence of a time-oscillatory or randomly fluctuating field in \cite{fletcher1988first, JavierBrey1994123, revelli2004diffusion, dhara2002coherent, dybiec2004resonant, pikovsky1997coherence}. In these cases, it was found that the average MFPT could be minimized by careful tuning of the characteristics of the field. In contrast, we tune characteristics of the trap motion in order to minimize the average MFPT. In this formulation with the rotating trap, by equivalence between \eqref{continuumpol} with \eqref{timedep}, minimization of the average MFPT is equivalent to minimizing the total mass

\begin{equation}
M(r_{0};\omega )=\int_{\Omega }\!v\,d\Omega \,, \label{mass}
\end{equation}%

\noI of the solution of \eqref{timedep} in the limit $t \to \infty$. We do not consider any transient effects in our analysis. In the rest of this paper, we adopt this interpretation instead of that of the MFPT, as it leads to results and calculations that are more easily interpreted from a physical standpoint.

This is one of the few configurations that is amenable to a full mathematical
analysis for a bounded domain in two dimensions. By taking advantage of the
radial symmetry, the problem becomes \textquotedblleft
stationary\textquotedblright\ in the co-rotating coordinate frame, making it
possible to apply a full range of techniques similar to those developed for
small stationary traps in \cite{holcman2014time, schuss2007narrow,
singer2006narrow, chevalier2011first, pillay2010asymptotic,
cheviakov2010asymptotic, kolokolnikov2005optimizing}. See \cite{holcman2014narrow} for a review of asymptotic methods used to study narrow escape problems.

Our main results are summarized in Figure \ref{opt_r_0_vs_omega_withflex}.
For a range of $\omega $, it shows the optimal radius of rotation $%
r_{0}^{opt}$ of the trap that minimizes $M(r_{0};\omega )$ with respect to $%
r_{0}$. The analysis shows four distinguished regimes, depending on the
relative sizes of $\omega $ and $\varepsilon ,$ as summarized in the
following table.

\begin{center}
\begin{tabular}{l|l}
Regime & Main result \\ \hline\hline
$\omega =\mathcal{O}(1)$ & \textquotedblleft bifurcation\textquotedblright\
near $\omega =\omega _{c}$ (\S \ref{omegaO1}, \S \ref{smallr0}) \\ \hline
$1\ll \omega \ll \mathcal{O}(\varepsilon^{-1} )$ & $r_{0}^{opt}\sim 1$ (\S \ref%
{largeomega}) \\ \hline
$\omega =\mathcal{O}(\varepsilon^{-1} )$ & transition region, optimal radius
depends only on $\omega _{0}=\varepsilon \omega $ (\S \ref{omegaeps}) \\ 
\hline
$\omega \gg \mO(\varepsilon^{-1}) .$ & $r_{0}^{opt}\sim 1/\sqrt{2}$ (\S \ref{2Dtrap})
\\ \hline
\end{tabular}
\end{center}

The left non-zero segment of the solid curve in Figure \ref{opt_r_0_vs_omega_withflex}%
, independent of $\varepsilon $, was generated by calculating $%
M(r_{0};\omega )$ in terms of an infinite series, which may be summed
numerically to determine the optimal value of $r_{0}$. The analysis, which
assumes $\omega \ll \mathcal{O}(\varepsilon ^{-1})$, is presented in \S \ref%
{omegaO1}. The circular points are the results of full numerically computed
solutions of \eqref{heateqn} with $\varepsilon =1\times 10^{-3}$. A notable feature seen in Figure \ref{opt_r_0_vs_omega_withflex} is the
presence of a bifurcation near $\omega _{c}\approx 3.026$, where for $\omega
<\omega _{c}$, the optimal radius of rotation is precisely 0. This result is due to the optimal location of a stationary trap being at the origin, which we show in \S \ref{omegaO1}. The presence of the bifurcation states that a rotating trap must rotate with rate above some critical speed in order to compensate for being located away from this otherwise optimal location. In \S \ref%
{smallr0}, we calculate the critical speed exactly. A typical solution in the regime 
$\omega \sim \mO(1)$ is shown in Figure \ref%
{flexcont_omega10} for $\omega =10$ and $r_{0}=0.6$. Note that solutions for 
$\omega \sim \mathcal{O}(1)$ lack radial symmetry.

\begin{empty}\begin{figure}
\begin{center}
	\includegraphics[width=.72\textwidth, height = .45\textwidth]{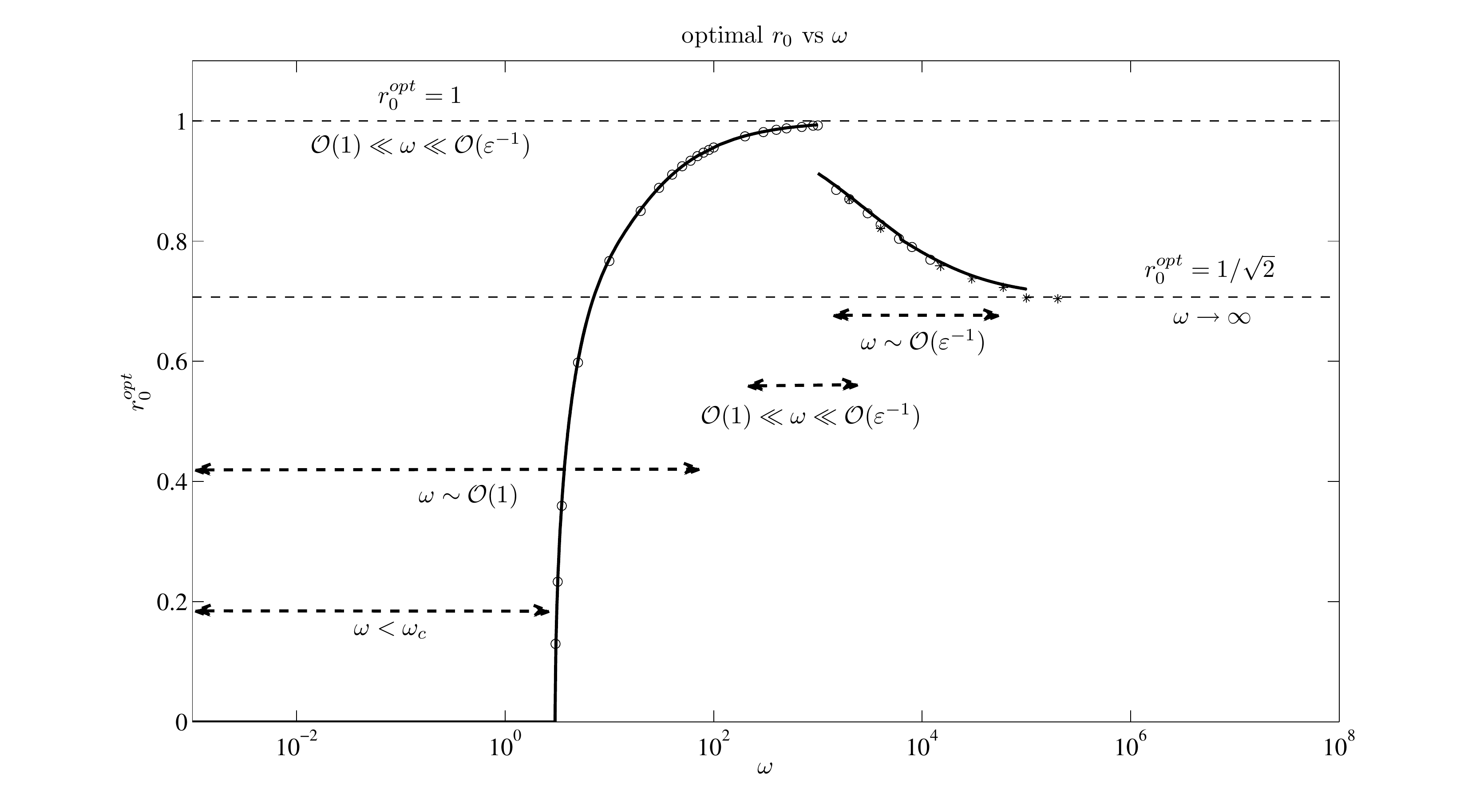}
	\caption{Asymptotic and numerical results for $r_0^{opt}$ in different regimes of $\omega$. Note that the scale of the horizontal axis is logarithmic. The left non-zero segment of the solid curve was obtained from a series solution of \eqref{heateqn} with $\omega \sim \mathcal{O}(1)$. The top thin dashed line is the result $r_0^{opt} \sim 1$ obtained from a leading order analysis in the regime $1 \ll \omega \ll \mathcal{O}(\varepsilon^{-1})$. The circles were obtained from full numerical solutions of \eqref{timedep} with $\varepsilon = 1\times 10^{-3}$. The right segment of the solid curve was obtained from a leading order calculation with $\omega \sim \mathcal{O}(\varepsilon^{-1})$. The overlaying circles represent results from numerical solutions with $\varepsilon = 1\times 10^{-3}$. In this regime, the relevant quantity is $\varepsilon\omega$, not  $\varepsilon$ and $\omega$ individually. As such, the stars, generated from the same computations with $\varepsilon = 5\times 10^{-3}$ and $\omega$ one-fifth of the value indicated on the horizontal axis, align closely with the circles. The lower thin dashed line indicates the result $r_0^{opt} \sim 1/\sqrt{2}$ for $\varepsilon \to 0$ and $\omega \to \infty$ with $\omega \gg \mathcal{O}(\varepsilon^{-1})$.} 
	\label{opt_r_0_vs_omega_withflex}
	\end{center}
\end{figure}\end{empty}

The top dashed line of Figure \ref%
{opt_r_0_vs_omega_withflex} at $r_{0}^{opt}=1$ indicates the value of $%
r_{0}^{opt}$ as $\omega \to \infty$ in the regime $1\ll \omega \ll \mathcal{O}(\varepsilon ^{-1})$.
In \S \ref{largeomega} for large $\omega $, we use boundary layers to
construct a leading order solution of \eqref{heateqn}. Whereas
the analysis of \S \ref{omegaO1} leads to an expression for $M(r_{0};\omega
) $ in terms of an infinite sum, the boundary layer analysis yields an
explicit leading order expression for $M(r_{0};\omega )$, from which we readily show
that $r_{0}^{opt}\rightarrow 1$ as $\omega \rightarrow \infty $ with $\omega
\ll \mathcal{O}(\varepsilon ^{-1})$. A typical solution in this regime is
shown in Figure \ref{flexcont_omega1000}. An internal layer develops in the
tail behind the trap, while away from this internal layer, the solution is nearly
radially symmetric.

The right segment of the solid curve of Figure \ref%
{opt_r_0_vs_omega_withflex} is calculated from a boundary layer solution
with $\omega =\omega _{0}/\varepsilon $ and $\omega _{0}\sim \mathcal{O}(1)$%
. A very delicate analysis of the boundary layer is required to derive the
asymptotic solution. This calculation is presented in \S \ref{omegaeps}. Unlike the $\omega \ll \mathcal{O}(\varepsilon^{-1})$ regimes, the
results in this regime depend on $\varepsilon$ through the quantity $\omega _{0}=\varepsilon
\omega $. Illustration of this dependence may be seen in Figure \ref{opt_r_0_vs_omega_withflex}. While the overlaying circles were determined from numerical
solutions with $\varepsilon =1\times 10^{-3}$ and $\omega$ given on the
horizontal axis, the stars were computed with $\varepsilon =5\times 10^{-3}$
with $\omega $ one-fifth the value on the horizontal axis. The dependence on
the product $\varepsilon \omega $ and not on $\varepsilon $ and $\omega $
individually may be inferred from the close agreement between the circles and
stars. A typical solution in this regime is shown in Figure \ref%
{flexcont_omega10000}. Compared to Figure \ref{flexcont_omega1000} with
smaller $\omega $, the internal layer in Figure \ref{flexcont_omega10000} is
considerably thinner. Away from the layer, the solution also exhibits a high
degree of radial symmetry.

\begin{empty}\begin{figure}[tbp]
  \begin{center}
    \mbox{
    \subfigure[$u(x,y)$ with $\omega = 10$] 
        {\label{flexcont_omega10}
        \includegraphics[width=.32\textwidth]{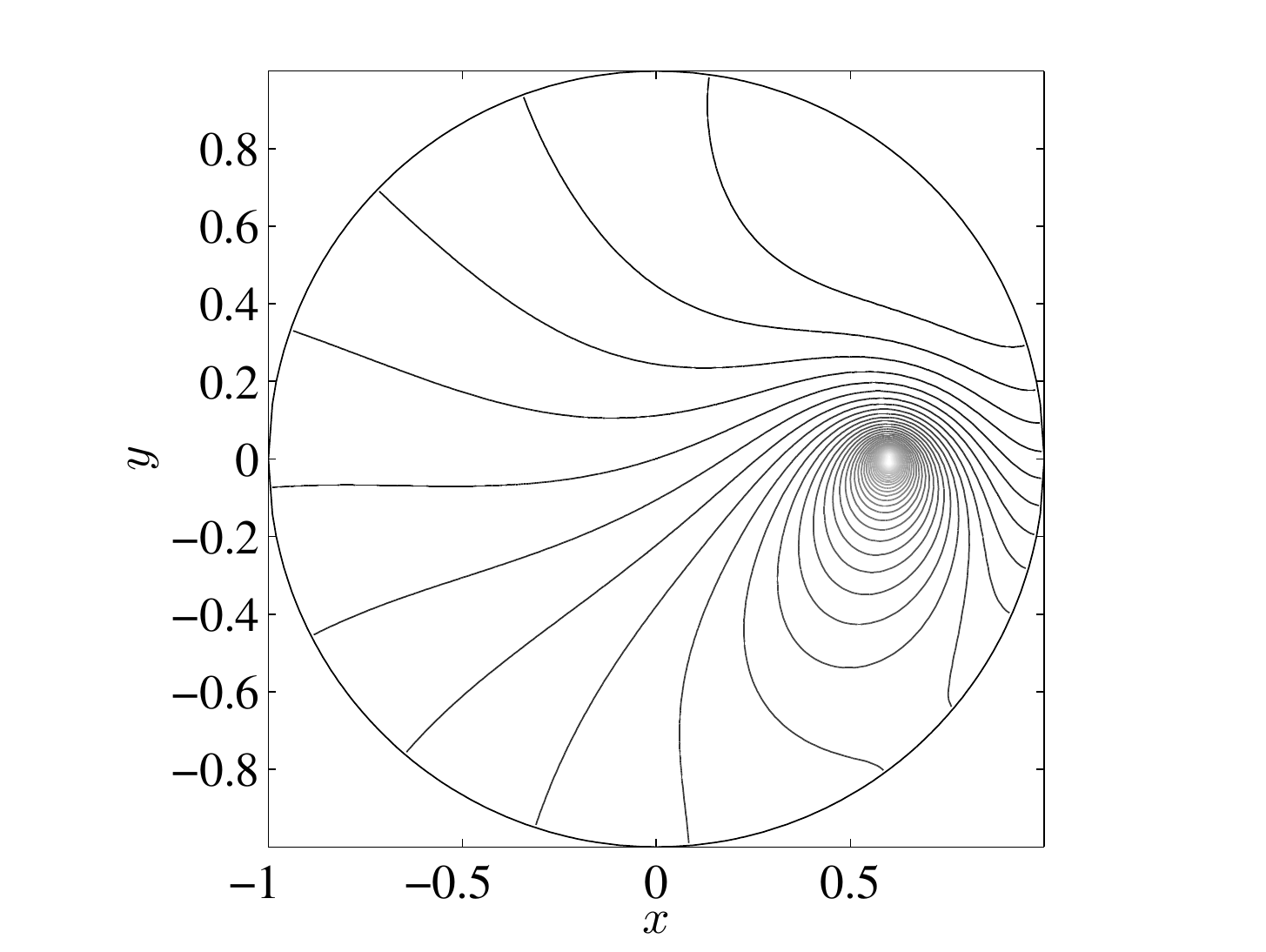}
        }   
    \subfigure[$u(x,y)$ with $\omega = 1000$] 
        {\label{flexcont_omega1000}
        \includegraphics[width=.32\textwidth]{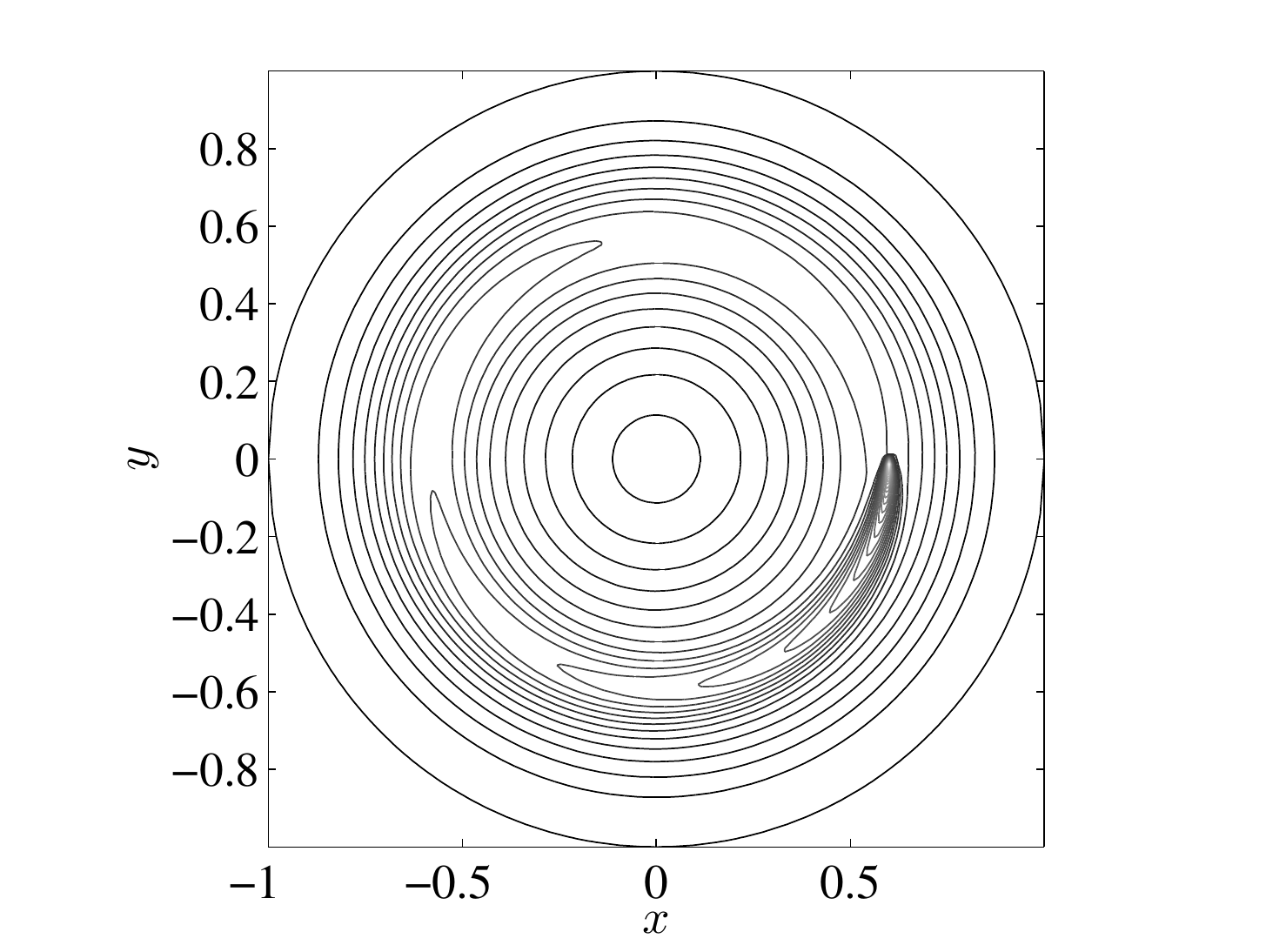}
        }
    \subfigure[$u(x,y)$ with $\omega = 1\times10^4$] 
        {\label{flexcont_omega10000}
        \includegraphics[width=.32\textwidth]{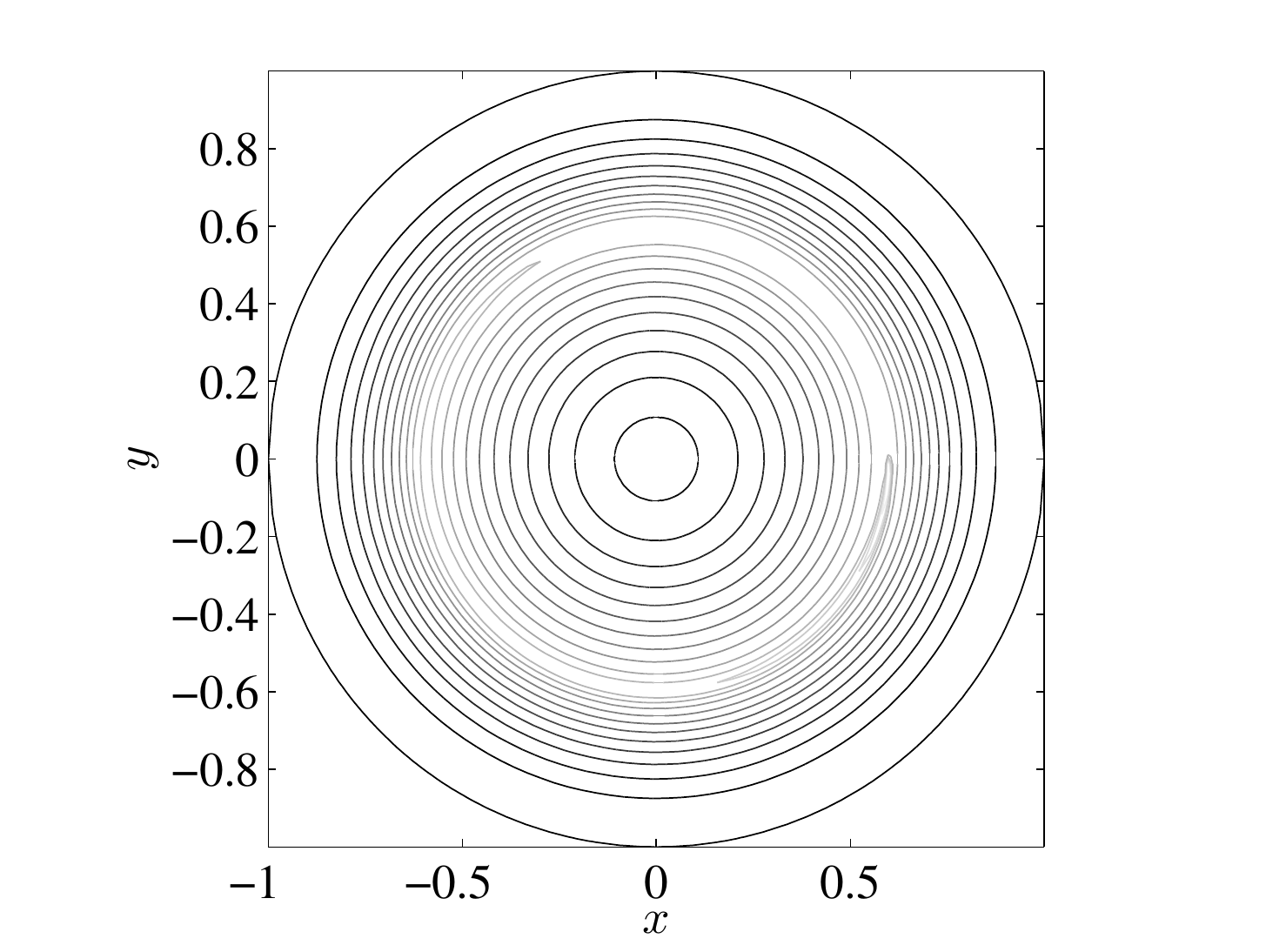}
        }}
    \caption{Contour plots of $u(x,y)$ obtained by numerically solving \eqref{heateqn} with $r_0 = 0.6$ and (a) $\omega = 10$, (b) $\omega = 1000$, and (c) $\omega = 1\times 10^{4}$. In (a), where $\omega \sim \mathcal{O}(1)$, the solution exhibits no radial symmetry. In (b) for larger $\omega$, an internal layer centered on the ring $r = r_0$ develops behind the trap. The solution is nearly radially symmetric in regions away from the layer. In (c), where $\omega \sim \mathcal{O}(\varepsilon^{-1})$, the layer becomes thinner and the solution exhibits greater radial symmetry. Here, $\varepsilon = 1\times10^{-4}$. FlexPDE \cite{flex} was used for numerical simulations.}
     \label{fig:flexcont}
  \end{center}
\end{figure}\end{empty}

Finally, for very large $\omega \gg \mathcal{O}(\varepsilon ^{-1})$, the
trap is rotating so fast that from the point of view of a particle in the domain, it
appears simultaneously everywhere along the circle of radius $r_{0}.$ In
this case the optimal radius asymptotes to $r_{0}^{opt}\sim 1/\sqrt{2}.$
This has a very nice geometric interpretation: the trap moving along such
radius divides the unit disk into two regions of equal area. This
calculation is presented in \S \ref{2Dtrap}.

The first step in the analysis is to transform \eqref{timedep} into the rotating frame of
the trap to obtain a time-independent problem. To do so, we first transform
to the polar coordinate system $(x,y)\rightarrow (r,\varphi )$ so that $%
x(r,\varphi )=r\cos \varphi $, $y(r,\varphi )=r\sin \varphi $ and $v(x,y,t)=%
\tilde{u}(r,\varphi ,t)$. The center of the trap is then given by $%
(r,\varphi )=(r_{0},\bmod(\omega t,2\pi ))$. Making the transformation into
the rotating frame $\theta =\varphi -\bmod(\omega t,2\pi )$ with $0<\theta
<2\pi $, and $\tilde{u}(r,\varphi ,t)=u(r,\theta (t))$, we obtain the
stationary problem

\begin{empty}\bes \label{heateqn}
\begin{equation} \label{heatpde}
	\Delta u + \omega u_\theta + 1 = 0 \,, \quad \mathbf{x} \in \Omega \setminus \Omega_\varepsilon(r_0) \,; 
\end{equation}
\begin{equation}	\label{heatbc}
	 u_r = 0 \,, \quad \mathbf{x} \in \partial \Omega \,;  \qquad u = 0\,, \quad \mathbf{x} \in \partial \Omega_\varepsilon(r_0) \,.
\end{equation}
\ees\end{empty}

\noI Here, $\Delta u$ denotes the Laplacian of $u(r,\theta )$ in radial
coordinates, $u_{\theta }$ and $u_{r}$ denote differentiation of $u$ with
respect to the angular and radial coordinates, respectively, and $\Omega
_{\varepsilon }(r_{0})$ denotes a circular hole of radius $\varepsilon $
centered at a distance $r_{0}$ from the origin located along the $\theta =0$
axis.

\setcounter{equation}{0}

\section{The regime $\protect\omega \gg \mathcal{O}(\protect\varepsilon %
^{-1})$}

\label{2Dtrap}

This is the simplest regime to analyze, as \eqref{heateqn} reduces to a radially symmetry problem for $u$. To see this, we let $\omega \to \infty$ in \eqref{heatpde} so that $u = u(r)$ to leading order when $(r,\theta) \neq (r_0, 0)$. In the inner region $\xi = \ve^{-1}(x-x_0)$, $\eta = \ve^{-1}y$, we have for $u = U(\xi,\eta)$ that $\ve^{-2}\Delta U + \ve^{-1}\omega r_0U_{\eta} + 1 = 0$, which suggests that $U = U(\xi)$ when $\omega \gg \mO(\ve^{-1})\,$. With $U = 0$ on the trap $|(\xi,\eta)| = 1$, and since it matches to a radially symmetric outer solution, we must have that $U = 0$. In this way, we obtain the limiting outer problem

\begin{empty}\bes \label{heateqnradsymm}
\begin{equation} \label{heatpderadsymm}
	\Delta u  + 1 = 0 \,, \quad \mathbf{x} \in \Omega \setminus \left\lbrace\mathbf{x}: r_0 - \varepsilon < |\mathbf{x}| < r_0 + \varepsilon \right\rbrace \,; 
\end{equation}
\begin{equation}	\label{heatbcradsymm}
	 u_r = 0 \,, \quad \mathbf{x} \in \partial \Omega \,;  \qquad u \enspace \mbox{bounded as} \enspace r \to 0 \,, \quad u = u_0\,, \quad |\mathbf{x}| = r_0-\ve \,, r_0 + \ve \,,
\end{equation}
\ees\end{empty}

\noI with $u_0 = 0$. The solution of \eqref{heateqnradsymm} with $u_0 = 0$ is

\begin{equation}  \label{us}
u(r) = \frac{r_0^2 + \varepsilon^2 - r^2}{4} + \left\{ 
\begin{array}{lr}
- \frac{\varepsilon r_0}{2} \,, & \quad 0 < r < r_0-\varepsilon \\ 
\frac{\varepsilon r_0}{2} + \frac{1}{2}\log \left(\frac{r}{r_0+\varepsilon}
\right) \,, & \quad r_0 + \varepsilon < r < 1%
\end{array}
\right. \,.
\end{equation}

\noI In this case, with $u$ given by \eqref{us}, the total mass as defined by \eqref{mass} is 

\begin{equation}  \label{masssym}
M(r_0;\omega) = M(r_0) = \pi \left\lbrack \frac{r_0^2}{2}-\frac{3}{8}-\frac{1%
}{2}\log(r_0+\varepsilon) + \varepsilon r_0(1-r_0^2) + \frac{1}{2}%
\varepsilon^2 - \varepsilon^3r_0 \right\rbrack \,.
\end{equation}

\noI The optimal radius of rotation $r_0^{opt}$ that minimizes $M$ satisfies 
$dM/dr_0 = 0$, yielding

\begin{equation}  \label{r0inf}
r_0^{opt} = \frac{1}{\sqrt{2}} - \frac{\varepsilon}{4} + \mathcal{O}%
(\varepsilon^2) \,.
\end{equation}

\noI The approach to a value of $r_0^{opt}$ slightly less than $1/\sqrt{2}$
as $\omega \to \infty$ with fixed $\ve$ was observed in obtaining the
numerical results presented in Figure \ref{opt_r_0_vs_omega_withflex}.

For $\varepsilon \rightarrow 0$, the optimal radius \eqref{r0inf} is the
same as that obtained in the limit of an analogous problem studied in \cite%
{kolokolnikov2005optimizing}. The objective of that work was to find
configurations for $N$ identical traps placed inside a unit disk that
optimized the fundamental Neumann eigenvalue of the Laplacian. For the
special case where the traps were restricted to lie on a ring of radius $r$,
it can be seen from Proposition 4.4 of \cite{kolokolnikov2005optimizing} that the optimal value of $r$ tends to $%
1/\sqrt{2}$ as $N\rightarrow \infty $.

In the following section, we solve \eqref{heateqn} in the regime $\omega
\sim \mathcal{O}(1)$ in terms of a series expansion. Calculating the mass,
we find that there exists a value $\omega _{c}>0$ independent of $%
\varepsilon $ for which $r_{0}^{opt}=0$ when $\omega <\omega _{c}$ and $%
0<r_{0}^{opt}<1$ when $\omega >\omega _{c}$. In \S \ref{smallr0}, we use the
results of \S \ref{omegaO1} to determine the exact value of $\omega _{c}$.

\setcounter{equation}{0}

\section{Asymptotic solution for $\protect\omega \sim \mathcal{O}(1)$}

\label{omegaO1}

For $\omega = \mathcal{O}(1)$, we solve \eqref{heateqn} using the method of matched
asymptotics as in \cite{kolokolnikov2005optimizing}. Near the trap, we make the change to the
inner variables

\begin{equation}  \label{innervars}
\mathbf{y} = \frac{\mathbf{x} - \mathbf{x}_0}{\varepsilon} \,, \quad u(%
\mathbf{x}) = U(\mathbf{y}) \,; \qquad \bx_0 = (r_0,0) \,,
\end{equation}

\noI so that the trap, in the inner region, is a circle of unit radius
denoted $\Omega _{1}$. Here, $\bx_{0}$ denotes the center of the trap in
Cartesian coordinates. With this scaling, we have in the inner region that $\ve^{-2}\Delta U + \ve^{-1}\omega r_0U_{\eta} + 1 = 0$. With $\omega \ll \mO(\ve^{-1})$, the inner problem for $U$ reduces to

\begin{empty}\bes \label{innereq}
	\begin{equation} \label{Ueq}
		\Delta_\mathbf{y} U = 0 \,, \quad \mathbf{y} \notin \Omega_1 \,; \qquad U = 0 \,, \quad |\mathbf{y}| = 1 \,,
	\end{equation}
	\begin{equation} \label{Uinf}
	  U \sim S\log|\mathbf{y}| \enspace \mbox{as} \enspace |\mathbf{y}| \to \infty \,.
	\end{equation}
\ees\end{empty}

\noI With $\by$ defined in \eqref{innervars}, the behavior of $u$ near the trap is determined by the far-field behavior in \eqref{Uinf} as

\begin{equation}  \label{outlog}
u \sim S\log|\mathbf{x} - \mathbf{x}_0| - S\log\varepsilon \,, \enspace %
\mbox{as} \enspace \bx \to \bx_0 \,.
\end{equation}

\noI The logarithmic behavior of $u$ as $\bx \to \bx_0$ suggests that

\begin{equation}  \label{uG}
u = -\pi G(\bx; \bx_0) + H \,,
\end{equation}

\noI where $G(\bx; \bx_0)$ is the Neumann Green's function satisfying

\begin{empty}\bes
\begin{equation} \label{Geq}
	\Delta G + \omega G_\theta = \frac{1}{\pi} - \delta(\bx - \bx_0) \,, \quad \bx \in \Omega \,; 
\end{equation}
\begin{equation} \label{Geqbc}
 \partial_r G = 0 \,, \quad \bx \in \partial \Omega \,; \qquad \int_\Omega \! G(\bx;\bx_0) \, d\Omega = 0 \,,
\end{equation}
\ees\end{empty}

\noI and $H$ is a constant obtained from matching the inner and outer
solutions.

The solution for $G(\bx; \bx_0)$ in \eqref{Geq} can be written as

\begin{equation}  \label{Greg}
G(\bx; \bx_0) = -\frac{1}{2\pi} \log|\bx - \bx_0| + R(\bx;\bx_0) \,,
\end{equation}

\noI where $R(\bx;\bx_0)$ remains finite as $\bx \to \bx_0$ and is referred
to as the regular part of $G(\bx; \bx_0)$. By \eqref{uG}, the behavior of $u$
as $\bx \to \bx_0$ is then

\begin{equation}  \label{ux0}
u \sim \frac{1}{2} \log |\bx - \bx_0| - \pi R(\bx_0;\bx_0) + H \,, \enspace %
\mbox{as} \enspace \bx \to \bx_0 \,.
\end{equation}

\noI Comparing \eqref{ux0} to \eqref{outlog}, we find that $S = 1/2$ and

\begin{equation}  \label{H}
H = \pi R(\bx_0; \bx_0) - \frac{1}{2} \log \varepsilon \,.
\end{equation}

\noI By \eqref{uG} and \eqref{Geq}, we have that the mass of $u$ in $\Omega$
is

\begin{equation}  \label{massH}
M(r_0;\omega) = \pi H \,,
\end{equation}

\noI with $H$ given in \eqref{H} and $r_0 = |\bx_0|$. The minimization of $M(r_0;\omega)$
is thus equivalent to the minimization of $R(\bx_0; \bx_0)$. In the case of a stationary trap located at $\bx = \bx_0$, an explicit formula for the regular part of the Neumann Green's function with $\omega = 0$ in \eqref{Geq} is given in \cite{kolokolnikov2005optimizing} as

\BE \label{regneumann}
	R_m(\bx_0; \bx_0) = \frac{1}{2\pi}\left\lbrack -\log\left|\bx_0|\bx_0| - \frac{\bx_0}{\bx_0} \right| + |\bx_0|^2 - \frac{3}{4} \right\rbrack \,.
\EE

\noI With \eqref{regneumann} for $R(\bx_0;\bx_0)$ in \eqref{massH}, a simply calculation shows that $M(r_0;0)$ is minimized when $r_0 = 0$. We show below that $M(r_0;\omega)$ is minimized at some $0 < r_0 < 1$ when $\omega > \omega_c$, where $\omega_c \approx 3.026$ is an $\mO(1)$ constant that we determine in \S \ref{smallr0}. When $\omega < \omega_c$ the optimal configuration is a stationary trap located at the origin.

For $\omega > 0$, we now compute $G(\bx; \bx_0)$ in the form of a Fourier series expansion. We first write the equation in \eqref{Geq} in polar coordinates as

\begin{equation}  \label{Geqpol}
G_{rr} + \frac{1}{r}G_r + \frac{1}{r^2}G_{\theta\theta} + \omega G_\theta = 
\frac{1}{\pi} - \frac{1}{r}\delta(r-r_0)\delta(\theta) \,,
\end{equation}

\noI where we have used that the location $\bx_0$ of the trap is along the $%
\theta = 0$ ray. We use separation of variables to write $G(\bx; \bx_0)$ as

\begin{equation}  \label{Gsep}
G(\bx; \bx_0) = G(r, \theta; r_0) = R_0(r) + \sum_{m > 0} e^{im\theta}R_m(r)
+ c.c. \,,
\end{equation}

\noI where $c.c.$ refers to the complex conjugate of the term involving the
summation. Substituting \eqref{Gsep} into \eqref{Geqpol} and recalling the
insulating boundary conditions in \eqref{Geqbc}, we obtain

\begin{empty}\bes \label{Req}
	\begin{equation} \label{m0}
		R_0^{\prime\prime} + \frac{1}{r}R_0^\prime = \frac{1}{\pi} - \frac{1}{2\pi r}\delta(r-r_0) \,, \quad R_0 \enspace \mbox{bounded as} \enspace r \to 0 \,, \quad R_0^\prime(1) = 0 \,,
	\end{equation}
	\begin{equation} \label{mn0}
		R_m^{\prime\prime} + \frac{1}{r}R_m^\prime + \left(i\omega m - \frac{m^2}{r^2} \right)R_m = -\frac{1}{2\pi r}\delta(r-r_0) \,,\quad m > 0 \,, \quad R_m \enspace \mbox{bounded as} \enspace r \to 0 \,, \quad R_m^\prime(1) = 0 \,.
	\end{equation}
\ees\end{empty}

\noI For $m > 0$, the homogeneous solution of \eqref{mn0} may be written as

\begin{equation}  \label{mn0sol}
R_m(r;\omega) = a_m I_m(c_mr) + b_m K_m(c_mr) \,; \qquad c_m \equiv -i 
\sqrt{i\omega m} \,,
\end{equation}

\noI where $I_m(r)$ and $K_m(r)$ are $m$-th order modified Bessel functions
of the first and second kind, respectively. Solving \eqref{mn0} separately
for $r < r_0$ and $r > r_0$, and applying appropriate continuity and jump
conditions at $r = r_0$, we obtain the solution for $R_m$,

\begin{empty}\bes \label{Gsol}
\begin{equation} \label{Rm}
	R_m(r;\omega) = \left\{
     \begin{array}{lr}
       \frac{1}{2\pi}\left \lbrack -\frac{K_m^\prime(c_m)}{I_m^\prime(c_m)}I_m(c_mr_0) + K_m(c_mr_0) \right\rbrack I_m(c_mr)\,, \quad &0 < r < r_0\\
       \frac{1}{2\pi}\left \lbrack -\frac{K_m^\prime(c_m)}{I_m^\prime(c_m)}I_m(c_m r) + K_m(c_m r) \right\rbrack I_m(c_mr_0)\,, \quad &r_0 < r < 1
     \end{array}
   \right. \,, \quad m > 0 \,,
\end{equation}

\noI where $I_m^\prime(c_m)$ and $K_m^\prime(c_m)$ denote the derivatives of $I_m$ and $K_m$ evaluated at $c_m$, respectively. In a similar way, we find that the solution to \eqref{m0} for $R_0(r)$ is

\begin{equation} \label{R0}
	R_0(r) = \frac{r^2}{4\pi} + a_0 - \left\{
     \begin{array}{lr}
       \frac{1}{2\pi}\log r_0 \,, \quad &0 < r < r_0\\
       \frac{1}{2\pi}\log r \,, \quad &r_0 < r < 1
     \end{array}
   \right. \,.
\end{equation}

\noI Note that the jump condition arising from the right-hand side of \eqref{m0} is automatically satisfied by \eqref{R0}. The constant $a_0$ is determined by the zero-mean condition in \eqref{Geqbc}, yielding

\begin{equation} \label{a0}
	a_0 = \frac{1}{8\pi}\lbrack 2r_0^2 - 3\rbrack \,.
\end{equation}
\ees\end{empty}

\noI The solution for $G(\bx;\bx_0)$ is then given by \eqref{Gsep} with %
\eqref{Gsol}.

To calculate $R(\bx_0;\bx_0)$, we use \eqref{Greg} to write

\begin{equation}  \label{reg}
R(\bx_0;\bx_0) = \lim_{\bx \to \bx_0} \left\lbrace G(\bx;\bx_0) + \frac{1}{%
2\pi}\log|\bx - \bx_0| \right\rbrace \,.
\end{equation}

\noI We next write $\log|\bx - \bx_0|$ in terms of its Fourier series as

\begin{equation}  \label{logseries}
\log|\bx - \bx_0| = \left\{ 
\begin{array}{lr}
\log r_0 - \frac{1}{2}\sum_{m > 0} \frac{1}{m}\left(\frac{r}{r_0}\right)^m
e^{im\theta} + c.c. \,, \quad & r < r_0 \\ 
\log r - \frac{1}{2}\sum_{m > 0} \frac{1}{m}\left(\frac{r_0}{r}\right)^m
e^{im\theta} + c.c. \,, \quad & r > r_0%
\end{array}
\right. \,.
\end{equation}

\noI Using the solution for $G(\bx;\bx_0)$ with $\theta \to 0$ and $r \to
r_0 $ as $\bx \to \bx_0$, we then use \eqref{logseries} to write \eqref{reg} as

\begin{equation}  \label{regseries}
R(\bx_0;\bx_0) = \frac{r_0^2}{2\pi} - \frac{3}{8\pi} + \sum_{m >
0}\left(R_m(r_0) - \frac{1}{4\pi m}\right) + c.c. \,.
\end{equation}

\noI Using \eqref{regseries}, we may then calculate the constant $H$ from %
\eqref{H}. The solution for $u$ is then given by \eqref{uG} with $G(\bx;\bx%
_0)$ given by \eqref{Gsep} with \eqref{Gsol} and $H$ given by \eqref{H} and %
\eqref{regseries}. A typical solution for $u$ with $\omega = 10$ and $r_0 =
0.6$ is shown in Figure \ref{usol}. The corresponding regular part of $u$ is
shown in Figure \ref{usolreg}. The contour plot of $u$ is shown in Figure %
\ref{usol_contour} and agrees with Figure \ref{flexcont_omega10}. Finally,
we calculate the mass $M$ in \eqref{massH} as

\begin{equation}  \label{massseries}
M(r_0;\omega) = \pi\left\lbrack \frac{r_0^2}{2} - \frac{3}{8} - \frac{1}{2}%
\log\varepsilon \right\rbrack + \pi^2 \sum_{m > 0}\left(R_m(r_0;\omega) - 
\frac{1}{4\pi m}\right) + c.c. \,.
\end{equation}

\noI Here, $c.c.$ represents the complex conjugate of the term involving the
summation, while the parametric dependence of $M$ on $\omega$ is through the
dependence of $R_m$ on $c_m$, defined in \eqref{mn0sol}.

\begin{empty}\begin{figure}[htbp]
  \begin{center}
    \mbox{
    \subfigure[$u(x,y)$] 
        {\label{usol}
        \includegraphics[width=.32\textwidth]{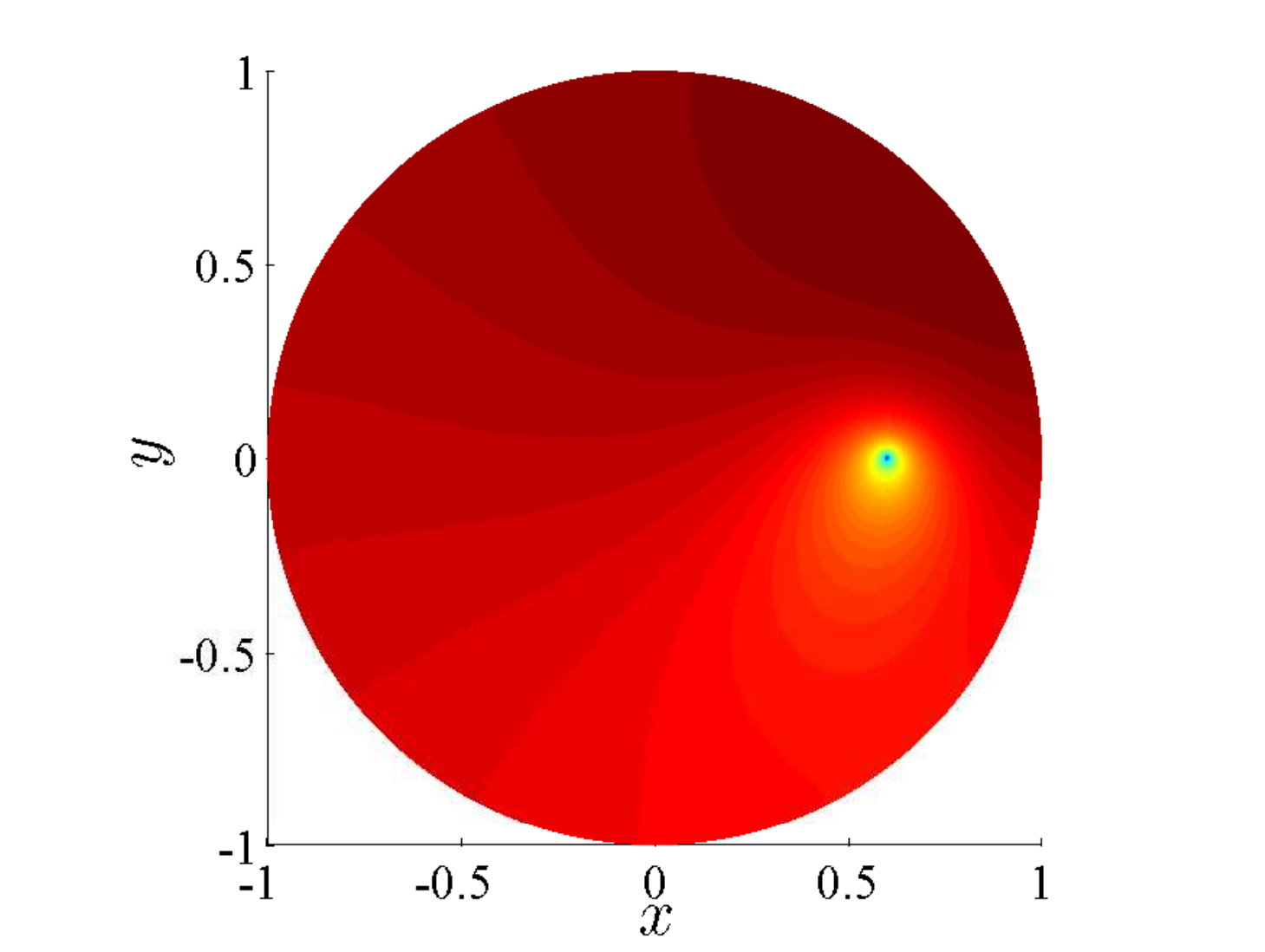}
        }   
    \subfigure[regular part of $u(x,y)$] 
        {\label{usolreg}
        \includegraphics[width=.32\textwidth]{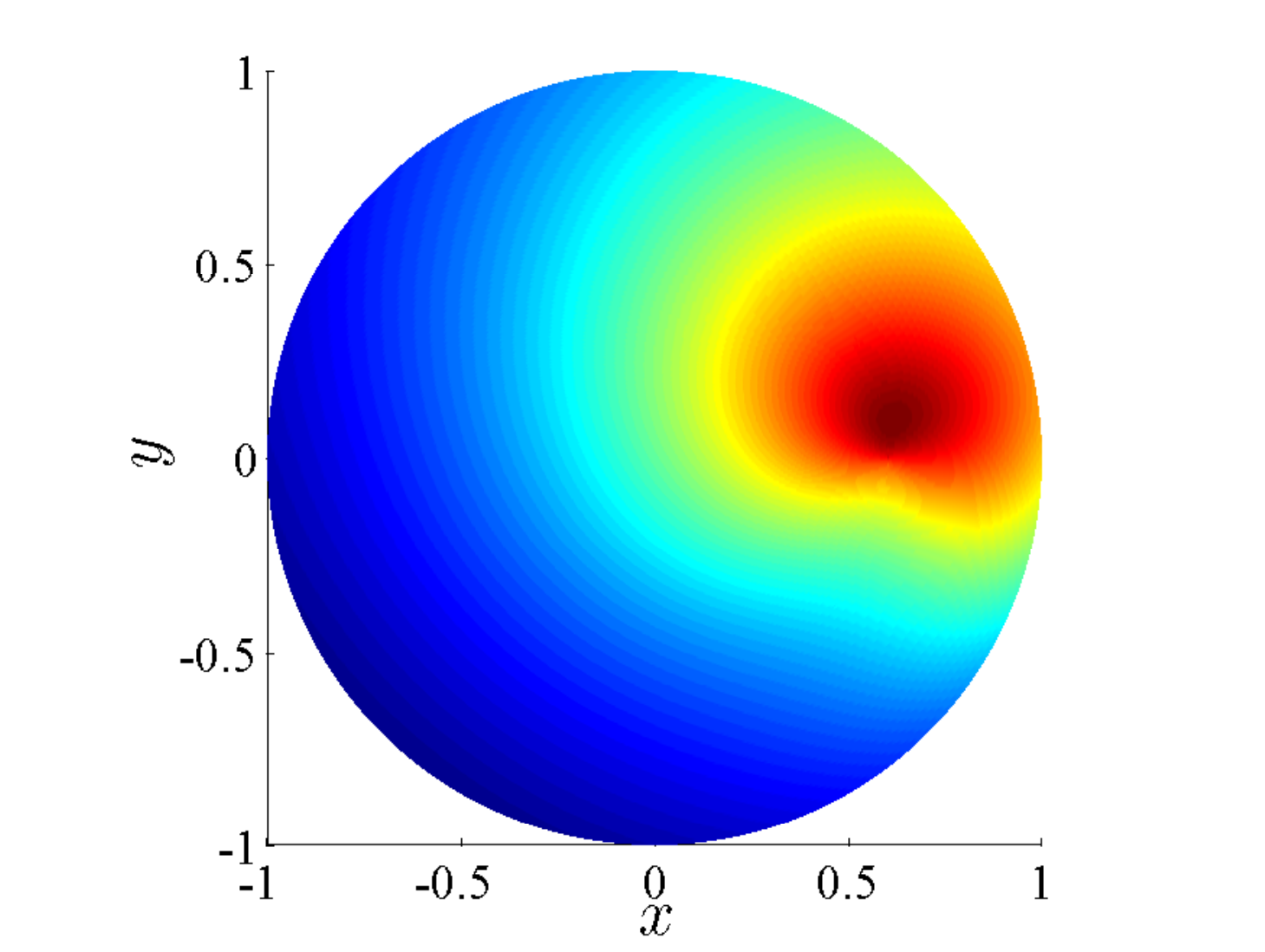}
        }
    \subfigure[contour plot of $u(x,y)$] 
        {\label{usol_contour}
        \includegraphics[width=.32\textwidth]{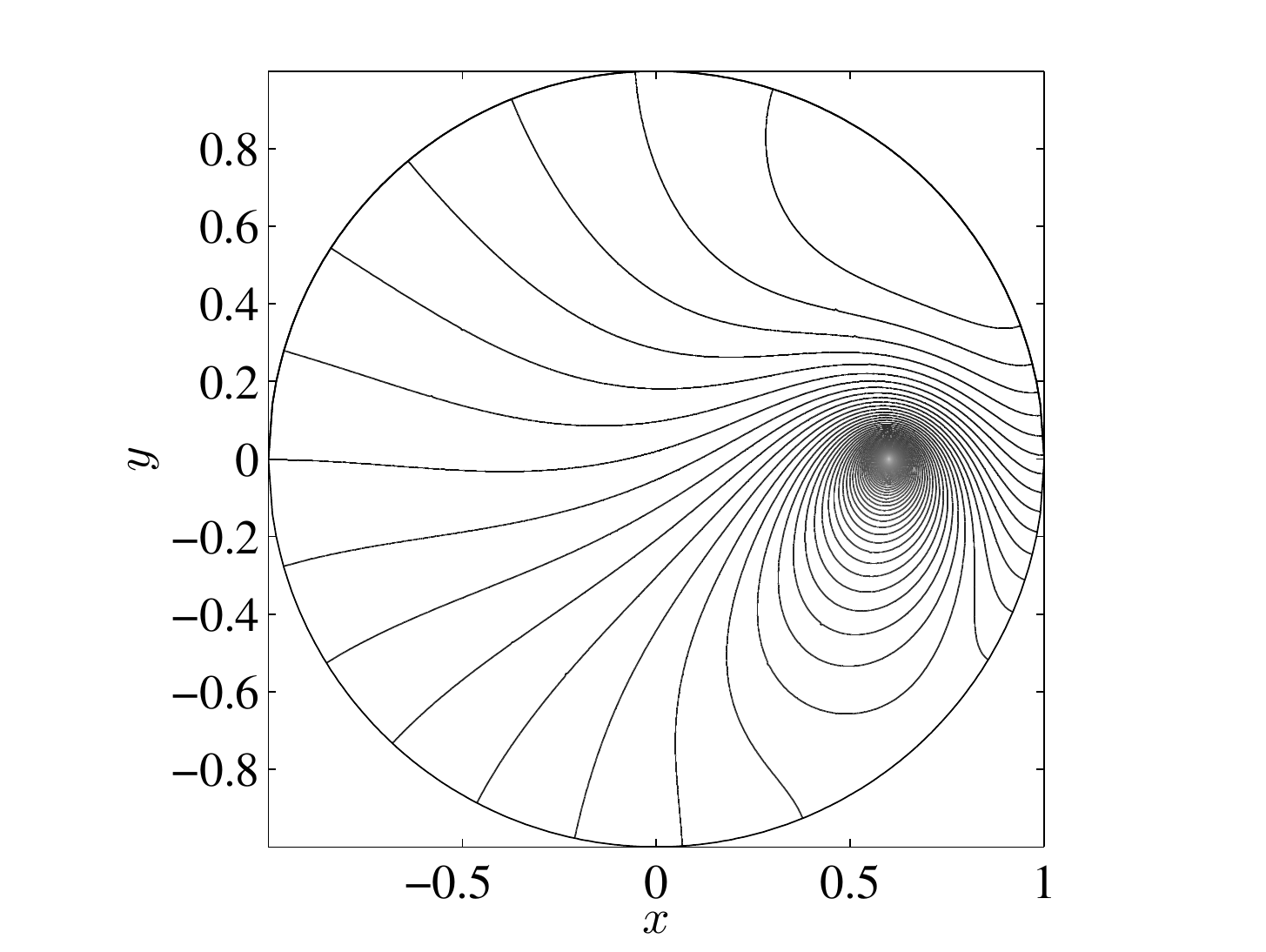}
        }}
    \caption{(a) Asymptotic solution $u(x,y)$ of \eqref{heateqn} with $\omega = 10$, $\varepsilon = 1\times 10^{-4}$, and $r_0 = 0.6$ as constructed from \eqref{uG}. (b) The corresponding regular part of $u(x,y)$. Red (blue) regions indicate large (small) values of $u$. (c) The contour plot of (a); compare with Figure \ref{flexcont_omega10} for the numerical solution with the same parameters.}
     \label{fig:usol}
  \end{center}
\end{figure}\end{empty}

For a range of $\omega \ll \mO(\varepsilon^{-1})$, we use \eqref{massseries}
to numerically determine the value of $r_0$ that minimizes $M$. The results
are presented in Figure \ref{fig:r0optw}. The first main feature of Figure %
\ref{r0optw} is the bifurcation that occurs near $\omega = \omega_c \approx
3.026$ (closeup in Figure \ref{r0optwbif}); for $\omega < \omega_c$, the
optimal radius of rotation remains zero. In \S \ref{smallr0} below, we
expand \eqref{massseries} for small $r_0 \ll 1$ to locate the exact value of 
$\omega_c$ at which the bifurcation occurs. The second main feature of
Figure \ref{r0optw} is the monotonic approach to $r_0^{opt} = 1$ for large $%
\omega$. In \S \ref{largeomega}, we construct a solution of \eqref{heateqn}
for $1 \ll \omega \ll \mathcal{O}(\varepsilon^{-1})$ to show that $r_0^{opt}
\to 1$ as $\omega \to \infty$ with $\omega \ll \mathcal{O}(\varepsilon^{-1})$%
. Note that this does not conflict with the result in \eqref{r0inf}, as the
analysis above, in particular the inner problem \eqref{innereq}, is valid only when $\omega \ll \mathcal{O}(\varepsilon^{-1})$%
. The regime $\omega = \mathcal{O}(\varepsilon^{-1})$ is a distinguished
limit and is discussed in \S \ref{omegaeps}.

\begin{empty}\begin{figure}[htbp]
  \begin{center}
    \mbox{
    \subfigure[$r_0^{opt}$ versus $\omega$] 
        {\label{r0optw}
        \includegraphics[width=.4\textwidth]{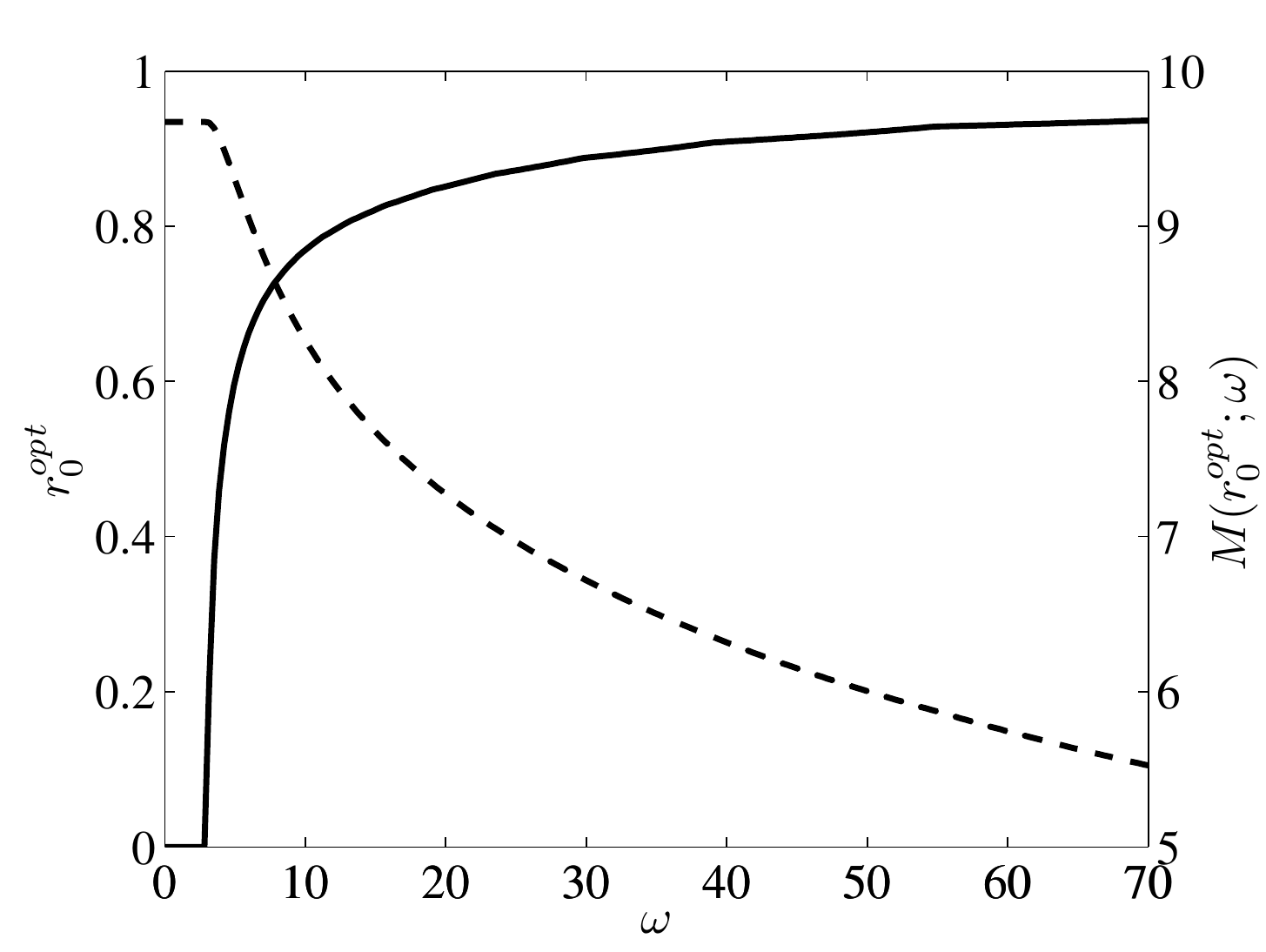}
        }   \hspace{1.5cm}
    \subfigure[$r_0^{opt}$ versus $\omega$ near bifurcation point] 
        {\label{r0optwbif}
        \includegraphics[width=.4\textwidth]{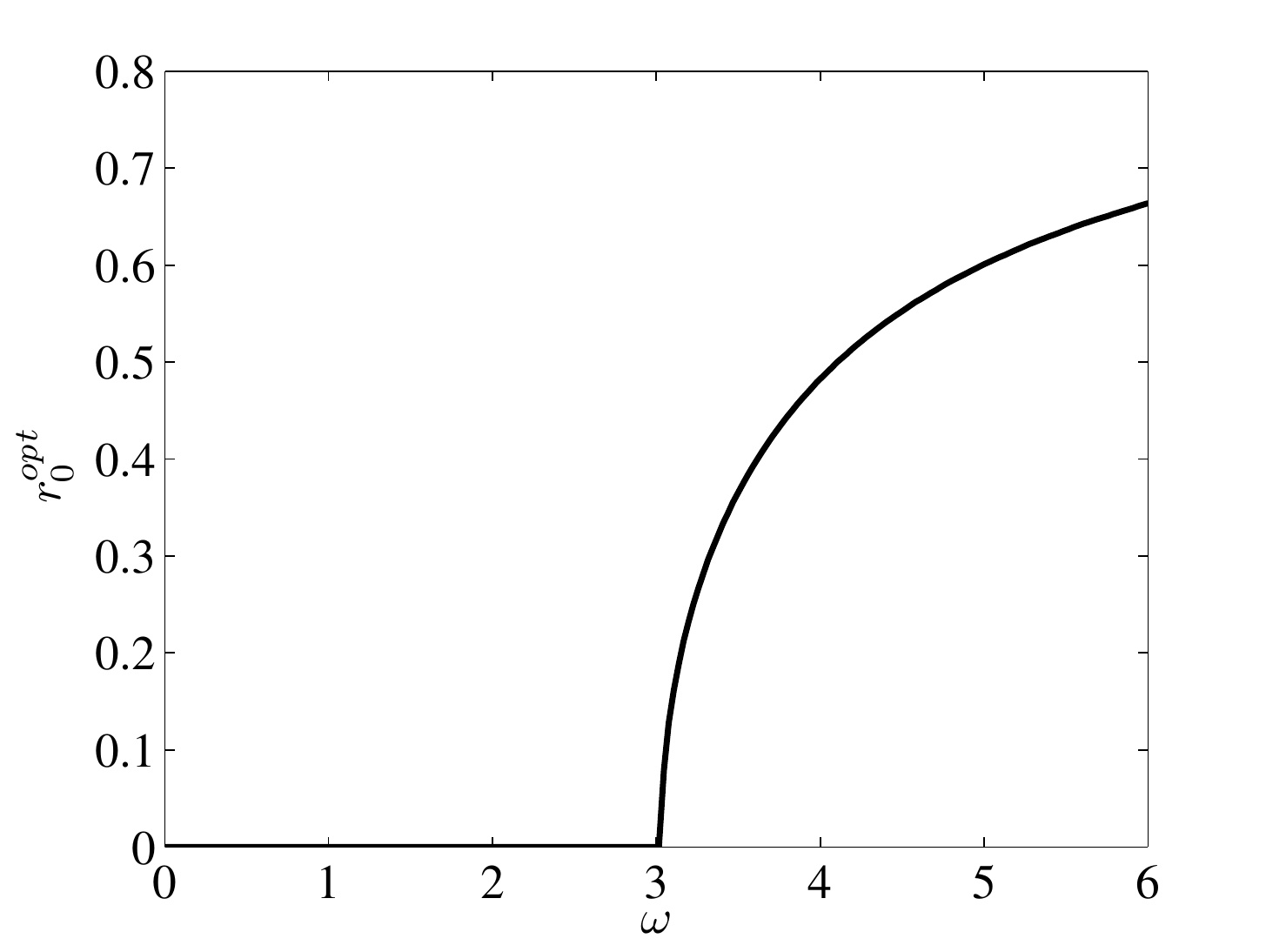}
        }}
    \caption{(a) Plot of $r_0^{opt}$ versus $\omega \ll \mathcal{O}(\varepsilon^{-1})$ (solid, left vertical axis) and the corresponding mass $M(r_0^{opt};\omega)$ (dashed, right vertical axis). The optimal radius remains zero for $\omega$ sufficiently small. (b) Closeup of the bifurcation point near $\omega \approx 3$ past which the optimal radius becomes non-zero.}
     \label{fig:r0optw}
  \end{center}
\end{figure}\end{empty}

\subsection{Bifurcation of $r_{0}^{opt}$ versus $\protect\omega $}

\label{smallr0}

The presence of a bifurcation of $r_{0}^{opt}$ near $\omega =3$ may be
confirmed by obtaining numerical solutions of \eqref{heateqn}. The computations were performed using the FlexPDE finite element PDE solver \cite{flex}. In
Figure \ref{fig:omegaopt}, we compare the mass $M(r_{0};\omega )$ as given
by \eqref{massseries} with that computed from numerical solutions of %
\eqref{heateqn}. In Figure \ref{omega02optimization}, we show that when $%
\omega =2$, the concavity at the point $r_{0}=0$ is positive with $%
M(r_0;\omega )$ increasing on the entire interval $0<r_{0}<1$, yielding $%
r_{0}^{opt}=0$. In Figures \ref{omega03p5optimization} and \ref%
{omega03p5optimization_zoom} with $\omega = 3.5$, we show that the concavity at $r_{0}=0$ has
become negative, thereby yielding $r_{0}^{opt}>0$. The bifurcation seen in
Figure \ref{r0optwbif} must then occur when the quadratic behavior of $%
M(r_{0};\omega )$ near $r_{0}=0$ changes from concave up to concave down. We
may thus determine the bifurcation point by expanding $M(r_{0};\omega )$ in %
\eqref{massseries} in powers of $r_{0}$ and calculating the value of $\omega 
$ at which the coefficient of $r_{0}^{2}$ changes sign. In the following analysis, we assume that $r_0 \gg \mO(\varepsilon)$.

\begin{empty}\begin{figure}[htbp]
  \begin{center}
    \mbox{
    \subfigure[$M(r_0)$ for $\omega = 2$] 
        {\label{omega02optimization}
        \includegraphics[width=.32\textwidth]{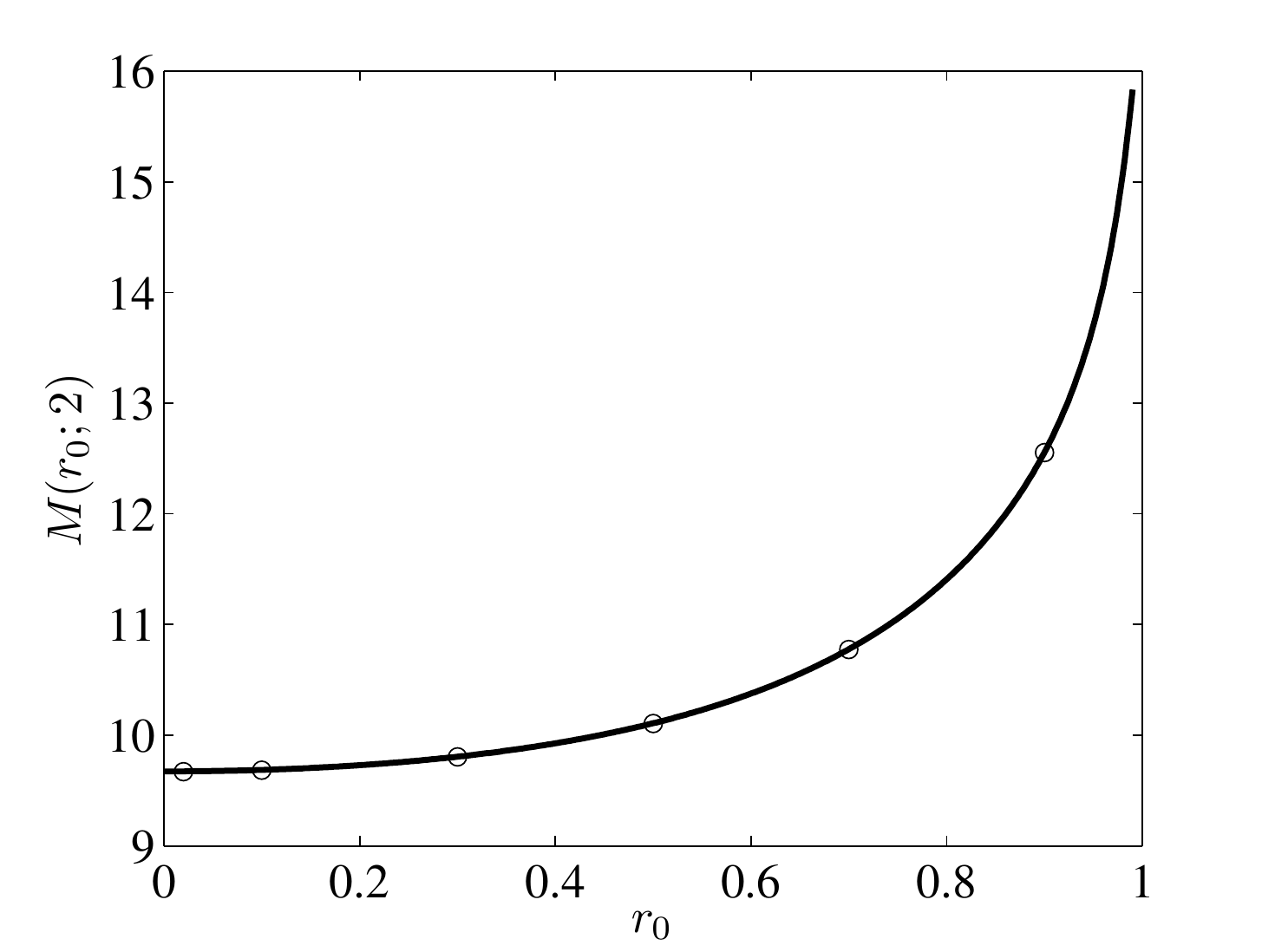}
        }   
    \subfigure[$M(r_0)$ for $\omega = 3.5$] 
        {\label{omega03p5optimization}
        \includegraphics[width=.32\textwidth]{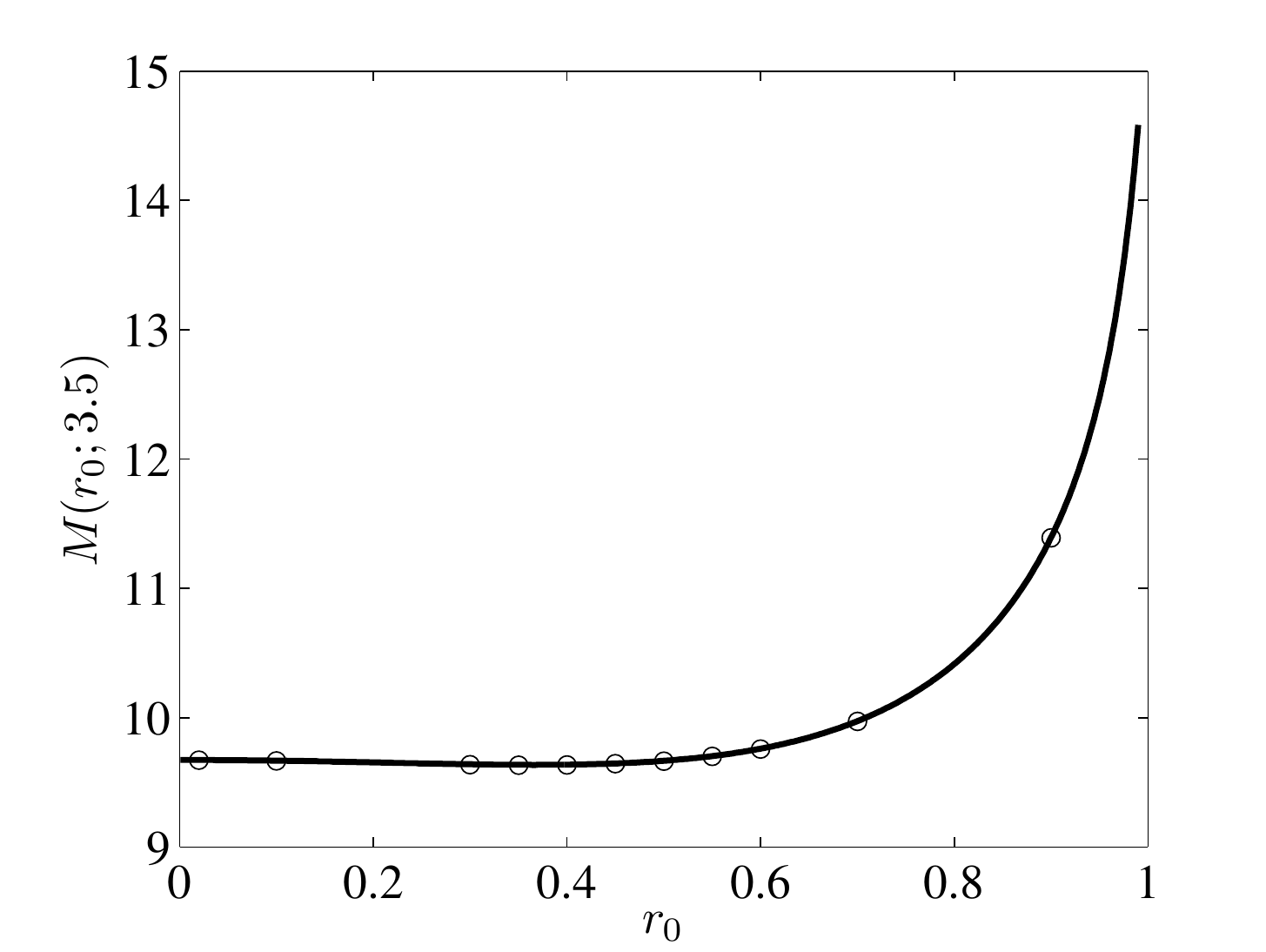}
        }
    \subfigure[$M(r_0)$ for $\omega = 3.5$ closeup] 
        {\label{omega03p5optimization_zoom}
        \includegraphics[width=.32\textwidth]{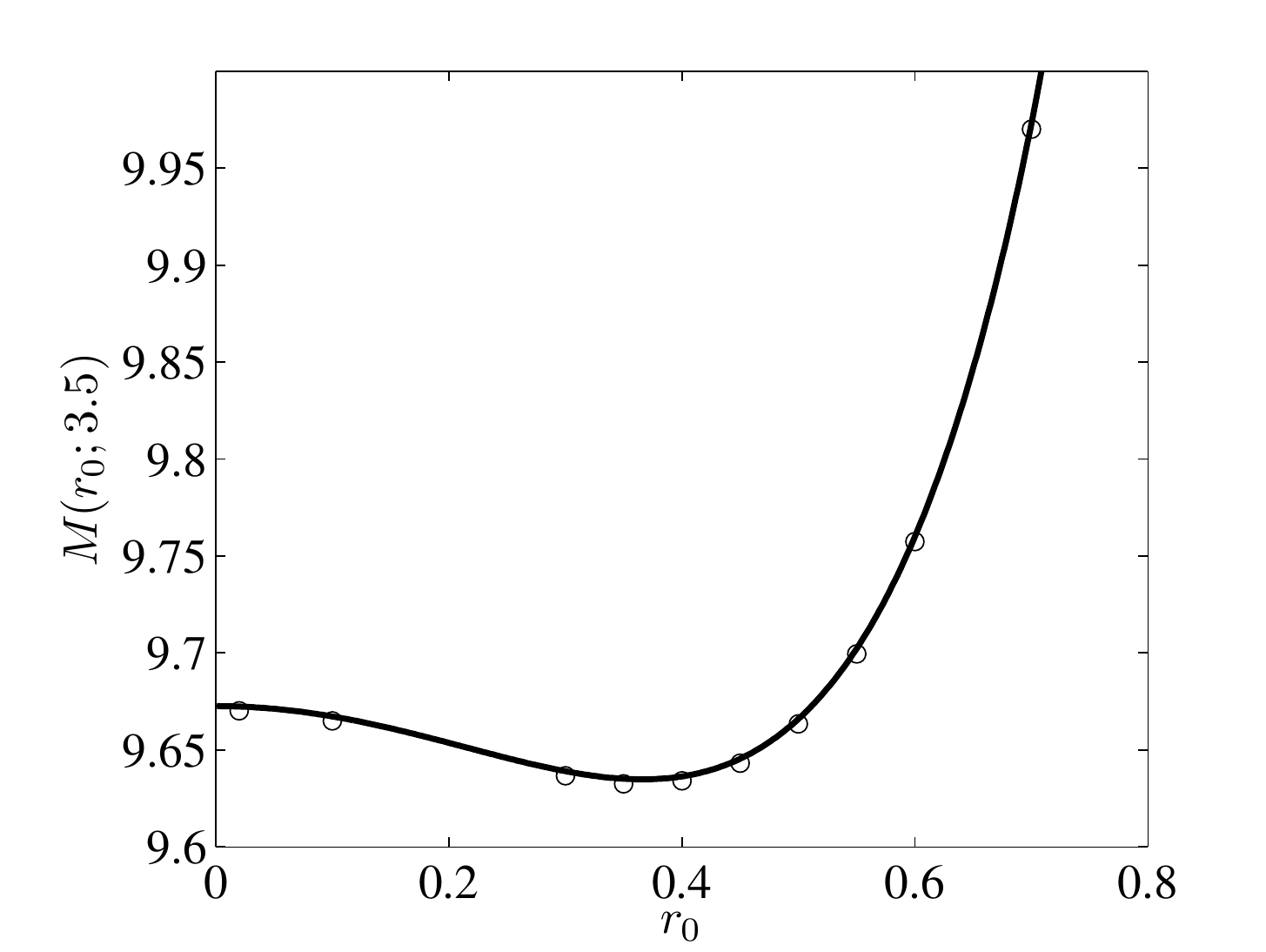}
        }}
    \caption{Plots of $M(r;\omega)$ for (a) $\omega = 2$ and (b),(c) $\omega = 3.5$. The solid curves are calculated from \eqref{massseries}, while the circles are obtained from numerical solutions of \eqref{heateqn}. In (a) with $\omega = 2$, the point $r_0 = 0$ is a global minimum so that $r_0^{opt} = 0$, while in (b) with $\omega = 3.5$, it is a local maximum (closeup in (c)), yielding $r_0^{opt} > 0$. Here, $\varepsilon = 1\times 10^{-3}$.}
     \label{fig:omegaopt}
  \end{center}
\end{figure}\end{empty}

To simplify calculations, we equivalently seek the leading order term of the
expansion in $r_0$ of the quantity

\begin{equation}  \label{S}
S = \frac{r_0^2}{2} - 2\Re\left\{ \sum_{m > 0}\left(-\pi R_m(r_0;\omega) + 
\frac{1}{4m} \right)\right\} \,,
\end{equation}

\noI where $R_m(r;\omega)$ is given in \eqref{Rm}. To do so, we write the
ascending series representation of $I_\nu(z)$ and $K_\nu(z)$ for $\nu > 0$
given in \cite{abramowitz1972handbook} as

\begin{empty}\bes \label{besseries}
	\begin{equation} \label{Iseries}
		I_\nu(z) = \left(\frac{z}{2}\right)^\nu \sum_{k=0}^{\infty} \frac{(z^2/4)^k}{k!\Gamma(\nu + k + 1)} \,,
	\end{equation}
	\begin{multline} \label{Kseries}
	 	K_\nu(z) = \frac{1}{2}\left(\frac{z}{2} \right)^{-\nu}\sum_{k=0}^{\nu-1}\frac{(n-k-1)!}{k!}\left(-\frac{z^2}{4} \right)^k + (-1)^{\nu+1}\log\left(\frac{z}{2}\right)I_\nu(z) \\ + (-1)^{\nu}\frac{1}{2}\left(\frac{z}{2} \right)^{\nu}\sum_{k=0}^{\infty}\left\lbrack \psi(k+1) + \psi(\nu+k+1) \right\rbrack \frac{(z^2/4)^k}{k!(n+k)!} \,,
	\end{multline}

\noI where $\gamma$ is Euler's constant, and $\psi(n)$ is given by

	\begin{equation} \label{psi}
		\psi(n) = \left\{
     \begin{array}{lr}
       -\gamma \,, \quad &n = 1\\
       -\gamma + \sum_{k=1}^{n-1}\frac{1}{k} \,, \quad &n > 1
     \end{array}
   \right. \,.
	\end{equation}
\ees\end{empty}

\noI With \eqref{besseries} and \eqref{Rm}, we find that

\begin{empty}\bes \label{Rseries}
\begin{equation} \label{R1series}
-\pi R_1(r_0;\omega) \sim -\frac{1}{4} + \frac{c_1^2}{8}\left\lbrack -\frac{1}{4} - \log\left(\frac{c_1r_0}{2}\right) + \frac{K_1^\prime(c_1)}{I_1^\prime(c_1)} + \frac{1}{2}(1-2\gamma)  \right\rbrack r_0^2 \,; \qquad c_1 \equiv -i\sqrt{i\omega}
\end{equation}
\begin{equation} \label{Rmseries}
-\pi R_m(r_0;\omega) \sim -\frac{1}{4m} + \frac{c_m^2}{8m(m^2-1)}r_0^2  \,; \qquad m > 1 \,, \quad c_m \equiv -i\sqrt{i\omega m} \,.
\end{equation}
\ees\end{empty}

\noI The $(4m)^{-1}$ term in \eqref{S} cancels the leading order constant
term in \eqref{R1series} and \eqref{Rmseries}. Further, since $c_m^2$ is pure imaginary, only the 
$m = 1$ term contributes to the leading order quadratic behavior of $S$. We
therefore have, for $\varepsilon \ll r_0 \ll 1$,

\begin{equation}  \label{Sleading}
S \sim a_2(\omega) r_0^2; \qquad a_2(\omega) \equiv \frac{1}{2} - 2\Re
\left\{\frac{c_1^2}{8}\left\lbrack -\frac{1}{4} - \log\left(\frac{c_1r_0}{2}%
\right) + \frac{K_1^\prime(c_1)}{I_1^\prime(c_1)} + \frac{1}{2}(1-2\gamma)
\right\rbrack \right\}\,,
\end{equation}

\noI where the dependence of $a_2(\omega)$ on $\omega$ is through $c_1$
defined in \eqref{R1series}. The value $\omega = \omega_c$ at which the
concavity of $S$ at $r_0 = 0$ changes sign is the value at which the
optimality of $r_0 = 0$ is lost. The bifurcation point therefore must satisfy

\begin{equation}  \label{omegac}
a_2(\omega_c) = 0 \,,
\end{equation}

\noI where $a_{2}(\omega )$ is defined in \eqref{Sleading}. Solving %
\eqref{omegac} numerically for $\omega _{c}$, we find that the bifurcation
in Figure \ref{r0optwbif} occurs at $\omega _{c}\approx 3.026$.%
\setcounter{equation}{0}

\section{Leading order solution for large $\protect\omega $ with $\protect%
\omega \ll \mathcal{O}(\protect\varepsilon ^{-1})$}

\label{largeomega}

As $\omega$ in \eqref{heatpde} becomes large with $\omega \ll \mathcal{O}%
(\varepsilon^{-1})$, an internal layer of width $\mathcal{O}(\omega^{-1/2})$
develops in a trail behind the rotating trap. An example of this is shown in
Figure \ref{flexcont_omega1000}, obtained by numerically solving %
\eqref{heateqn} with $\omega = 1000$ and $\varepsilon = 1\times 10^{-4}$. An
asymptotic solution with the same parameters is shown in Figure \ref%
{usol_lomega}. The internal layer centered on the ring $r = r_0$ may be
clearly seen in the corresponding contour plot in Figure \ref%
{usol_lomega_contour}. Away from the internal layer, the solution is nearly
radially symmetric. We now construct this solution and derive an
approximation to the mass $M(r_0;\omega)$ in \eqref{massseries} for large $%
\omega$. We then show that as $\omega \to \infty$ with $\omega \ll \mO%
(\varepsilon^{-1})$, the optimal radius $r_0^{opt} \to 1$. The analysis
below assumes that $1-r_0\sim \mO(1)$ and $r_0\omega \gg 1$.

\begin{empty}\begin{figure}[htbp]
  \begin{center}
    \mbox{
    \subfigure[$u(x,y)$] 
        {\label{usol_lomega}
        \includegraphics[width=.5\textwidth]{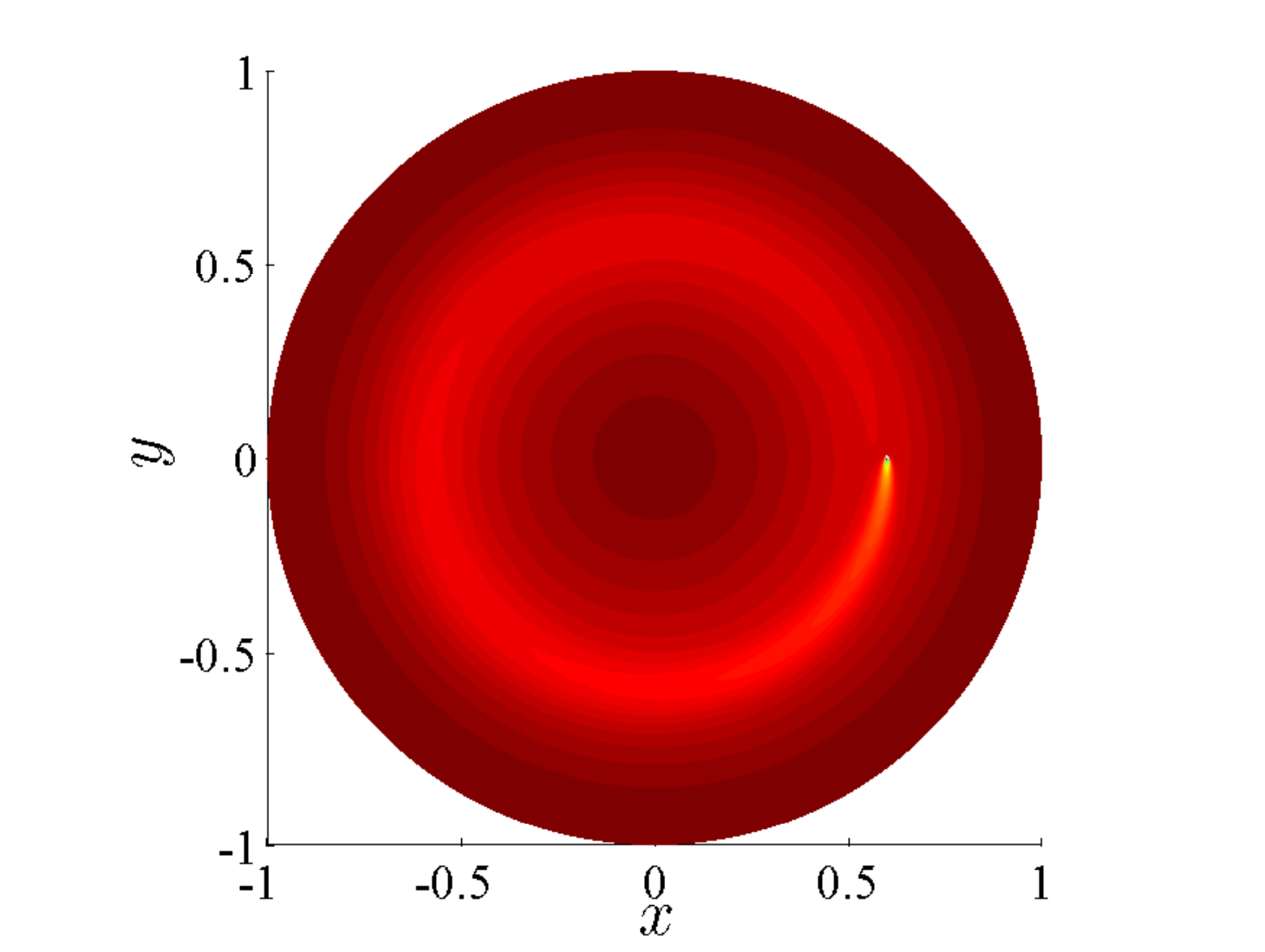}
        }   
    \subfigure[contour plot of $u(x,y)$] 
        {\label{usol_lomega_contour}
        \includegraphics[width=.5\textwidth]{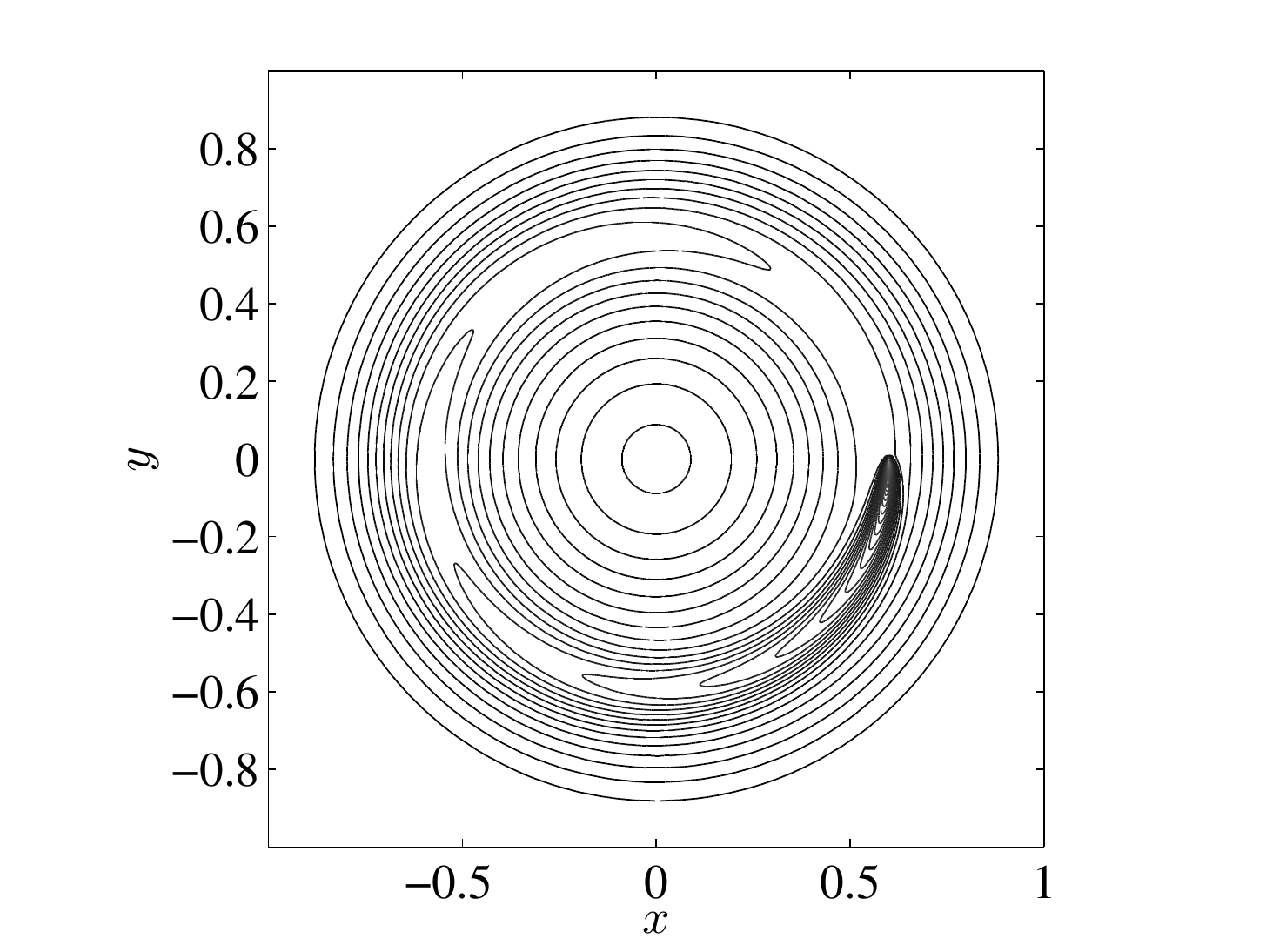}
        }}
    \caption{(a) Asymptotic solution $u(x,y)$ of \eqref{heateqn} with $\omega = 1000 \gg 1$, $\varepsilon = 1\times 10^{-4}$, and $r_0 = 0.6$ as constructed from \eqref{uG}. Red (blue) regions indicate large (small) values of $u$. (b) The corresponding contour plot of $u(x,y)$. An internal layer of width $\mathcal{O}(\omega^{-1/2})$ centered on the ring $r = r_0$ is clearly seen. The solution is nearly radially symmetric away from the internal layer. Compare with Figure \ref{flexcont_omega1000} for the numerical solution with the same parameters.}
     \label{fig:usol_lomega}
  \end{center}
\end{figure}\end{empty}

To construct a solution of \eqref{heatpde}, we first identify three distinct
regions of the solution of \eqref{Geq}. In addition to the $\mathcal{O}%
(\varepsilon)$ region identified in \eqref{innereq} and the $%
\mathcal{O}(\omega^{-1/2}) $ internal parabolic layer seen in Figure \ref%
{usol_lomega_contour}, there is an elliptic layer of extent $\mathcal{O}%
(\omega^{-1})$ surrounding the $\mathcal{O}(\varepsilon)$ region. The
solution will be constructed by matching the elliptic layer to the $\mathcal{%
O}(\varepsilon)$ inner region, and then the parabolic layer to the elliptic
layer.

For the elliptic layer, the Cartesian coordinate system is convenient. In
the $\mathcal{O}(\omega^{-1})$ vicinity of the trap, we make the change of
variables

\begin{equation}  \label{elayer}
\xi = \omega(x-r_0) \,, \quad \eta = \omega y \,; \qquad G(r,\theta) = \hat{G%
}(\xi,\eta) \,.
\end{equation}

\noI Substituting \eqref{elayer} into \eqref{Geqpol} with $G_\theta = xG_y -
yG_x$ and using the scaling property of the delta function $\delta(ax) =
\delta(x)/|a|$, the leading order equation for $\omega \gg 1$ becomes

\begin{equation}  \label{Geqlomega}
\hat{G}_{\xi\xi} + \hat{G}_{\eta\eta} + r_0 \hat{G}_{\eta} =
-\delta(\xi)\delta(\eta) \,; \qquad -\infty <\xi \,, \eta < \infty \,, \quad 
\hat{G} \enspace \mbox{bounded as} \enspace |\xi| \,, |\eta| \to \infty \,.
\end{equation}

\noI The condition at infinity in \eqref{Geqlomega} is required to match the
elliptic intermediate layer to the parabolic internal layer. To solve %
\eqref{Geqlomega}, we proceed as in \cite{hahnozisikheatcond}
\, and write

\begin{equation}  \label{Gansatz}
\hat{G}(\xi, \eta) = \mathcal{G}(\rho)e^{-\frac{r_0}{2}\eta} + \hat{H} \,;
\qquad \rho^2 = \xi^2 + \eta^2 \,, \quad -\infty < \xi\,, \eta < \infty \,, \quad 0
< \rho < \infty \,,
\end{equation}

\noI where $\hat{H}$ is a constant to be determined from the zero
mean-condition in \eqref{Geqbc}. Substituting \eqref{Gansatz} into %
\eqref{Geqlomega}, we calculate

\begin{empty}\bes
\begin{equation} \label{mGeq}
	\mathcal{G}_{\rho\rho} + \frac{1}{\rho}\mathcal{G}_\rho - \left(\frac{r_0}{2} \right)^2\mathcal{G} = -\frac{1}{2\pi \rho}\delta(\rho)\,; \qquad 0 < \rho < \infty \,,
\end{equation}
\begin{equation} \label{mGbc}
	\mathcal{G} \enspace \mbox{bounded as} \enspace \rho \to \infty \,.
\end{equation}
\ees\end{empty}

\noI The homogeneous solution of \eqref{mGeq} is given by a linear
combination of modified Bessel functions

\begin{equation}  \label{bess0}
\mathcal{G}(\rho) = c_1I_0\left(\frac{r_0\rho}{2}\right) + c_2K_0\left(\frac{%
r_0\rho}{2}\right) \,.
\end{equation}

\noI In \eqref{bess0}, $c_1 = 0$ by the boundedness condition in \eqref{mGbc}%
, while $c_2$ is determined by integrating \eqref{mGeq} over a circle of
radius $\delta \to 0$,

\begin{equation}  \label{intcond}
\lim_{\delta\to0} \left.2\pi\delta c_2 \frac{d}{d\rho} K_0\left(\frac{r_0\rho%
}{2}\right) \right|_{\rho = \delta} = -1 \,.
\end{equation}

\noI Using the small argument asymptotics for $K_0(z)$,

\begin{equation}  \label{K0asymp}
K_0(z) \sim -\log\frac{z}{2} - \gamma \,,
\end{equation}

\noI we calculate from \eqref{intcond} that $c_2 = (2\pi)^{-1}$ in \eqref{bess0}. The solution of %
\eqref{Geqlomega} for the elliptic layer is then given by

\begin{equation}  \label{elayersol}
\hat{G}(\xi,\eta) = \frac{1}{2\pi}K_0\left(\frac{r_0}{2}\sqrt{\xi^2 + \eta^2}%
\right)e^{-\frac{r_0}{2}\eta} + \hat{H} \,,
\end{equation}

\noI where $\hat{H}$ is a constant to be computed, while $\xi$ and $\eta$
are defined in \eqref{elayer}.

For the parabolic layer of thickness $\mathcal{O}(\omega^{-1/2})$, we
introduce the scaled variables

\begin{equation}  \label{player}
\tilde{\theta} = 2\pi-\theta \,, \quad \tilde{r} = \sqrt{\omega}(r-r_0) \,;
\qquad G(r,\theta) = \tilde{G}(\tilde{r}, \tilde{\theta}) \,.
\end{equation}

\noI Substituting \eqref{player} into \eqref{Geqpol} and collecting terms of 
$\mathcal{O}(\omega)$, we obtain the parabolic equation

\begin{equation}  \label{playereq}
\tilde{G}_{\tilde{\theta}} = \tilde{G}_{\tilde{r}\tilde{r}} \,; \qquad 0<%
\tilde{\theta}<2\pi \,, \quad -\infty < \tilde{r} < \infty \,.
\end{equation}

\noI We require boundedness of $\tilde{G}$ as $|\tilde{r}| \to \infty$ in
order to match to the outer region. We now compute a solution of %
\eqref{playereq} that matches the behavior of the elliptic layer %
\eqref{elayersol} as $\eta \to -\infty$. To do so, we first use the large
argument asymptotic form $K_0(z) \sim \sqrt{\frac{\pi}{2z}}e^{-z}$ as $z \to \infty$, to calculate



\begin{equation}  \label{Geta}
\hat{G}(\xi,\eta) \sim \frac{1}{\sqrt{4\pi r_0|\eta|}}\,e^{-\frac{r_0\xi^2}{%
4|\eta|}} + \hat{H}\,, \qquad \eta \to -\infty \,.
\end{equation}

\noI To write $\xi$ and $\eta$ in terms of $\tilde{r}$ and $\tilde{\theta}$,
we first note that, near $r = r_0$ and $\tilde{\theta} = 0^+$, we have that $%
x \sim r_0 + \tilde{r}$ and $y \sim -r_0\tilde{\theta}$. With $\xi$ and $%
\eta $ defined in \eqref{elayer}, we obtain

\begin{equation}  \label{xieta}
\xi \sim \sqrt{\omega}\tilde{r} \,, \quad \eta \sim -\omega r_0 \tilde{\theta%
} \,; \qquad \tilde{\theta} > 0 \,.
\end{equation}

\noI Substituting \eqref{xieta} into \eqref{Geta}, we obtain the solution
for the parabolic layer

\begin{equation}  \label{playersol}
\tilde{G}(\tilde{r},\tilde{\theta}) = \frac{1}{2r_0\sqrt{\pi\omega\tilde{%
\theta}}} \,e^{-\frac{\tilde{r}^2}{4\tilde{\theta}}} + \hat{H} \,.
\end{equation}

\noI The solution \eqref{playersol} may also be obtained in a similar way by
explicitly calculating the initial condition for \eqref{playereq} in terms
of a weighted delta function

\begin{equation}  \label{playericdeltaH}
\tilde{G}(\tilde{r},0) = \frac{1}{r_0\sqrt{\omega}}\delta(\tilde{r}) + \hat{H%
} \,.
\end{equation}

\noI The solution to \eqref{playereq} with initial conditions given by %
\eqref{playericdeltaH} may then be written in terms of the fundamental
solution of the diffusion equation, yielding \eqref{playersol}.

With \eqref{elayersol} and \eqref{playersol}, the inner solution for $G$ in %
\eqref{Geqpol} near the ring $r = r_0$ is then given by the composite
solution $G_i(r,\theta) = \hat{G} + \tilde{G} - c_p$, where $c_p$ is the
common part given by \eqref{Geta}. We thus calculate

\begin{equation}  \label{Gi}
G_i(r,\theta) = \frac{1}{2\pi}K_0\left(\frac{r_0}{2}\sqrt{\xi^2 + \eta^2}%
\right)e^{-\frac{r_0}{2}\eta} + \frac{1}{2r_0\sqrt{\pi\omega(2\pi-\theta)}}%
e^{-\frac{\omega(r-r_0)^2}{4(2\pi-\theta)}} - \frac{1}{2\sqrt{\pi r_0|\eta|}}%
\,e^{-\frac{r_0\xi^2}{4|\eta|}}\Theta(-\eta) + \hat{H} \,,
\end{equation}

\noI where $\xi = \xi(r,\theta)$ and $\eta = \eta(r,\theta)$ are defined in %
\eqref{elayer}. For the outer solution $G_0$ of \eqref{Geqpol}, we note
that, to leading order in $\omega$, $G_{0\theta} = 0$. For $G_0 = G_0(r)$
radially symmetric, we integrate both sides of \eqref{Geqpol} from $\theta:
0\to 2\pi$ to obtain

\begin{empty}\bes \label{Geqpolradsymmboth}
\begin{equation} \label{Geqpolradsymm}
	G_{0rr} + \frac{1}{r}G_{0r}  = \frac{1}{\pi} - \frac{1}{2\pi r}\delta(r-r_0) \,; \qquad 0<r<1 \,, \quad G_{0r}(1) = 0 \,.
\end{equation}

\noI A unique solution of \eqref{Geqpolradsymm} may be obtained by imposing the matching condition

\begin{equation} \label{Geqpolmatch}
	G_0(r_0) = \hat{H} \,,
\end{equation}
\ees\end{empty}

\noI obtained from letting $\omega \to \infty$ in \eqref{Gi} with $|r-r_0|$
remaining of $\mathcal{O}(1)$. The solution of \eqref{Geqpolradsymmboth} is
then

\begin{equation}  \label{G0}
G_0(r) = \frac{r^2-r_0^2}{4\pi} - \frac{1}{2\pi}\Theta(r-r_0)\log\left(\frac{%
r}{r_0} \right) + \hat{H}\,,
\end{equation}

\noI where $\Theta(r)$ is the Heaviside step function. The leading order
composite solution of \eqref{Geqpol} for $\omega \gg 1$ is then given by $G
= G_0 + G_i - \hat{H}$, yielding

\begin{empty}\bes \label{Gcompfull}
\begin{multline}\label{Gcomp}
	G(r,\theta) = \frac{r^2-r_0^2}{4\pi} - \frac{1}{2\pi}\Theta(r-r_0)\log\left(\frac{r}{r_0} \right) + \frac{1}{2\pi}K_0\left(\frac{r_0}{2}\sqrt{\xi^2 + \eta^2}\right)e^{-\frac{r_0}{2}\eta} + \\ + \frac{1}{2r_0\sqrt{\pi\omega(2\pi-\theta)}}e^{-\frac{\omega(r-r_0)^2}{4(2\pi-\theta)}} - \frac{1}{2\sqrt{\pi r_0|\eta|}}\,e^{-\frac{r_0\xi^2}{4|\eta|}}\Theta(-\eta) + \hat{H} + \mathcal{O}(\omega^{-1}) \,,
\end{multline}

\noI where we have used \eqref{Gi} and \eqref{G0} for $G_i$ and $G_0$. The constant $\hat{H}$ is determined by the zero-mean condition in \eqref{Geqbc}. Since the solution in \eqref{Gcomp} omits terms of order $\mO(\omega^{-1})$, and with inner layer terms contributing a mean of $\mathcal{O}(\omega^{-1})$, we need only account for the mean of the first two terms in \eqref{Gcomp}. That is,

\begin{equation} \label{Hhat}
	\hat{H} = -\frac{1}{\pi}\left\lbrack -\frac{r_0^2}{2}+\frac{3}{8}+\frac{1}{2}\log r_0 \right\rbrack + \mathcal{O}(\omega^{-1})\,.
\end{equation}
\ees\end{empty}

\noI The solution to $u$ is then given by \eqref{uG} with $G$ and $\hat{H}$
defined in \eqref{Gcompfull}.

We now calculate the constant $H$ in \eqref{uG} by the matching condition
given in \eqref{outlog} with $S = 1/2$. To determine the asymptotic behavior
of $G$ as $\bx \to \bx_0$, we first note that the second and third terms in %
\eqref{Gi}, by construction, cancel near the trap, while $G_0(r) \to \hat{H}$%
. Therefore, using the small argument asymptotics for $K_0(z)$ in %
\eqref{K0asymp}, we calculate that

\begin{equation}  \label{Gasymp}
G \sim \frac{1}{2\pi}\left\lbrack -\log|\bx - \bx_0| - \log\left(\frac{%
r_0\omega}{4}\right) - \gamma \right\rbrack + \hat{H} \,, \quad \mbox{as} %
\enspace \bx \to \bx_0 \,,
\end{equation}

\noI where we have used \eqref{elayer} to write $\xi$ and $\eta$ in terms of 
$x$ and $y$. With the asymptotics for $G$ in \eqref{Gasymp}, \eqref{outlog} and \eqref{uG} yield the matching condition at the trap

\begin{equation}  \label{matchlomega}
\frac{1}{2}\left\lbrack \log|\bx - \bx_0| + \log\left(\frac{r_0\omega}{4}%
\right) + \gamma \right\rbrack -\pi \hat{H} + H \sim \frac{1}{2}\log|\bx - %
\bx_0| - \frac{1}{2}\log\varepsilon \,.
\end{equation}

\noI Solving for $H$ in \eqref{matchlomega}, we obtain

\begin{equation}  \label{Hlomega}
H = \pi\hat{H} - \frac{1}{2}\left\lbrack \log\left(\frac{r_0\omega\varepsilon}{4}%
\right) + \gamma \right\rbrack \,.
\end{equation}

\noI In Figures \ref{usol_lomega_asymp} and \ref{usol_lomega_asymp_contour},
we show a solution constructed with $G$ and $H$ as given in \eqref{Gcompfull}
and \eqref{Hlomega}. The parameters are the same as those used in Figures %
\ref{flexcont_omega1000} and \ref{fig:usol_lomega}. In Figure \ref%
{ring_lomega}, we show the corresponding value of $u$ along the ring $r =
r_0 $. The solid curve is computed numerically from the series expansion of 
\S \ref{2Dtrap}, while the dashed curve is computed from the asymptotic
construction \eqref{Gcompfull} and \eqref{Hlomega}. The figure indicates
excellent agreement between the two results.

\begin{empty}\begin{figure}[htbp]
  \begin{center}
    \mbox{
    \subfigure[$u(x,y)$] 
        {\label{usol_lomega_asymp}
        \includegraphics[width=.32\textwidth]{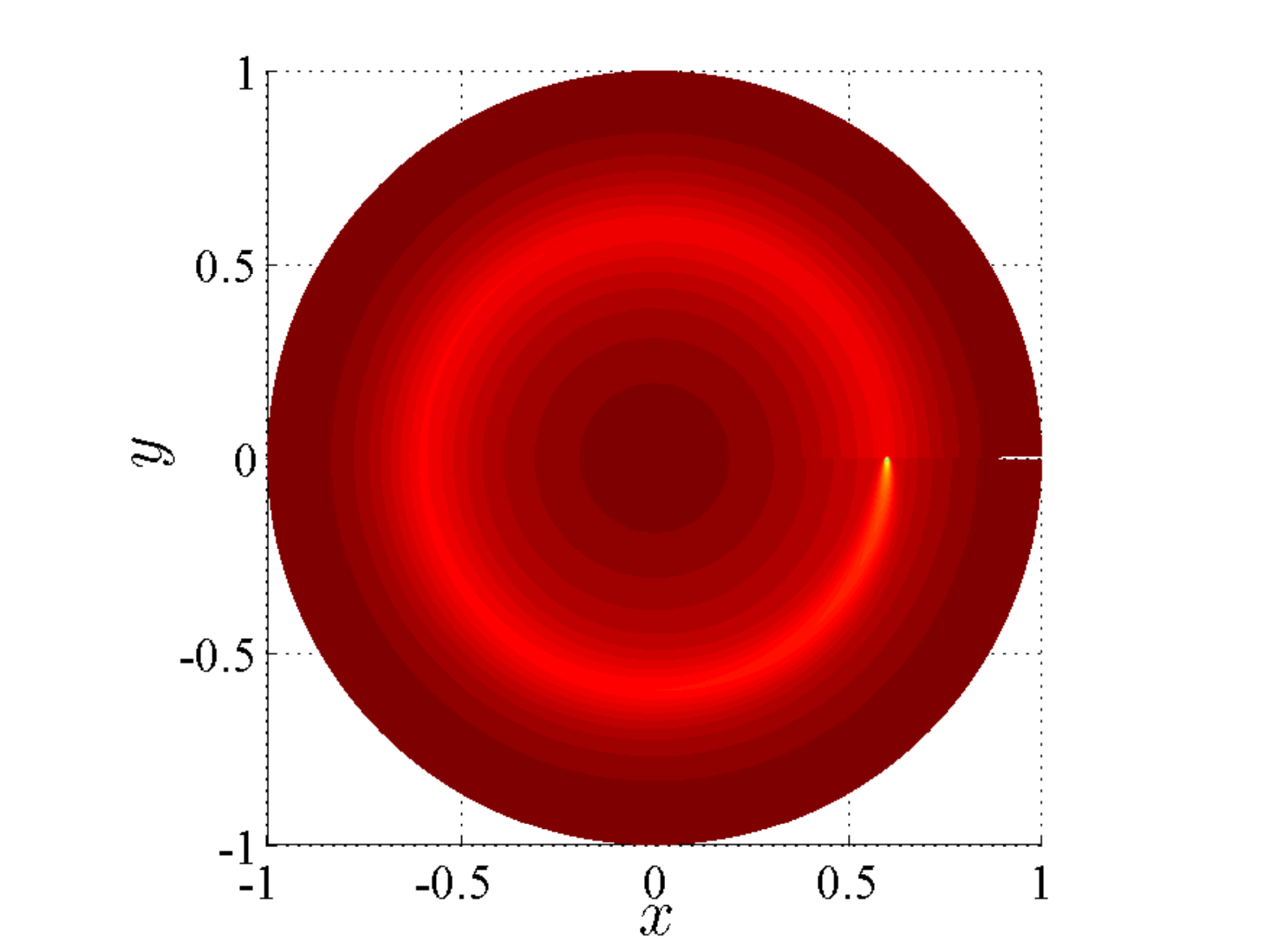}
        }   
    \subfigure[contour plot of $u(x,y)$] 
        {\label{usol_lomega_asymp_contour}
        \includegraphics[width=.32\textwidth]{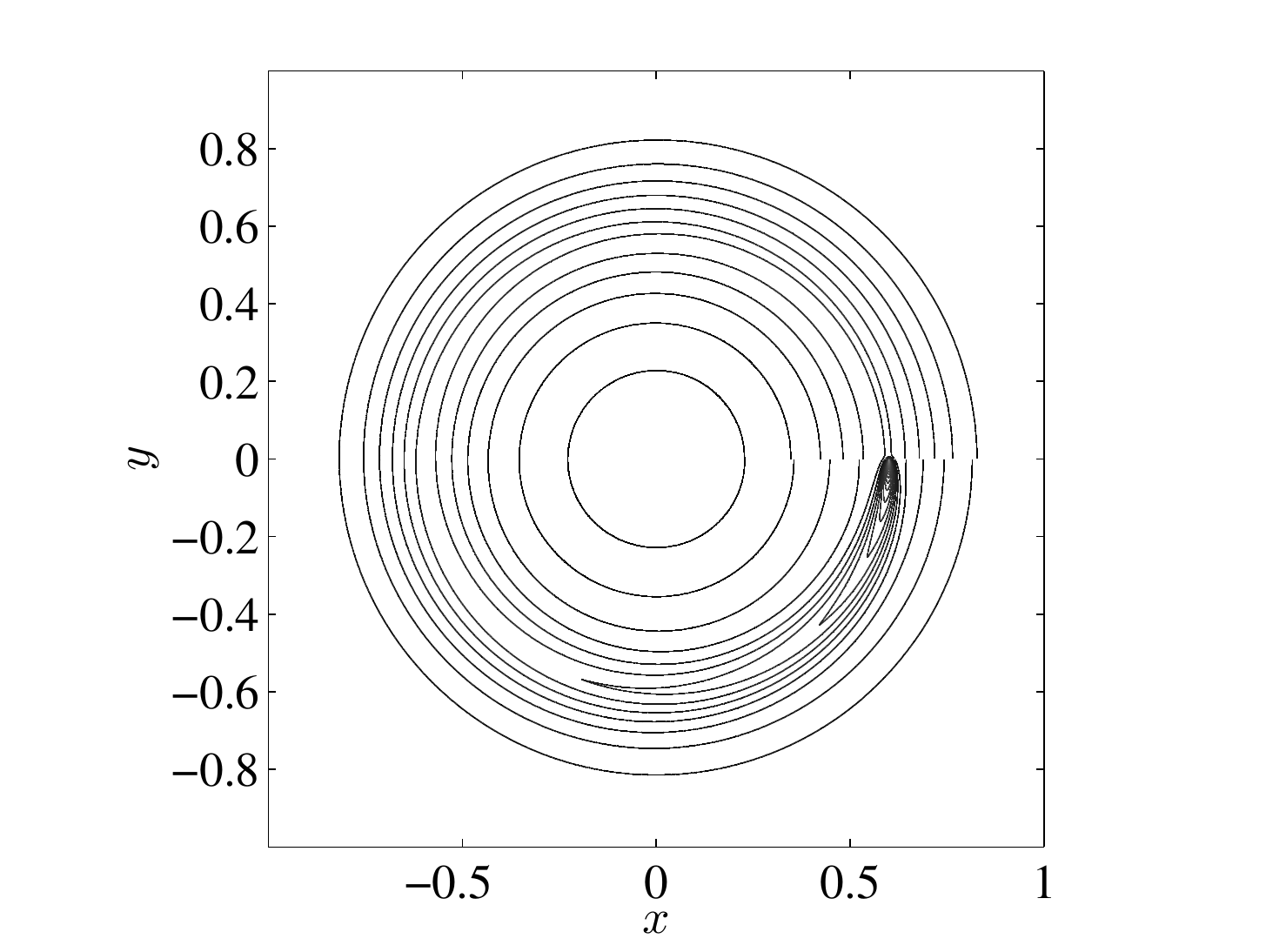}
        }
    \subfigure[value of $u$ on $r = r_0$] 
        {\label{ring_lomega}
        \includegraphics[width=.32\textwidth]{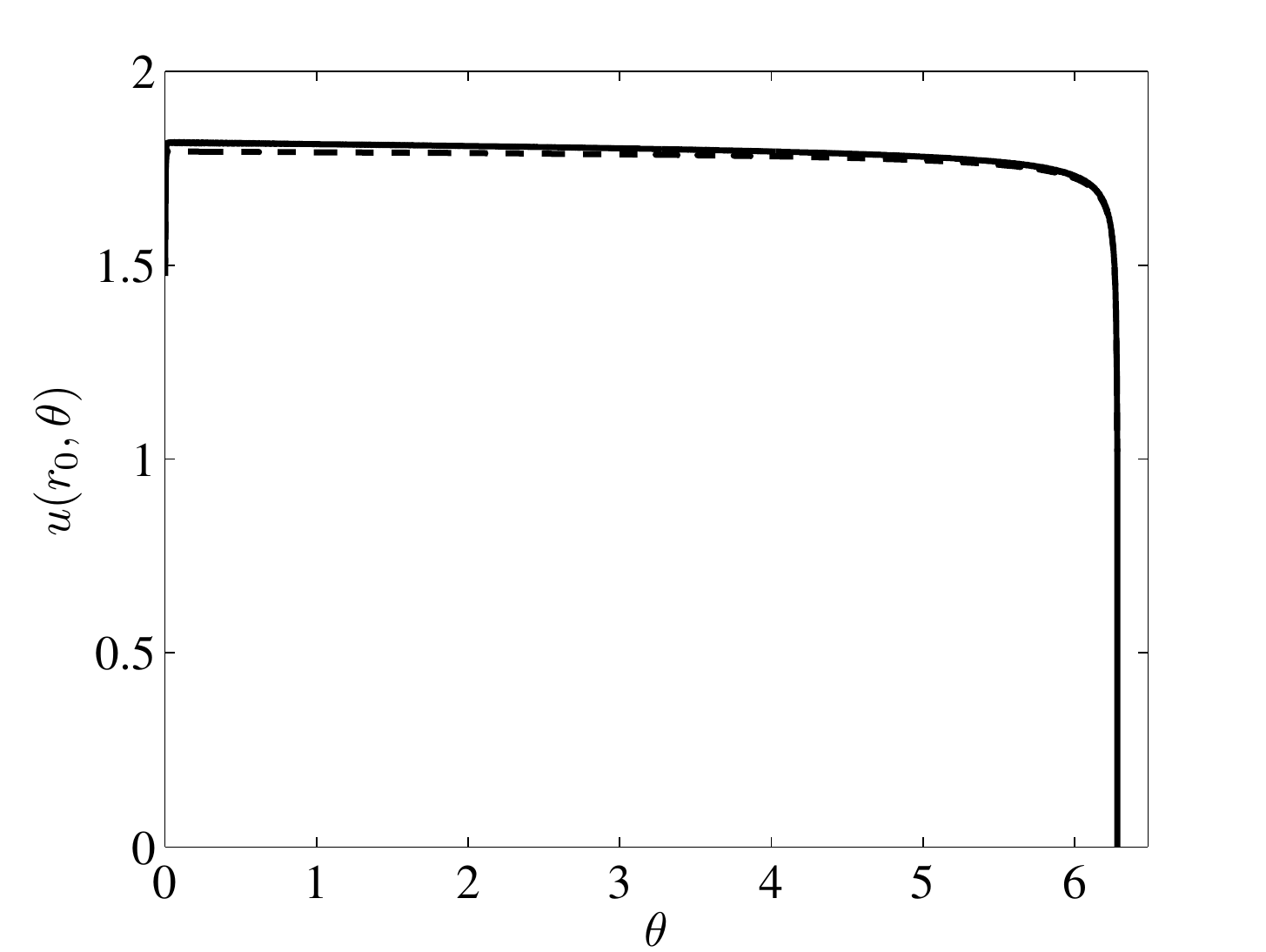}
        }}
    \caption{(a) Leading order asymptotic solution $u(x,y)$ of \eqref{heateqn} with $\omega = 1000$, $\varepsilon = 1\times 10^{-4}$, and $r_0 = 0.6$ as constructed from \eqref{Gcompfull} and \eqref{Hlomega}. The parameters are the same as those used in Figures \ref{flexcont_omega1000} and \ref{fig:usol_lomega}. Red (blue) regions indicate large (small) values of $u$. (b) The corresponding contour plot of $u(x,y)$. (c) The value of $u$ along the ring $r = r_0$. The solid curve is computed numerically from the series expansion, while the dashed curve is computed from \eqref{Gcompfull} and \eqref{Hlomega}.}
     \label{fig:usol_lomega_asymp}
  \end{center}
\end{figure}\end{empty}

Finally, with $H$ given in \eqref{Hlomega} and $\hat{H}$ defined by \eqref{Hhat}, we use \eqref{massH} to
calculate the mass

\begin{equation}  \label{masslomega}
M(r_0;\omega) = \pi \left\lbrack \frac{r_0^2}{2} - \log r_0 - \frac{3}{8} - 
\frac{1}{2}\log \left(\frac{\varepsilon\omega}{4}\right) - \frac{\gamma}{2}%
\right\rbrack + \mathcal{O}(\omega^{-1}) \,,
\end{equation}

\noI valid for $r_0\omega \gg 1$. Differentiating \eqref{masslomega} by $r_0$, we find that $r_0^{opt} =
1 $ as $\omega \to \infty$ with $\omega \ll \mathcal{O}(\varepsilon^{-1})$,
consistent with the results of Figure \ref{opt_r_0_vs_omega_withflex}. In Figure \ref%
{mass_lomega} for $\omega = 1000$ and $\varepsilon = 1\times 10^{-4}$, we
show a plot of the total mass as computed by \eqref{massseries} (solid) and %
\eqref{masslomega} (dashed). The circles are data from full numerical
solutions of \eqref{heateqn}. In the case of the former, the optimal value
of $r_0$ is slightly less than one, while the latter case indicates that $%
r_0=1$ is optimal. The discrepancy is likely due to the $\mathcal{O}%
(\omega^{-1})$ terms neglected in \eqref{masslomega}, and violation of the
assumption $1-r_0 \sim \mathcal{O}(1)$.

\begin{empty}\begin{figure}
\begin{center}
	\includegraphics[width=.5\textwidth]{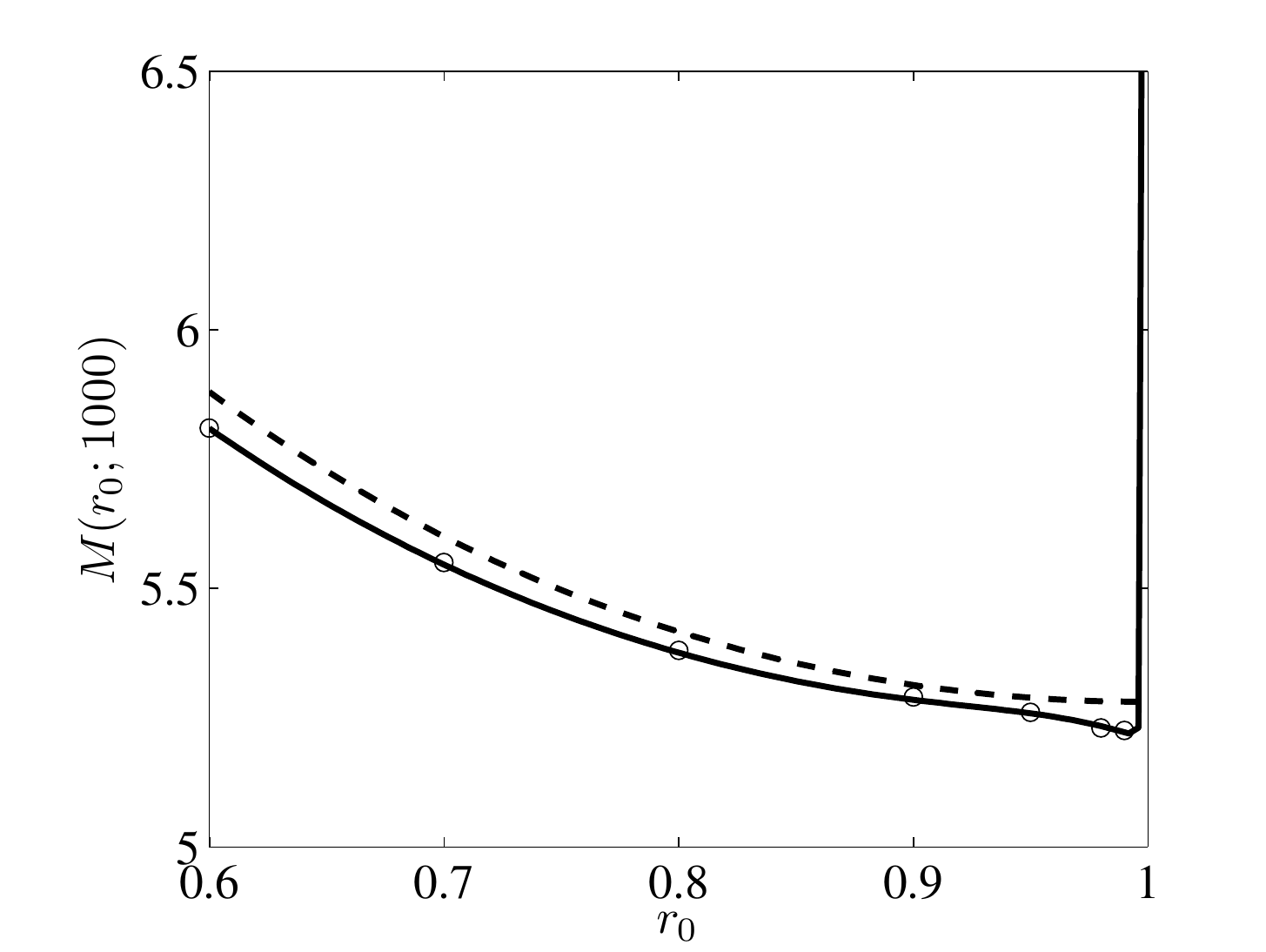}
	\caption{Total mass versus $r_0$ as computed from \eqref{massseries} (solid), \eqref{masslomega} (dashed), and the full numerical solution of \eqref{heateqn} (circles). The discrepancy near $r_0 = 1$ is likely due to the $\mathcal{O}(\omega^{-1})$ terms neglected in \eqref{masslomega} and violation of the assumption $1-r_0 \sim \mathcal{O}(1)$. Here, $\omega = 1000$ and $\ve = 1 \times 10^{-4}$. }
	\label{mass_lomega}
	\end{center}
\end{figure}\end{empty}\setcounter{equation}{0}

\section{The regime $\protect\omega \sim \mathcal{O}(\protect\varepsilon ^{-1})$}

\label{omegaeps}

In Figure \ref{opt_r_0_vs_omega_withflex}, we observe a transition from $r_0^{opt} \sim 1$ to $r_0^{opt} \sim 1/\sqrt{2}$ in the regime $\omega = \mO(\ve^{-1})$. In this section, we analyze this transition. Unlike the analysis of \S \ref{largeomega} in which we constructed an explicit solution for $u$ in the inner region, here we employ a boundary integral method to extract only the essential information required to determine the optimal radius of rotation. We first note from \S \ref%
{largeomega} that the elliptic layer of extent $\mathcal{O}(\omega^{-1})$
coincides with the inner layer of extent $\mathcal{O}(\varepsilon)$ when $%
\omega \sim \mathcal{O}(\varepsilon^{-1})$. The regime $\omega \sim \mO(\ve%
^{-1})$ is thus a distinguished regime not contained in the analysis of \S \S %
\ref{omegaO1} or \ref{largeomega}. Indeed, with $\omega = \ve^{-1}\omega_0$ with $\omega_0 = \mO(1)$, the inner equation for \eqref{heateqn}
becomes the radially asymmetric problem

\begin{equation}  \label{inneqasym}
u_{\xi\xi} + u_{\eta\eta} + \omega_0r_0u_\eta = 0 \,, \quad (\xi, \eta)
\notin \Omega_1 \,, \qquad u = 0 \,, \quad (\xi,\eta) \in \partial \Omega_1 \,, \qquad u \sim u_0(s_0) \enspace  \mbox{as} \enspace |(\xi,\eta)| \to \infty \,.
\end{equation}

\noI Here, $(\xi,\eta) = \ve^{-1}(x-x_0, y)$ and $\partial\Omega_1$ is the unit circle. Our goal is to compute the value of $u_0(s_0)$, which determines the value of $u$ on the ring $r = r_0$ in the outer problem. To do so, we first let $u = u_0 \mu(\xi,\eta) + u_0$ in \eqref{inneqasym} so that we have

\begin{equation}  \label{inneqasymmu}
\Delta \mu + s_0 \mu_\eta = 0 \,, \quad (\xi, \eta)
\notin \Omega_1 \,, \qquad \mu = -1 \,, \quad \pmb{\xi} \in \partial \Omega_1 \,, \qquad \mu \sim 0 \enspace  \mbox{as} \enspace |\pmb{\xi}| \to \infty \,; \quad s_0 \equiv r_0\omega_0 \,.
\end{equation}

\noI In \eqref{inneqasymmu}, $\pmb{\xi} = (\xi, \eta)$ and $\Delta$ denotes the Laplacian with respect to the $\pmb{\xi}$. To calculate $u_0(s_0)$, we use a boundary integral method to reformulate \eqref{inneqasymmu} as an integral equation for the normal derivative $\partial \mu / \partial n$ on $\partial \Omega_1$ (the same approach for solving \eqref{inneqasymmu} with different a geometry is adopted in \cite{choi2005steady} in the context of diffusion in the presence of steady two-dimensional potential flow around a finite absorber). The constant $u_0(s_0)$ is then calculated by integrating \eqref{heateqn} over $\Omega$ and imposing the solvability condition that the flux on $\partial \Omega_1$ must be equal to $-\pi$. That is, $u_0(s_0)$ is given by

\BE	\label{u0}	
	u_0(s_0) = \frac{-\pi}{\Phi(s_0)} \,; \qquad \Phi(s_0) \equiv \int_{\partial \Omega_1} \! \frac{\partial \mu}{\partial n} \, dS\,.
\EE

\noI The dependence of $u_0$ on $s_0$ is due to the appearance of $s_0$ as a parameter in the equation for $\mu$ in \eqref{inneqasymmu}.

We first consider the Green's function for the adjoint problem

\BE \label{Gadjeq}
	\Delta \tilde{G} - s_0 \tilde{G}_\eta = \delta(\pmb{\xi} - \mathbf{z}) \,, \quad \tilde{G} \to 0 \enspace \mbox{as} \enspace |\pmb{\xi}| \to \infty \,; \qquad \mathbf{z} = (z_1, z_2) \,.
\EE

\noI The solution of \eqref{Gadjeq} is given by

\BE
 	\tilde{G}(\pmb{\xi};\mathbf{z}) = -\frac{1}{2\pi} e^{\frac{s_0}{2} (\eta - z_2)} K_0\left(\frac{s_0}{2}|\pmb{\xi} - \mathbf{z}|\right) \,.
\EE

\noI Next, we multiply \eqref{inneqasymmu} by $\tilde{G}$, integrate by parts, use Green's second identity, and apply the boundary conditions to obtain that

\BE \label{mu}
	\mu(\rho\cos\phi, \rho\sin\phi) = \int_0^{2\pi} \! \left. \frac{\partial \tilde{G}}{\partial \tilde{r}} \right |_{\tilde{r}=1}  d\tilde{\theta} + \int_0^{2\pi} \! \left(\tilde{G} \, \frac{\partial \mu}{\partial \tilde{r}} \right)_{\tilde{r} = 1}  d\tilde{\theta} - s_0  \int_0^{2\pi} \! \tilde{G} \, \Big|_{\tilde{r} = 1} \sin\tilde{\theta} \, d\tilde{\theta} \,.
\EE

\noI In \eqref{mu}, we have made the substitutions $(\xi, \eta) \to (\tilde{r}\cos\tilde{\theta}, \tilde{r}\sin\tilde{\theta})$ and $(z_1, z_2) \to (\rho\cos\phi, \rho\sin\phi)$. To obtain an integral equation for $\sigma(\tilde{\theta}) \equiv \partial \mu/\partial \tilde{r}$ on $\partial \Omega_1$, we impose the condition in \eqref{inneqasymmu} that $\mu = -1$ on $(\rho,\phi) = (1, \phi)$ for $\phi \in \lbrack 0, 2\pi)$. Imposing this condition in \eqref{mu}, we obtain the integral equation for $\sigma(\tilde{\theta})$

\BE \label{inteq}
	-\int_0^{2\pi} \! \left. \tilde{G} \right |_{\tilde{r} = \rho = 1} \sigma(\tilde{\theta}) \,d\tilde{\theta} = \frac{1}{2} + \int_0^{2\pi} \! \left. \frac{\partial \tilde{G}}{\partial \tilde{r}} \right |_{\tilde{r} = \rho =1}\!d\tilde{\theta} \,- s_0  \int_0^{2\pi} \! \tilde{G} \, \Big|_{\tilde{r} = \rho = 1} \sin\tilde{\theta} \, d\tilde{\theta} \,; \qquad \sigma(\tilde{\theta}) \equiv \left. \frac{\partial \mu}{\partial \tilde{r}}\right |_{\tilde{r} = 1} \,.
\EE

\begin{figure}[htbp]
  \begin{center}
    \mbox{
    \subfigure[$u_0$ and $u_0^\prime$] 
        {\label{u0andu0prime}
        \includegraphics[width=.32\textwidth]{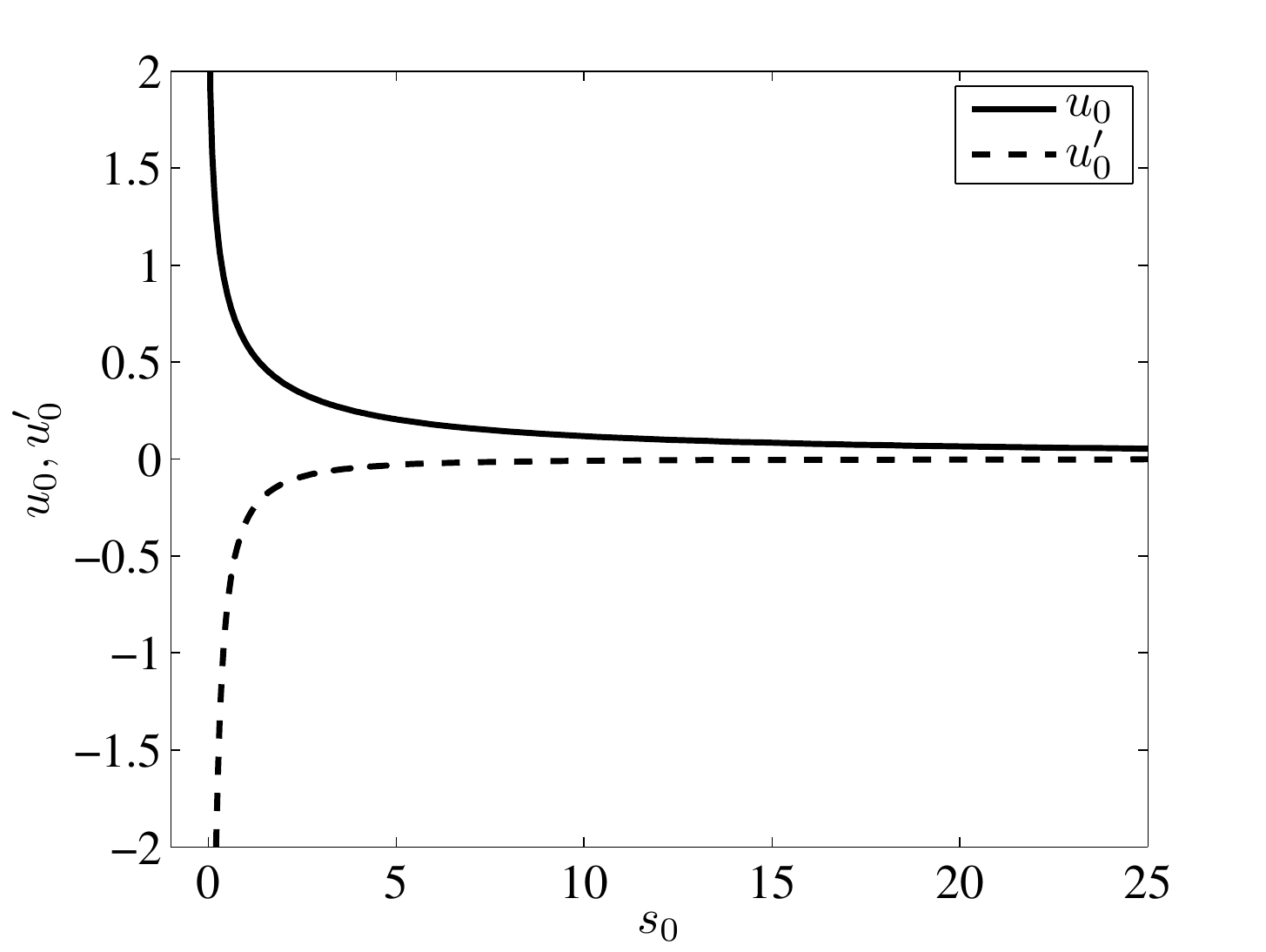}
        }   
    \subfigure[mass versus $r_0$ with $\omega_0 = 4$] 
        {\label{omegaeps_mass}
        \includegraphics[width=.32\textwidth]{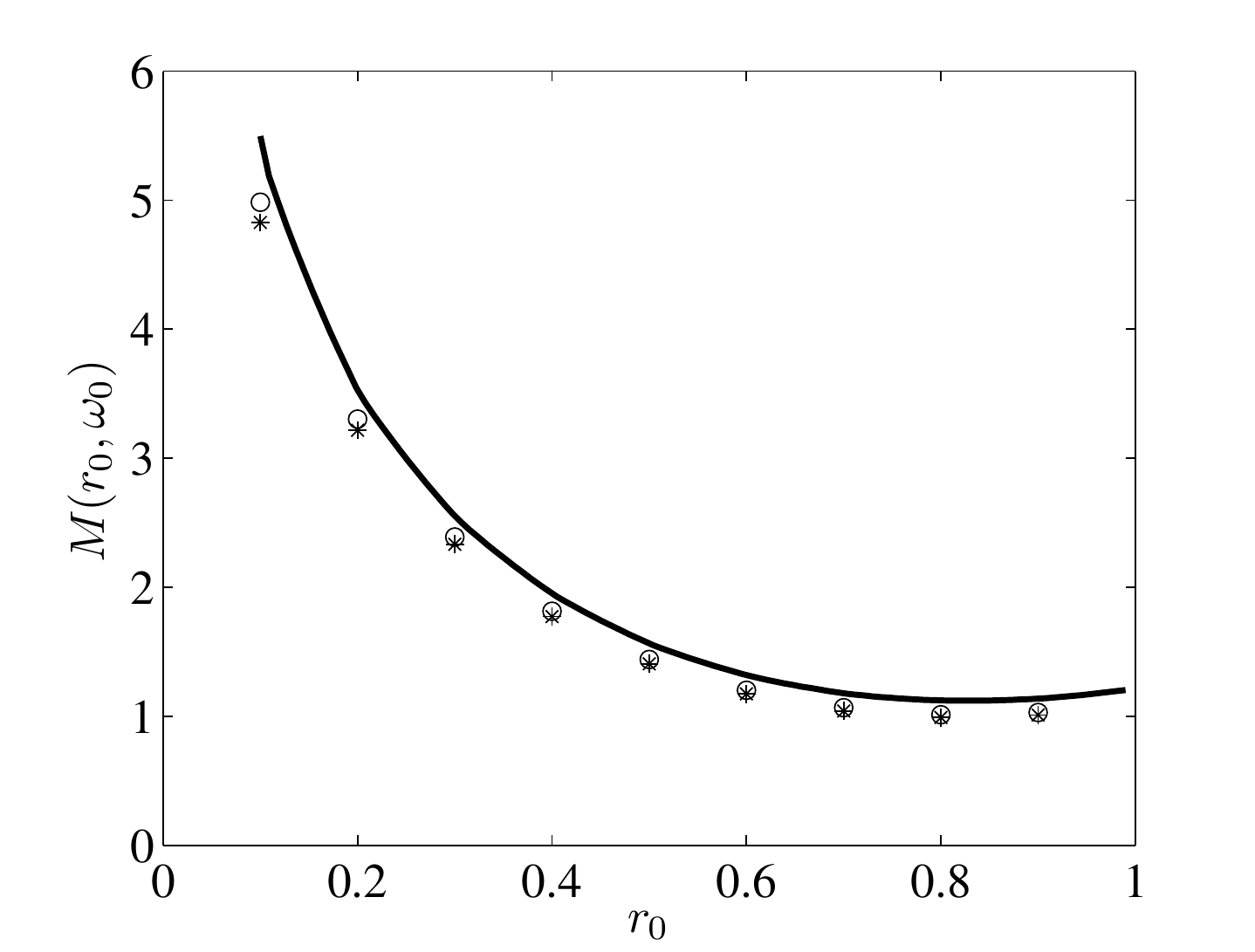}
        }
    \subfigure[$r_0^{opt}$ versus $\omega_0$] 
        {\label{omegaeps_r0opt}
        \includegraphics[width=.32\textwidth]{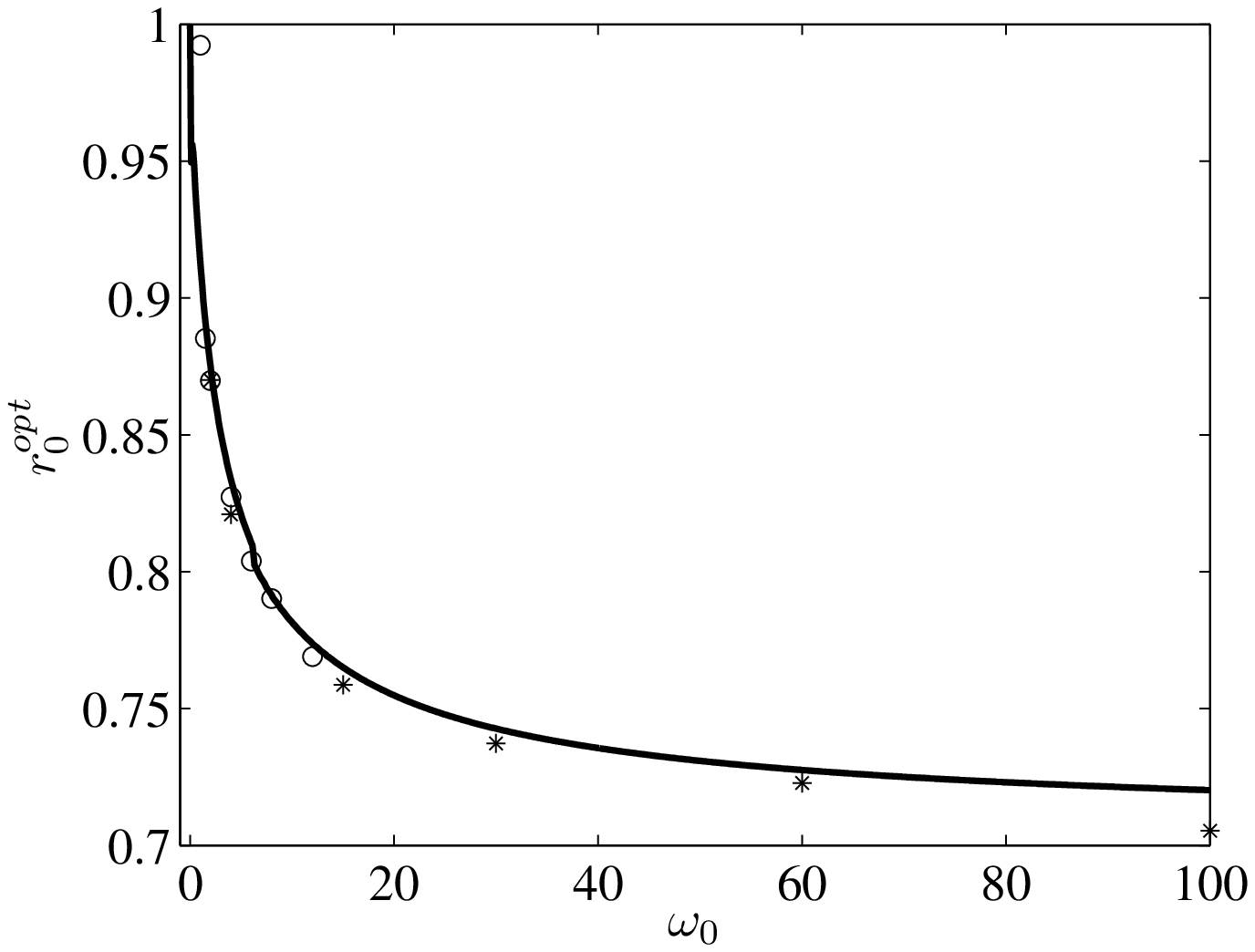}
        }}
    \caption{(a) Numerically computed values of $u_0(s_0)$ and $u_0^\prime(s_0)$ for a range of $s_0$. Here, $s_0$ is defined in \eqref{inneqasymmu}. (b) Comparison of $M(r_0;\omega)$ as given by \eqref{Meps} with $\omega_0 = 4$ (solid curve) and the numerical solution of \eqref{heateqn} with $\ve = 2.5\times 10^{-3}$, $\omega = 1600$ (circles) and $\ve = 5\times 10^{-3}$, $\omega = 800$ (stars). (c) Comparison of the asymptotic prediction of $r_0^{opt}$ obtained by solving \eqref{dMdr0} (solid curve), and results from numerical solutions of \eqref{heateqn} with $\ve = 1\times 10^{-3}$ (circles) and $\ve = 5\times10^{-3}$ (stars). Here, $\omega_0 = \ve\omega$. }
     \label{fig:inneru0}
  \end{center}
\end{figure}

\noI The $1/2$ term on the right-hand side of \eqref{inteq} is a result of evaluating $\mu$ on the boundary $\Omega_1$ and thus integrating over only half of the delta function in its Green's function representation. For a range of $s_0$, we finally solve \eqref{inteq} as a linear system for $\sigma(\tilde{\theta})$ at discrete values of $\tilde{\theta}$ and use \eqref{u0} to calculate $u_0(s_0)$. The results of this calculation are shown in Figure \ref{u0andu0prime}, where we show both $u_0$ (solid) and $u_0^\prime$ (dashed).

The leading order far-field behavior of $\mu$ in \eqref{inneqasymmu} must be independent of the geometry of $\Omega_1$ so that in the far field, $\mu$ must be a constant multiple of $F = K_0(s_0|\pmb{\xi}|/2)e^{-s_0\eta/2}$, which satisfies \eqref{inneqasymmu} in the case where $\Omega_1$ is the contour on which $F = -1$. Then, for the internal parabolic layer of extent $\mO(\sqrt{\ve})$, we introduce the rescaled variables $\hat{r} = \sqrt{\omega_0}(r-r_0)/\sqrt{\ve}$ and $\hat{\theta} = 2\pi - \theta$ and follow the analysis of \S \ref{largeomega} to obtain from \eqref{heatpde}



\begin{equation}  \label{playereqasym}
u_{\hat{r}\hat{r}} - u_{\hat{\theta}} = 0 \,, \qquad u(\hat{r},0) = \sqrt{\ve}c\delta(\hat{r}) + u_0 \,,
\end{equation}

\noI where $c$ is an $\mO(1)$ constant and $u_0$ is given by \eqref{u0}. The solution in the internal layer
then follows directly from \eqref{playersol}. As discussed in \S \ref%
{largeomega}, the only term in the inner and internal layers relevant to the
leading order expression for $M(r_0;\omega)$ is the constant term $u_0$, which is
required to uniquely determine the leading order outer solution.

For the outer equation with $\omega = \omega_0/\ve$ in \eqref{heatpde}, the
leading order behavior of the solution must be radially symmetric. With the matching condition $u = u_0$ on the ring $r = r_0$, we
therefore obtain the radially symmetric problem

\begin{empty}\bes \label{heatout}
\begin{equation} \label{heatoutpde}
	u_{rr} + \frac{1}{r}u_r  + 1 = 0 \,, \quad \mathbf{x} \in \Omega \setminus \left\lbrace\mathbf{x}: |\mathbf{x}| = r_0 \right\rbrace \,; 
\end{equation}
\begin{equation}	\label{heatoutbc}
	 u_r = 0 \,, \quad \mathbf{x} \in \partial \Omega \,;  \qquad u \enspace \mbox{bounded as} \enspace r \to 0 \,; \quad u = u_0\,, \quad |\mathbf{x}| = r_0 \,,
\end{equation}
\ees\end{empty}

\noI with $u_0$ determined empirically from Figure \ref{u0andu0prime}. The dependence of $u$ on $\omega_0$ is through that of $u_0$ on $s_0 = r_0\omega_0$. The solution to \eqref{heatout} is

\begin{equation}  \label{usoleps}
u(r) = \frac{r_0^2 - r^2}{4} + u_0(s_0) + \frac{1}{2}\Theta(r-r_0)\log\left(%
\frac{r}{r_0}\right) \,,
\end{equation}

\noI where $\Theta(r)$ is the Heaviside step function. Integrating $u$ in \eqref{usoleps} over the domain $\Omega$, the leading order expression for the mass $M(r;\omega)$ may then be written

\begin{equation}  \label{Meps}
M(r_0;\omega) = M(r_0;\omega_0) = \pi \left\lbrack \frac{r_0^2}{2}-\frac{3}{8%
}-\frac{1}{2}\log(r_0) + u_0(r_0\omega_0) \right\rbrack \,; \qquad \omega_0
\equiv \ve\omega \,.
\end{equation}

\noI A comparison of $M(r_0;\omega)$ as given by \eqref{Meps} versus numerical results as computed from \eqref{heateqn} is shown in Figure \ref{omegaeps_mass} for $\omega_0 = 4$. The solid curve is calculated from \eqref{Meps} with $u_0$ given in Figure \ref{u0andu0prime}, while the circles and stars are from numerical solutions of \eqref{heateqn} with $\ve = 2.5\times 10^{-3}$ and $\ve = 5\times 10^{-3}$, respectively. The agreement between the circles and stars confirms the analytical result that $M$ is independent of $\ve$ for fixed $\omega_0$.

To calculate the optimal radius $r_0^{opt}$, we set to zero the derivative of $M(r_0;\omega_0)$ in \eqref{Meps} with respect to $r_0$. That is, the optimal radius $r_0^{opt}$ satisfies

\begin{equation}  \label{dMdr0}
r_0^{opt} - \frac{1}{2r_0^{opt}} + \omega_0u^\prime(s_0)= 0 \,.
\end{equation}

\noI Solving \eqref{dMdr0} numerically for various $\omega_0$, we obtain the
solid curve in Figure \ref{omegaeps_r0opt}. The circles and stars in Figure \ref{omegaeps_r0opt} are results obtained from numerical
solutions of \eqref{heateqn} with $\ve = 1\times 10^{-3}$ (circles) and $\ve %
= 5\times 10^{-3}$ (stars) and $\omega = \omega_0/\ve$. We make several
remarks. First, the agreement between the circles and the stars confirms
that $r_0^{opt}$ is a function only of the product $\ve\omega \equiv
\omega_0 $, not $\ve$ or $\omega$ individually. The size of the trap and the
frequency of rotation may then be said to be in balance, as doubling one
parameter has the same effect on the optimal radius as halving the other. This is
in contrast to the $\omega \sim \mO(1)$ regime in which $r_0^{opt}$ depends
only on $\omega$ and not $\ve$.

Second, the numerical results appear to diverge from the asymptotics for large $\omega
_{0}$. This may be due to the fact that the analysis assumes $\omega = \ve^{-1}\omega_0$ with $\omega_0 = \mO(1)$; for $\omega _{0}\gg 1$, we observe numerically that $%
r_{0}^{opt} $ asymptotes to a value slightly below the line $r_{0}^{opt}=1/%
\sqrt{2}$, as predicted by \eqref{r0inf}. Lastly, we illustrate in Figure %
\ref{fig:mass_vs_r0_omegaeps} the transition from $r_{0}^{opt}\approx 1$ in
the $1\ll \omega \ll \mO(\ve^{-1})$ regime to that shown in Figure \ref%
{omegaeps_r0opt} for the $\omega \sim \mO(\ve^{-1})$ regime. Figures \ref%
{mass_vs_r0_omega1000} and \ref{mass_vs_r0_omega1500}, generated from
numerical solutions of \eqref{heateqn} with $\omega _{0}=1$ (left) and $%
\omega _{0}=1.5$ (right), each show two local minima in the relationship $%
M(r_{0};\omega _{0})$. The minimum located near $r_{0}=1$ is that which has
persisted from the $1\ll \omega \ll \mO(\ve^{-1})$ regime, while the one
located away from $r_{0}=1$ is formed as $\omega $ enters the $\omega \sim %
\mO(\ve^{-1})$ regime. The results in Figure \ref{fig:mass_vs_r0_omegaeps}
then suggest that the transition occurs at some $\omega _{0}^{(c)}\in
(1,1.5) $ at which the value of $M$ at the left local minimum dips below
that of $M$ at the right minimum. The location of the left minimum continues
to decrease in $r_{0}$ for increasing $\omega _{0}$, as illustrated by
Figure \ref{omegaeps_r0opt}. The leading order expression for $%
M(r_{0};\omega )$ in \eqref{Meps} does not capture the right minimum, as its
derivation relies on an $\mO(1)$ distance between the boundaries of the trap
and domain.

\begin{empty}\begin{figure}[htbp]
  \begin{center}
    \mbox{
    \subfigure[mass versus $r_0$ with $\omega_0 = 1$] 
        {\label{mass_vs_r0_omega1000}
        \includegraphics[width=.5\textwidth]{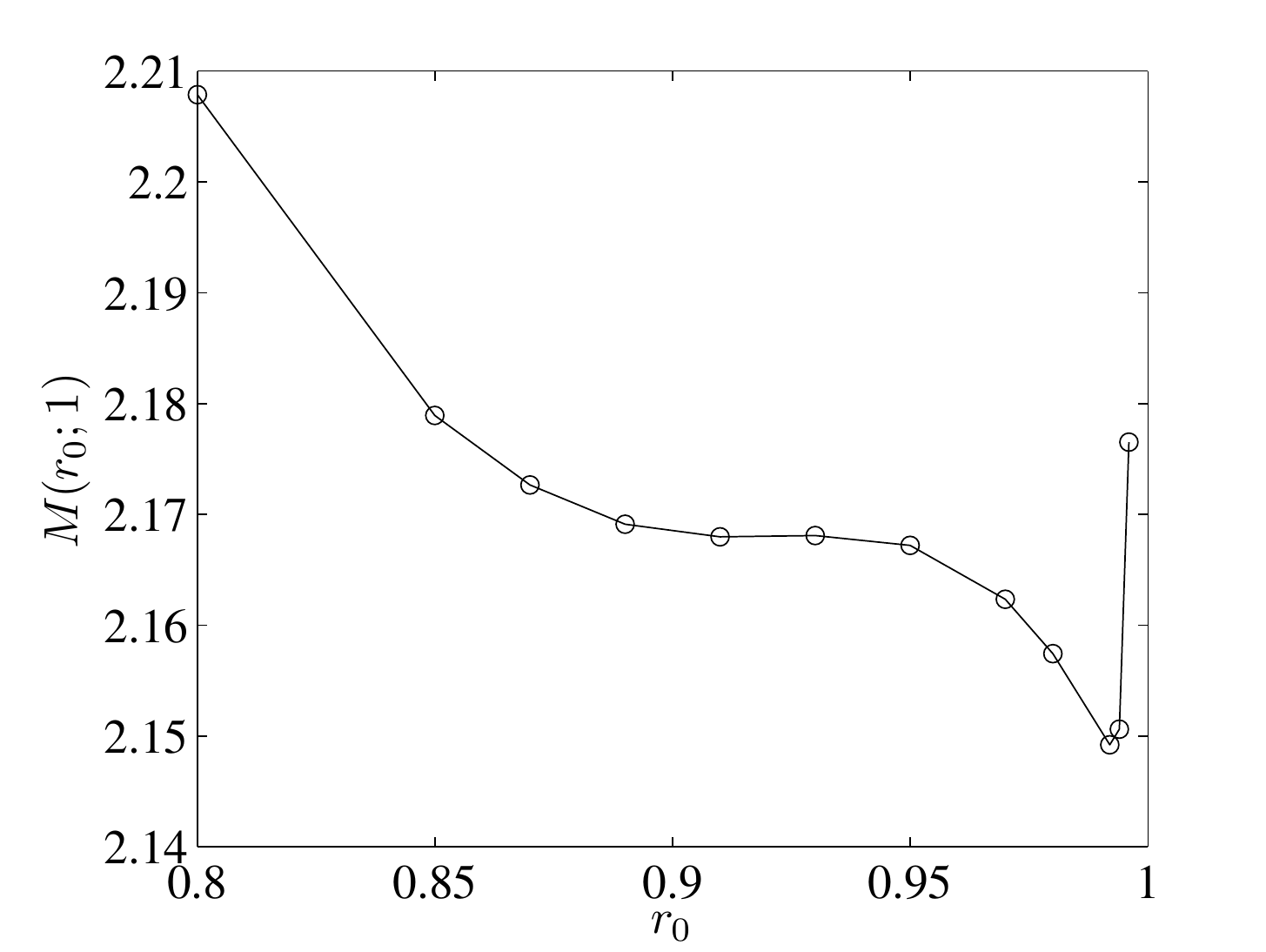}
        }   
    \subfigure[mass versus $r_0$ with $\omega_0 = 1.5$] 
        {\label{mass_vs_r0_omega1500}
        \includegraphics[width=.5\textwidth]{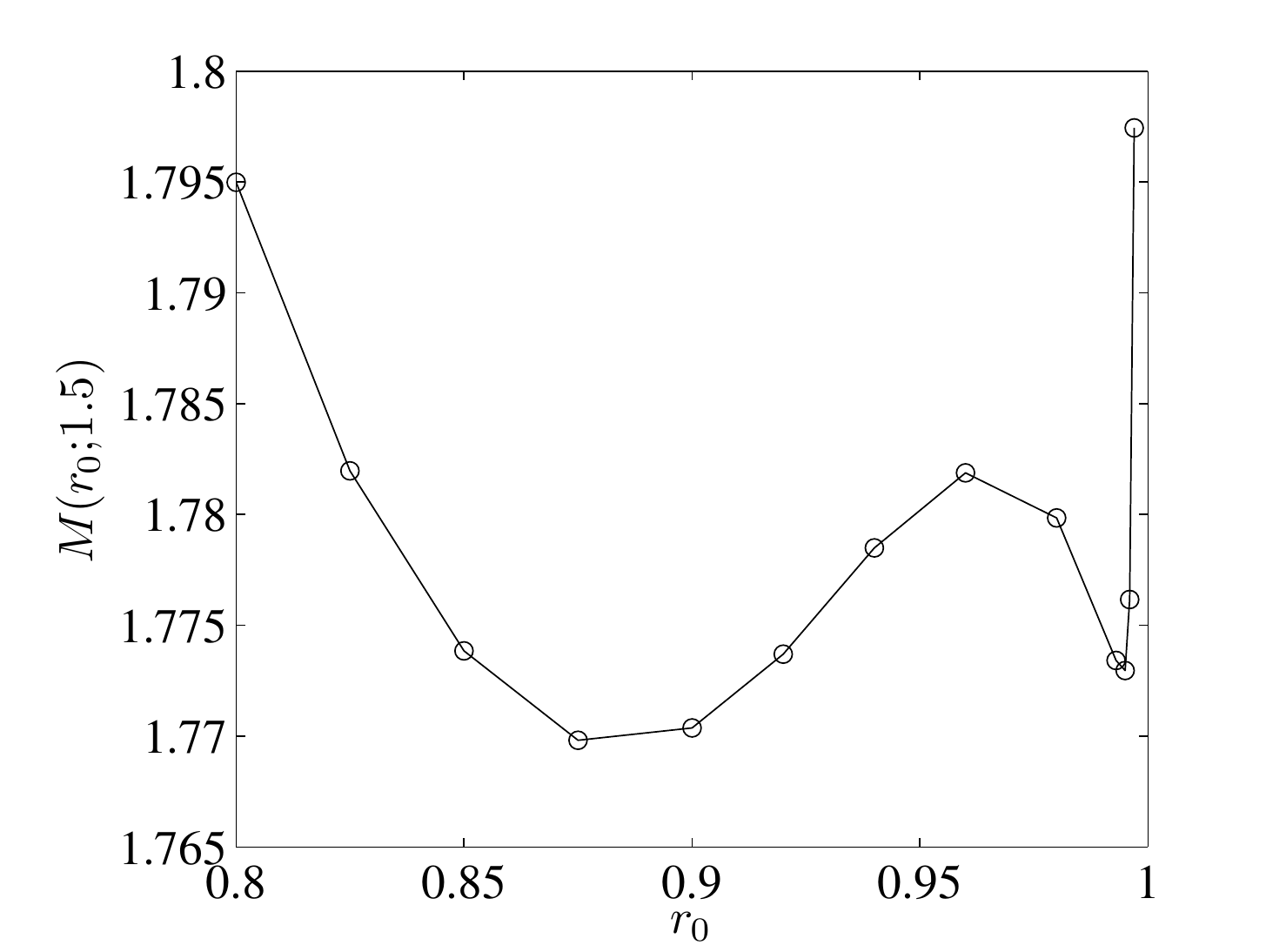}
        }}
    \caption{The relationship $M(r_0;\omega_0)$, generated from numerical solutions of \eqref{heateqn} with (a) $\omega_0 = 1$ and (b) $\omega_0 = 1.5$. Here, $\varepsilon = 1\times 10^{-3}$. In (a), the local minimum away from $r_0 = 1$ is less optimal than that near $r_0 = 1$. In (b), the situation reverses whereby the left local minimum dips below that at the right. The location of the left minimum continues to decrease in $r_0$ for increasing $\omega_0$, as illustrated by Figure \ref{omegaeps_r0opt}.}
     \label{fig:mass_vs_r0_omegaeps}
  \end{center}
\end{figure}\end{empty}

\section{Discussion}

\label{discussion}

\begin{empty}\begin{figure}[tbp]
  \begin{center}
    \mbox{
    \subfigure[$r_0^{opt}$ versus $r_0\omega$] 
        {\label{opt_r_0_vs_speed}
        \includegraphics[width=.32\textwidth]{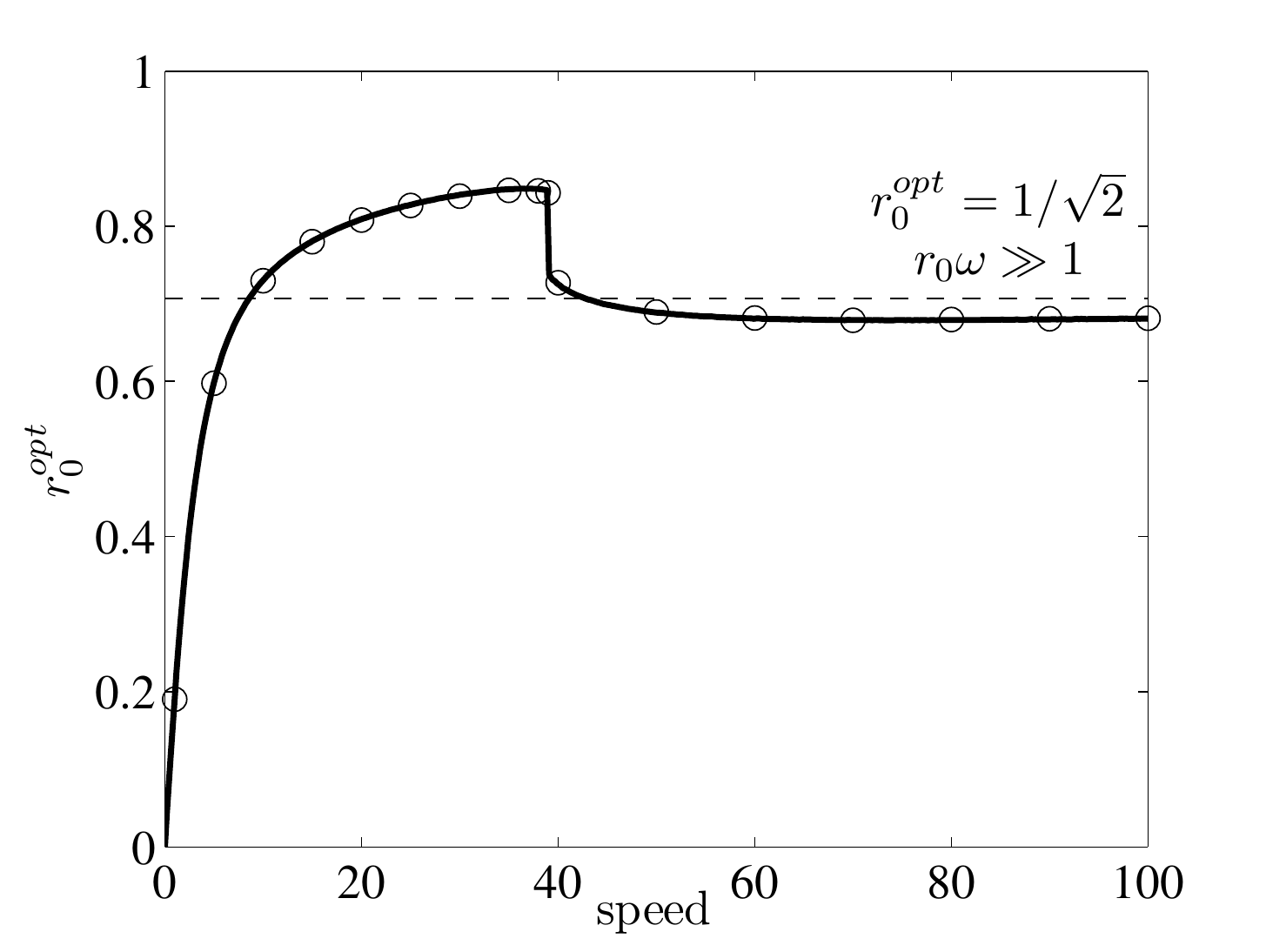}
        }   
    \subfigure[mass versus $r_0$ with $r_0\omega = 39$] 
        {\label{mass_vs_r0_speed39}
        \includegraphics[width=.32\textwidth]{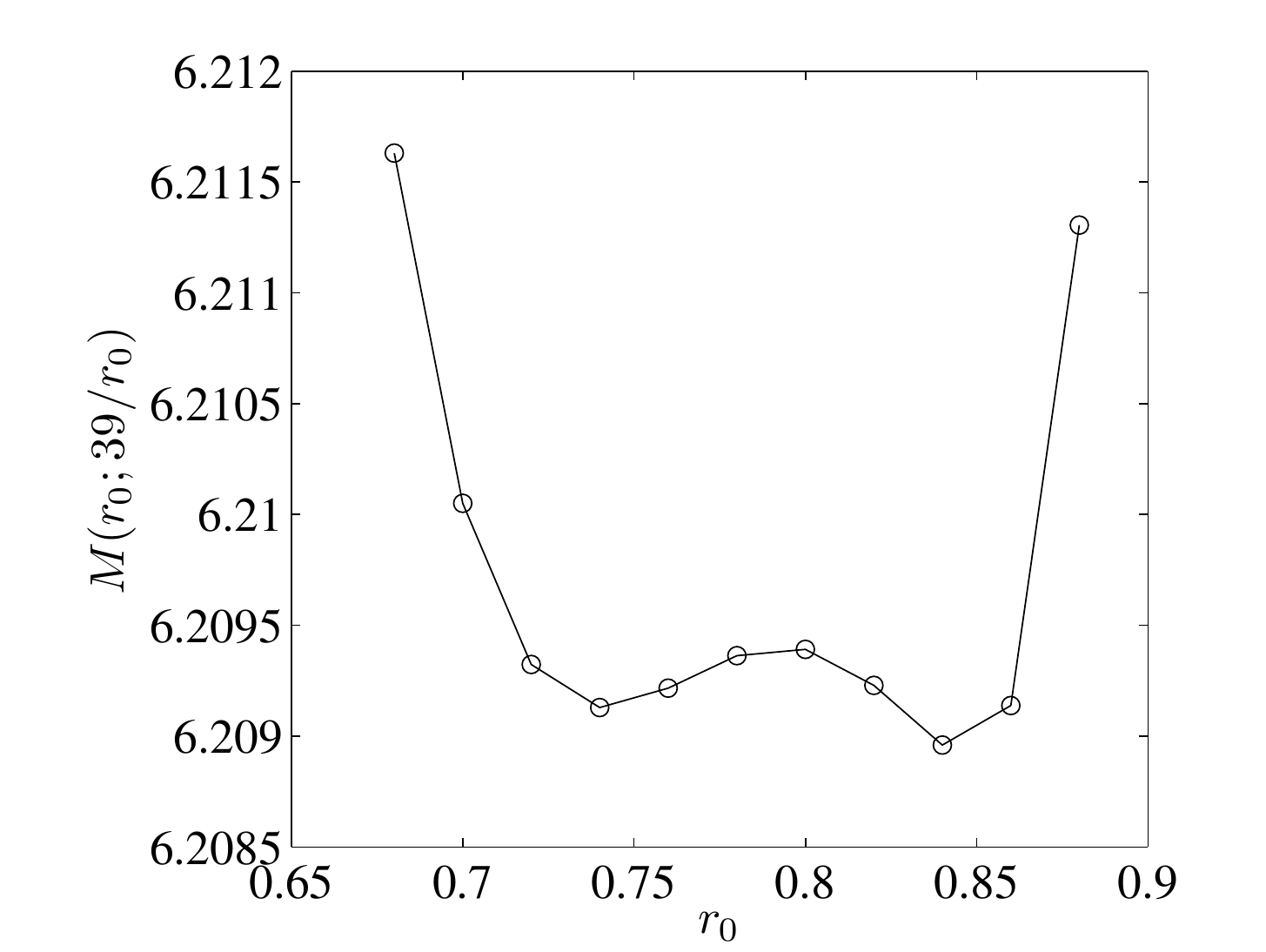}
        }
    \subfigure[mass versus $r_0$ with $r_0\omega = 40$] 
        {\label{mass_vs_r0_speed40}
        \includegraphics[width=.32\textwidth]{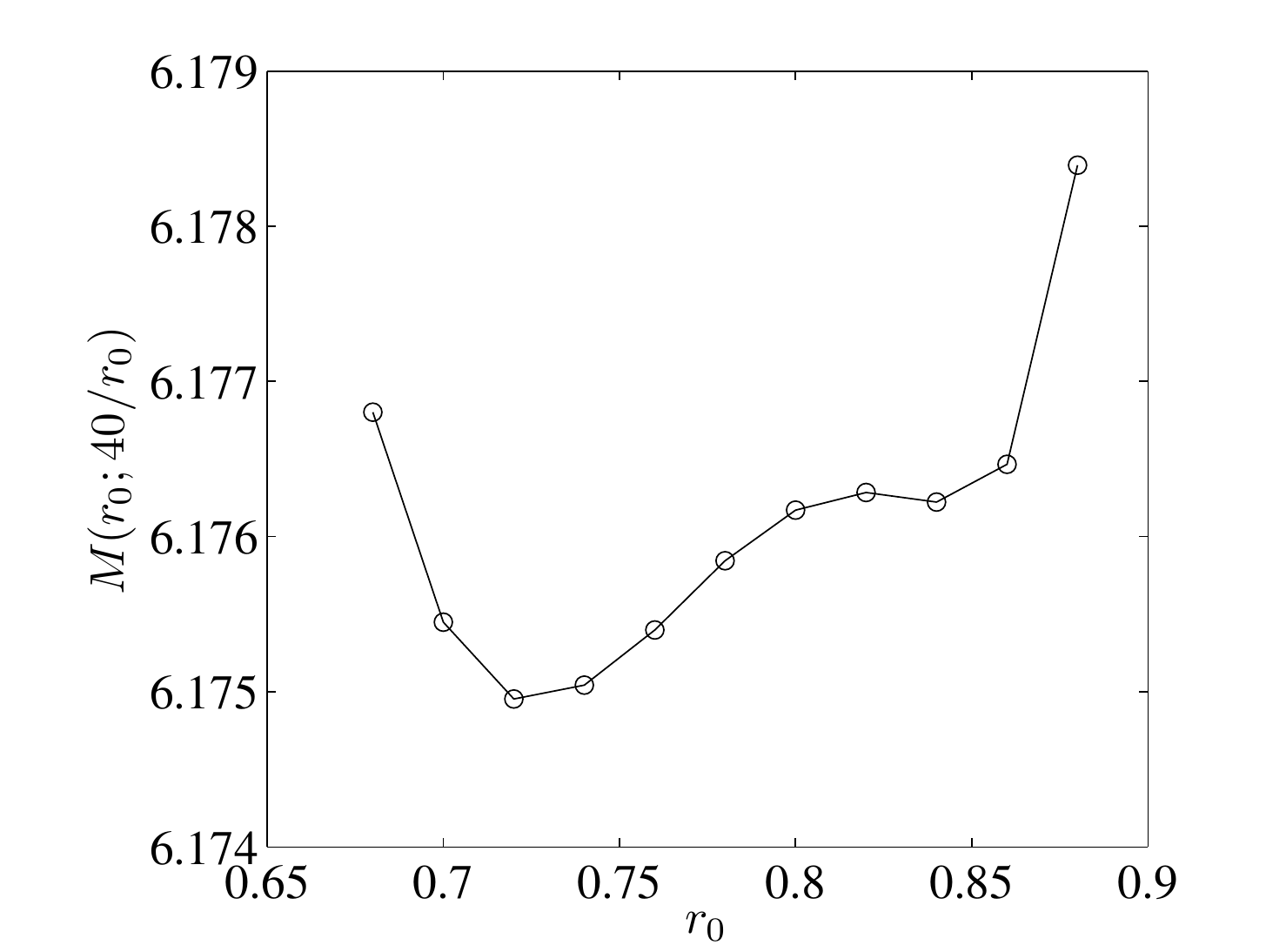}
        }}
    \caption{(a) Asymptotic (solid) and numerical (circles) results for $r_0^{opt}$ for a range of speed $r_0\omega$. Unlike the case with constant $\omega$, no bifurcation is observed so that $r_0^{opt} > 0$ for any $r_0\omega > 0$. The optimal radius reaches a maximum of $r_0^{opt} \approx 0.85$ when $r_0\omega \approx 39$ before a transition occurs to a smaller optimal radius. The transition is illustrated in the mass versus $r_0$ plots shown in (b) and (c) for $r_0\omega = 39$ and $r_0\omega = 40$, respectively. Two local minima are present. As $r_0\omega$ increases, the left minimum dips below that at the right. The results were obtained from numerical solutions of \eqref{heateqn} with $\varepsilon = 1\times10^{-3}$. The same transition may also be observed from asymptotic results.}
     \label{fig:opt_r_0_vs_speed}
  \end{center}
\end{figure}\end{empty}

We have studied the average MFPT\ over a unit disk domain with a small
rotating trap. By taking advantage of the radial geometry, we were able to
extend the asymptotic techniques that were developed for stationary traps
to the problem of a moving trap. With this radial symmetry, we showed that minimizing the average MFPT was equivalent to minimizing the steady-state mass of a simple diffusive system with uniform feed and a rotating Dirichlet trap.

Several surprising \textquotedblleft bifurcations\textquotedblright\ emerge.
For small angular velocities $\left( 0\leq \omega <\omega _{c}\approx
3.026\right) ,$ the trap should be located at the center of the disk in
order to minimize the average MFPT. When $\omega $ is large but fixed with $%
\varepsilon \rightarrow 0$ (that is, $1\ll \omega \ll \mO(\varepsilon^{-1})$), the
trap should be located very close to the boundary of the disk. On the other
hand when $\varepsilon $ is small but fixed with $\omega \rightarrow \infty $
(that is, $\omega \gg \mO(\varepsilon^{-1}) ),$ the optimal trap radius approaches $1/%
\sqrt{2}.$ In this case, the path taken by the trap subdivides the unit disk
into two regions of equal area. Because $\omega $ is so large, such a regime
is equivalent to having a trapping boundary all along the length of the path:\ that
is, from the particle point of view, the trap appears to be simultaneously
present all along its path. Most interestingly, there is a discontinuous
``jump''\ in the optimal radius (at around $\omega \approx 10^{3}$ in Figure %
\ref{opt_r_0_vs_omega_withflex}) as $\omega $ is increased. This "jump"\
occurs due to the presence of two local minima, one of which overtakes
the other as $\omega $ is increased; see also Figure \ref%
{fig:mass_vs_r0_omegaeps}.

The most intricate regime is precisely the transition regime $\omega
=\mathcal{O}(\varepsilon^{-1} )$ where the "jump" occurs. In \S \ref{omegaeps} we used a
boundary integral method approach to compute the asymptotics of the optimal radius in the regime $\omega =\mathcal{O}(\varepsilon^{-1})$. By doing so, we captured the transition between the regime $1\ll \omega \ll \mO(\varepsilon^{-1})$ in which $r_0^{opt} \to 1$, and the regime $\Omega \gg \mO(\varepsilon^{-1})$ in which $r_0^{opt} \to 1/\sqrt{2}$. 

The moving trap is very closely related to problems involving moving sources
for the diffusion equation; see for example \cite{akbari2011geometrical} and
references therein. Some applications include welding \cite{cline1977heat},
calculation of heat flux generated by friction in a pin-on-disc tribometer 
\cite{laraqi2009temperature}, and welding with CO$_{2}$ lasers \cite%
{akbari2011geometrical}.

Throughout the paper, we considered the problem of computing the optimal
radius as a function of angular velocity $\omega.$ Equally, it is
interesting to see how the optimal radius depends on the \emph{speed }$%
s=r_{0}\omega.$ This dependence is shown in Figure \ref{fig:opt_r_0_vs_speed}%
. As with Figure \ref{opt_r_0_vs_omega_withflex}, note that the optimal
radius approaches $r_{0}\sim 1/\sqrt{2}$ for large $\omega , $ as well as
the presence of the ``jump''\ near $\omega \approx 40$ independent of $\ve$. Two notable differences are that $r_0^{opt}$ does not make an asymptotic approach to $1$ for large speed, and there is also no ``bifurcation''\ near the origin:\ the optimal $r_{0}$ is strictly
positive regardless of how small $s$ is. Note that the $r_{0}^{opt} \sim 1/\sqrt{2}$ result in both the $1 \ll \omega \ll \mO(\varepsilon^{-1})$ and $\omega \sim \mO(\varepsilon^{-1})$ regimes may be inferred from \eqref{masslomega} and \eqref{Meps} by replacing $\omega$ and $\omega_0$ by $s/r_0$ and $\varepsilon s/r_0$, respectively, and differentiating the resulting expression with $s$ held constant. The same result for the $\omega \gg \mathcal{O}(\varepsilon^{-1})$ regime is immediate from \eqref{masssym}.

Of course, the problem we studied has a very special geometry and it is an
open question to consider the obvious generalizations:\ a non-circular
domain, more complex trap motion (with or without a stochastic component), multiple traps, etc. On the other hand,
this simple setting allows for a detailed analysis which shows that even a
very simple situation has a surprisingly rich structure, with several
different transitions depending on the relative strengths of the trap radius 
$\varepsilon $ and its rotation rate $\omega$. As such, it provides a good
test case for future studies of MFPT with moving traps.

\subsubsection*{Acknowledgments}

J. C. Tzou was supported by an AARMS Postdoctoral Fellowship. T.
Kolokolnikov is supported by NSERC discovery and NSERC accelerator grants. We thank Michael Ward for useful discussions and suggestions.

\bibliographystyle{elsart}
\bibliography{moving_arx}

\end{document}